\documentclass[10pt]{amsart}
\usepackage[backend=biber,
            isbn=false,
            doi=false,
            maxbibnames=5,
            style=alphabetic,
            citestyle=alphabetic]{biblatex}
\usepackage{url}
\renewbibmacro{in:}{}
\bibliography{refs.bib}

\usepackage{macros}
\usepackage{multirow}

\usepackage[foot]{amsaddr}

\makeatletter
\def\paragraph{\@startsection{paragraph}{4}%
  \z@\z@{-\fontdimen2\font}%
  {\normalfont\bfseries}}
\makeatother

\title{Twists of Superconformal Algebras}
\author{Chris Elliott \and Owen Gwilliam \and Matteo Lotito}

\begin{document}

\begin{abstract}
We take first steps toward a theory of ``conformal twists'' for superconformal field theories in dimension 3 to 6,
extending the well-known analysis of twists for supersymmetric theories. 
A conformal twist is a square-zero odd element in the superconformal Lie algebra,
and we classify all twists and describe their orbits under the adjoint action of the superconformal group.
We work mostly with the complexified superconformal algebras, unless explicitly stated otherwise; 
real forms of the superconformal algebra may have important physical implications, but we only discuss these subtleties in a few special cases.
Conformal twists can give rise to interesting subalgebras and protected sectors of operators in a superconformal field theory,
with the Donaldson--Witten topological field theory and the vertex operator algebras of 4-dimensional $\mathcal{N}{=}\hspace{.1em}2$ SCFTs being prominent examples.
To obtain mathematical precision, we explain how to extract vertex algebras and $\mathbb E_n$ algebras from a twisted superconformal field theory using factorization algebras.  
\end{abstract}

\keywords{Super Lie algebra, conformal geometry, superconformal field theory, nilpotence variety, associated variety, vertex algebra, factorization algebra}

\maketitle

\tableofcontents

\section{Introduction}

In this paper we explore the notion of supersymmetric twisting in the context of a superconformal field theory,
with a focus on what can be seen directly from superconformal algebras.
The questions we pursue in this paper are motivated by physics but connect with contemporary work in representation theory,
and in this introduction we aim to situate our work in both subjects.
  
The idea of twisting a supersymmetric theory goes back to Witten \cite{Witten_1988},
who introduced it as a recipe for constructing topological quantum field theories,
and it has provided a number of examples with rich consequences in physics and mathematics.
We wish to emphasize a more abstract formulation of the twisting procedure: 
given a field theory with an action of a super Lie algebra $\mf g$ and an odd element $\mc Q \in \mf g$ satisfying $[\mc Q, \mc Q] = 0$, 
there is a deformation of the theory by the action of $\mc Q$.  
For example, if $H_{\mc Q}$ is the Hamiltonian associated to $\mc Q$,
we may replace the algebra of local operators $\mr{Op}_{\mr{loc}}$ by its $\mc Q$-cohomology
\[\mr{Op}_{\mr{loc}}^{\mc Q} = \mr H^\bullet(\mr{Op}_{\mr{loc}}, \{H_{\mc Q},-\}),\]
where the condition $[\mc Q, \mc Q] = 0$ ensures that the differential does indeed square to zero.
For us the pertinent super Lie algebra $\mf g$ will be a superconformal algebra (i.e., the relevant spacetime symmetries for a superconformal theory).

In the last two decades, representation theorists have analyzed the space of square-zero odd elements of super Lie algebras (see especially the work of Jenkins and Nakano~\cite{JenkinsNakano}).
Of particular relevance to us is the work of Duflo and Serganova \cite{DufloSerganova} (see also \cite{GHSS}), 
who studied {\it simple} super Lie algebras over the complex numbers and described the adjoint orbits of such square-zero odd elements of such algebras.  
Many superconformal algebras, after complexification, are simple, as we discuss in Section~\ref{sec: background}.
For example, for spacetime dimensions three and six, the superconformal algebras correspond to the super Lie algebras $\mr{osp}(n|m)$,
which were studied in detail by Gruson \cite{Gruson} prior to Duflo and Serganova. 
In this paper we explain how these structural mathematical results interface with the role of these super Lie algebras as symmetries of superconformal field theories.

The more familiar case of supersymmetry (i.e. of a supersymmetric extension of the group of isometries rather than of conformal transformations) motivates our approach.
As mentioned, Witten introduced twisting to obtain a topological quantum field theory out of a supersymmetric theory:
if the stress-energy tensor $T_{\mu \nu}$ of a supersymmetric theories is exact for the action of $\mc Q$, 
then this stress-energy tensor vanishes in the $\mc Q$-cohomology of observables \footnote{Note that such theories are sometimes called ``Witten-type'' topological theories, as opposed to ``Schwarz-type'' topological theories where $T_{\mu \nu}$ is exactly zero, as for instance in \cite{BBRT}.},
and the associated twisted theory is topological.  
Thus, it is natural to study topological twisting in two stages:
\begin{itemize}
\item identify the square zero odd elements (``twists'') of each super Poincar\'e algebra, possibly up to the adjoint action of the super Poincar\'e group, and
\item describe the $\mc Q$-cohomology of observables for supersymmetric theories, looking for those where the stress-energy tensor vanishes.
\end{itemize}
Recently, the twists in super Poincar\'e algebras that would produce topological theories were all identified and the topological twists of super Yang--Mills theories (formulated in the Batalin--Vilkovisky formalism) were described in  \cite{ElliottSafronov, ElliottSafronovWilliams}.
Note that there are twists that lead to {\em non}-topological theories,
where subtle structures emerge, such as theories that look holomorphic \cite{Johansen, Baulieu} or look topological in some directions and holomorphic in others \cite{KapustinHolo, KallenZabzine, ACMV}.
In the present work our goal is to pursue analogues for \emph{superconformal} field theories by working with the superconformal algebras that extend super Poincar\'e algebras.

Working in the superconformal case is not simply a matter of reproducing the calculations from earlier work in yet another example.
There are substantially different features that allow us to construct interesting and novel algebras of operators upon twisting,
and superconformal transformations produce fascinating relationships between different twists.
An important motivation for us --- demonstrating some novelties not apparent in the supersymmetric case --- is the work of Beem, Lemos, Liendo, Peelaers, Rastelli and van Rees \cite{Beem_2015}, who showed how such superconformal twists lead to novel constructions of vertex algebras.
Combinations of these authors wrote a series of subsequent papers that address more examples and explore their consequences  \cite{Beem_2015b, Beem_2015c, Beem_2017, Beem_2018}.
In this paper we do not focus on specific theories, 
but rather analyze formal aspects of superconformal actions on algebras of observables and the structures that may arise upon twisting.
Our paper might be seen as trying to provide some systematic explanation for the kinds of results found by Beem {\it et al}.
In Section~\ref{sec:twistedobsphys} we provide an overview with a physical flavor;
in Section~\ref{sec:twistedobsmath} we state precise mathematical results using the language of (pre)factorization algebras.
(We discuss again the results of Beem {\it et al} in Section~\ref{Beemetal}.)

We now turn to describing our questions and efforts in more detail.

\subsection{What Twisting Does to Superconformal Theories}

Let us provide a gloss for the geometric context before discussing field theories and their observables.

Let $\RR^{p,q}$ denote the real vector space of dimension $p+q = n$ with a pseudo-Riemannian structure of signature $(p,q)$.
Euclidean space is denoted $\RR^{n,0}$ and Minkowski space is denoted $\RR^{n-1,1}$ (or $\RR^{1,n-1}$, depending on preference).
There is a group $\mr{Conf}(p,q)$ of all conformal transformations, containing the isometries but also more.
There is a natural conformal compactification $C(\RR^{p,q})$
on which $\mr{Conf}(p,q)$ acts transitively, with $\RR^{p,q}$ as a dense open submanifold,
akin to the Riemann sphere compactifying the complex line.
(For a fuller review, see Section~\ref{conformal_section}.)
Let $\mathfrak{conf}(p,q)$ denote the real Lie algebra,
and let $\mathfrak{conf}(p+q,\CC)$ denote its complexification,
which only depends on the total dimension, not the signature.
 
Almost tautologically, a conformal field theory lives on $C(\RR^{p,q})$,
but to compare with other theories (and for the sake of familiarity),
one often restricts the theory to a dense open subset $\RR^{p,q} \subset C(\RR^{p,q})$,
which one views as the relevant ``spacetime.''
This choice of embedding is somewhat arbitrary, 
as any element $g$ of the conformal group $\mr{Conf}(p,q)$ moves the ``defining'' $\RR^{p,q}$ to another copy~$g \RR^{p,q}$.
At first glance, a theory may look different on these two embeddings, especially when written in explicit coordinates.

A {\em super}\/conformal field theory has a super Lie group $\mr{SConf}(p,q|S)$ of symmetries, 
extending the action of $\mr{Conf}(p,q)$.
(Here $S$ just denotes data chosen to characterize the super extension.)
For the most part we will focus on the super Lie algebra~$\mathfrak{sconf}(p,q|S)$,
which we call a {\em superconformal algebra}.

Let $\obs$ denote the algebra of observables of a superconformal field theory (classical or quantum).  
The most natural place for a superconformal field theory to live is the conformal compactification $C(\RR^{p,q})$, so our algebra $\obs$ can be thought of as living over the conformal compactification.  
That is, for any open subset $U \sub C(\RR^{p,q})$ there is a vector space $\obs(U)$ of observables supported on the subset~$U$.
(The notion of prefactorization algebra encodes the dependence of observables on open sets.)

As a superconformal theory, 
the superconformal group acts equivariantly:
for every $g \in \mr{SConf}(p,q|S)$ and open set $U$, there is an isomorphism $\obs(U) \to \obs(gU)$.
Similarly, the super Lie algebra $\mathfrak{sconf}(p,q|S)$ acts by derivations on every $\obs(U)$.
As the observables form a complex vector space, this action extends to an action by the complexified super Lie algebra,
which we will denote~$\mf{sconf}(n|S,\CC)$ for now.

If $\mc Q$ is an odd element of the complexified superconformal Lie algebra such that $[\mc Q, \mc Q] = 0$,
we may study the algebra of observables $\obs^{\mc Q}$ in the theory \emph{twisted} by $\mc Q$.  
This twisted algebra of observables still lives over $C(\RR^{p,q})$, 
and a goal in this paper is to develop a systematic understanding of the properties of this twisted algebra 
determined purely by the superconformal algebra.  
To this end, let us describe a few natural structures determined by $\mc Q$ and independent of the choice of theory and ruminate on what they mean for the twisted theory.

A central fact is that not all symmetries of the superconformal theory are symmetries of a twisted theory.
Instead, there is an interesting relationship between the symmetries of the superconformal theory and its twist.
\begin{itemize}

\item \textbf{Closed elements of the Lie algebra}: Let
\[\mf z_{\mc Q}^\CC = \{A \in \mf{sconf}(n|S,\CC)_0 \colon [A, \mc Q] = 0\},\]
where $\mf{sconf}(n|S,\CC)_0$ denotes the even part of the superconformal algebra.
The Lie subalgebra $\mf z_{\mc Q}^\CC$ consists of even symmetries that commute with the supercharge $\mc Q$.  
These symmetries remain as symmetries of the twisted theory $\obs^Q$.  
The action of non-closed symmetries is broken when we pass to the twist.

\item \textbf{Closed elements of the Lie group}: Let
\[Z_{\mc Q}^\CC \sub \mr{SConf}(n|S, \CC)_0 \]
denote the Lie group exponentiating $\mf z_{\mc Q}^\CC$, 
and let $Z_{\mc Q}$ denote its intersection with the real form $\mr{SConf}(p,q|S)_0$.
This Lie subgroup consists of the even symmetries that continue to act on the twisted theory $\obs^{\mc Q}$ at the group level, by genuine conformal transformations.

\begin{ex}
If $Z_{\mc Q}$ contains the group of isometries of an affine subspace $\RR^{a,b} \sub C(\RR^{p,q})$ then we may study the observables in $\obs^{\mc Q}$ supported along this subspace: this algebra of restricted observables is now isometry-invariant.
\end{ex}

\item \textbf{Exact elements of the Lie algebra}: Let
\[\mf b_{\mc Q}^\CC = \{A \in \mf{sconf}(n|S,\CC)_0 \colon A = [\mc Q, \mc Q'] \text{ for some } \mc Q'\}.\]
These provide even symmetries of the twisted theory that are explicitly trivialized.  
That is to say, the conserved current for a symmetry from $\mf b_{\mc Q}^\CC$ is exact in $\obs^{\mc Q}$, 
so it vanishes in cohomology.

\item \textbf{Exact elements of the Lie group}: Let
\[B_{\mc Q}^\CC \sub \mr{SConf}(n|S, \CC)_0 \]
denote the Lie group exponentiating $\mf b_{\mc Q}^\CC$, 
and let $B_{\mc Q}$ denote its intersection with the real form $\mr{SConf}(p,q|S)_0$.
This Lie subgroup provides the even symmetries of the twisted theory that act on spacetime by genuine conformal transformations but are trivialized on the twisted theory.
\end{itemize}
We can draw conclusions from this analysis most readily for specific types of symmetry occurring in the groups $Z_{\mc Q}$ and~$B_{\mc Q}$.

\begin{ex}[$\bb E_k$-algebras]
Suppose there is an affine subspace $\RR^{k} \sub C(\RR^{p,q})$ with definite signature so that the group of isometries of the subspace lies in $B_{\mc Q}$.  
Then if we restrict observables of $\obs^Q$ to those supported along this affine subspace, 
we have an isometry action that is trivialized up to homotopy.  
If the dilation of $\RR^k$ also lies in $B_{\mc Q}$, 
then these restricted observables have the structure of a \emph{framed $\bb E_k$-algebra}: 
an algebra over the operad of framed little $k$-disks.  
This structure encodes many properties of a $k$-dimensional topological field theory,
and such a twisted theory can reasonably be viewed as topological on this affine subspace.
\end{ex}

\begin{rmk}
When doing this calculation carefully, we will need the group of isometries of $\RR^k$ not only to be exact, 
but also we will need to choose the potentials coherently: 
it is necessary to have a subspace of odd symmetries (the potentials) that makes exact the full algebra of isometries, as in Definition~\ref{potential_def}.
\end{rmk}

\begin{rmk}
One could just as well consider not only the subalgebra of \emph{even} closed and exact elements, but the full sub super Lie algebra of \emph{all} closed and exact elements.  For instance, the full superalgebra of $\mc Q$-closed elements will continue to act on the $\mc Q$-twisted theory, and it could be useful to take advantage of the continued presence of certain odd symmetries.  For simplicity we will not consider this further in the present paper.
\end{rmk}

\begin{ex}[Vertex Algebras]
Suppose that there is a subspace $\RR^2 \sub C(\RR^{p,q})$ so that the full group of isometries of $\RR^2$ lies in $Z_{\mc Q}$, 
but only the complexified translation $\ol d_{z}$  lies in $\mf b^\CC_{\mc Q}$,
where $z$ denots a complex coordinate on $\RR^2 \iso \CC$.  
Then if we restrict observables of $\obs^Q$ to those supported along this subspace, 
we have an isometry action so that the complexified antiholomorphic translation is trivialized up to homotopy.  
In this case (and under some additional technical assumptions), 
the restricted observables can be equipped with the structure of a \emph{vertex algebra}.  
This construction was originally considered by Beem {\it et al} \cite{Beem_2015} in their derivation of vertex algebras from 4d $\mathcal N=2$  superconformal field theory.
\end{ex}

Our analysis also allows for a situation that lies in between these two extremes, 
where we extract a field theory on $\RR^k \times \CC^\ell$ whose observables are topological along the $k$ real directions and holomorphic along the $\ell$ complex directions.

\begin{rmk}
The Lie group $\mr{SConf}(p,q|\mc N)_0$ acts simultaneously on the space of nilpotent supercharges $\mc Q$ by the adjoint action and on the spacetime $C(\RR^{p,q})$ by conformal transformations.  
The resulting twisted theories are equivalent.  
We observe, however, that if we study the algebra of operators after restricting to an affine subspace, 
the conformal transformations will move this subspace around.  
Thus for instance, the intersection with a choice of subspace $\RR^{p,q} \sub C(\RR^{p,q})$ may look quite different for different points in the same orbit.
\end{rmk}

In the main body of this paper, we investigate the following questions.
\begin{enumerate}
 \item For a given superconformal group $\mr{SConf}(p,q|\mc N)$, what is the locus $\mc{N}\mathrm{ilp}$ of square-zero odd elements (i.e., twists) of the complexified Lie algebra?  What are the orbits for the adjoint action of the even part of the superconformal group?
 \item For each orbit, what do the Lie groups and Lie algebras $\mf z_{\mc Q}^\CC, Z_{\mc Q}, \mf b_{\mc Q}^\CC, B_{\mc Q}$ look like?
 \item Can we locate subspaces whose isometry groups are $\mc Q$-closed?  
 If so we can obtain nice isometry-invariant algebras of observables on $\RR^k$ from the $\mc Q$-twisted theory.
 \item What does the subalgebra of exact isometries of such a subspace look like --- both real and complexified --- and can we find potentials for these subalgebras?  
 If so the algebra of twisted observables will have nice invariance properties that we can analyze.
\end{enumerate}
Notice that all of these questions are totally independent from the specific choice of superconformal theory, and can be applied to draw conclusions in a completely uniform way.
As a specific case study, we revisit the results of Beem {\it et al} in Section~\ref{Beemetal} and relate the approach of Saberi and Williams to ours in Section~\ref{sec: Saberi Williams}.

\subsection*{Acknowledgements}

We thank Philip Argyres, Mitch Weaver, and Brian Williams for helpful discussions on this topic, 
and particularly Richard Eager, who offered detailed comments on the manuscript, especially in guiding us through the literature.
We also thank an insightful, patient, thorough referee whose feedback substantially improved the paper.
The National Science Foundation supported O.G. through DMS Grant 2042052.
ML is supported in part by the National Research Foundation of Korea (NRF) Grant 2021R1A2C2012350.
C.E. and O.G. thank their REU students at Amherst College (Osha Jones and Ziji Zhou) and UMass Amherst (Nikos Orginos, Nathaniel Reid, and Sam Silver) for work on two- and three-dimensional versions of these problems,
which gave helpful complementary insights.

\section{Superconformal Algebras and Twists}
\label{sec: background}

We rapidly review the conformal and superconformal groups and their associated (super) Lie algebras.
For the conformal situation, we follow Schottenloher's conventions (see Chapter 2 of~\cite{Schottenloher}).

\subsection{The Conformal Case} \label{conformal_section}

Throughout we will work with a pair $(p,q)$ of natural numbers such that $p+q > 2$.
The case $p+q = 2$ is strikingly different and so deserves a separate examination.

Let $\RR^{p,q}$ denote the vector space $\RR^{p+q}$ equipped with the nondegenerate symmetric bilinear pairing of signature $(p,q)$.
We say it has dimension $D = p+q$.
We denote the pairing by
\[
\<v,w\> = v_1 w_1 + \cdots + v_p w_p - v_{p+1} w_{p+1} - \cdots - v_{p+q} w_{p+q},
\]
and use $|v|^2 = \<v,v\>$ for the associated ``norm.''
A crucial role for us is played by a closely related space,
as it leads to a clean definition of the conformal group and -- later -- will be the spacetime relevant to twisted superconformal field theories.
Let $\RR\PP^{p+q+1}$ denote the $p+q+1$-dimensional real projective space.

\begin{dfn}
Let $C(\RR^{p,q})$ denote the {\em conformal compactification} of $\RR^{p,q}$.
It is the projectivization of the real variety given by the vectors $v \in \RR^{p+1,q+1}$ such that $|v|^2 = 0$.
In other words, it is the hypersurface in the projective space $\RR\PP^{p+q+1}$ cut out by the quadratic equation
\[
0 = x_0^2 + \cdots + x_p^2 - x_{p+1}^2 - \cdots - x_{p+q+1}^2
\]
where $(x_0: \cdots : x_{p+q+1})$ is a point given in homogeneous coordinates.
\end{dfn}

In concrete terms,
there is an embedding $\iota \colon \RR^{p,q} \hookrightarrow \RR^{p+1,q+1}$ by
\[
\iota(v_1, \ldots, v_{p+q}) 
= \left(\frac{1-|v|^2}{2},v_1, \ldots, v_{p+q},\frac{1+|v|^2}{2}\right).
\]
Note that $\iota(v)$ is always a vector of norm zero, and that it is never the zero vector.
We can thus map $\RR^{p,q}$ into the projective space $\RR\PP^{p+q+1}$
by sending $v$ to
\[
\left(\frac{1-|v|^2}{2}:v_1:\ldots: v_{p+q}:\frac{1+|v|^2}{2}\right) \in \bb{RP}^{p+q+1},
\]
using homogeneous coordinates.
Then $C(\RR^{p,q})$ is the natural completion of this copy of $\RR^{p,q}$ in projective space.

It is straightforward to describe the symmetries of this compactification:
we take the linear automorphisms of $\RR^{p+1,q+1}$ that preserve the subspace of norm-zero vectors and then projectivize.

\begin{dfn}
The {\em conformal group} $\Conf(p,q)$ is the connected component of the identity in~$\mathrm{O}(p+1,q+1)/\{\pm 1\}$.
\end{dfn}

There are familiar classes of transformations within this group:
\begin{itemize}
\item the orthogonal transformations $\mathrm{SO}(p,q)$,
\item the translations $\RR^{p+q}$,
\item the dilations $\RR_{>0}$, and
\item the ``special conformal transformations.''  The subgroup consisting of such elements is isomorphic to a second copy of the additive group~$\RR^{p+q}$.
\end{itemize}
In fact, these subgroups collectively generate the full conformal group;
in other words, we can always factor a conformal transformation as some composition of transformations of those four types.
We refer to \cite[\S 2.2]{Schottenloher} and \cite[\S 4.1]{dFMS} for a thorough discussion \footnote{Note that the action of the special conformal transformations on a point in $\RR^{p,q}$ can be interpreted as an inversion (about the origin), followed by a translation, followed by another inversion.}

\begin{rmk} \label{conformal_embedding_rmk}
 The identification of these classes of generators of conformal transformations is associated to the chosen conformal embedding $\iota \colon \RR^{p,q} \to C(\RR^{p,q})$.  For instance the orthogonal transformations, translations and dilations generate the subgroup of conformal transformations that preserve the image of $\iota$.  We obtain distinct but conjugate subgroups by choosing alternative conformal embeddings -- from our construction of the conformal compactification we obtain such an embedding from every codimension two linear subspace of the form $\RR^{p,q} \subseteq \RR^{p+1,q+1}$, hence an embedding $\SO(p,q) \hookrightarrow \SO(p+1,q+1)$.
\end{rmk}

The Lie algebra $\mathfrak{conf}(p,q)$ of conformal symmetries is thus isomorphic to $\mathfrak{so}(p+1,q+1)$.

\begin{lmm}
As a vector space, there is a direct sum decomposition
\[
\mathfrak{conf}(p,q) \cong \mathfrak{so}(p,q) \oplus \RR^{p+q} \oplus \RR \oplus  \RR^{p+q},
\]
where the summands correspond, in order, to the Lie algebras of the four subgroups listed above.
\end{lmm}

This Lie algebra has a representation in vector fields on $\RR^{p,q}$, by differentiating the action of the group on the compactification,
and the infinitesimal translations, orthogonal transformations, and dilations act in the familiar way,
which we write out below in explicit detail.
This infinitesimal action plays a central role in characterizing the superconformal Lie algebra below.

\subsubsection{A more physical formulation}

We now recall a standard notation in physics for generators of $\mathfrak{conf}(p,q)$ and write the commutation relations.
It explicitly realizes $\mathfrak{conf}(p,q)$ as a Lie subalgebra of the vector fields on~$\RR^{p+q}$.

We use the customary notation from the physics literature where repeated indices are summed over, 
e.g., $\partial^\mu x_\mu = \sum_{\mu=1}^{d=p+q} \partial^\mu x_\mu$.
Furthermore, indices are ``raised'' and ``lowered'' by contracting with the metric (tensor) $g_{\mu\nu}$ or its inverse.
For example, 
$\partial_\mu = g_{\mu\nu} \partial^\nu = \sum_{\nu=1}^{d} g_{\mu\nu} \partial^\nu$.
The metric is given by $g_{\mu\nu} = \text{diag}(\mathbb{1}_p, -\mathbb{1}_q)$.

\begin{notation}
Let $\mu, \nu$ denote indices running from 1 to $p+q$, associated to the standard basis for $\RR^{p+q}$.
The generators of $\mathfrak{conf}(p,q)$ are
\[
\begin{array}{ccc}
M_{\mu\nu} & = & x_\mu \partial_\nu - x_\nu \partial_\mu\\
P_{\mu} & = & - \partial_\mu\\
D & = & - x_\mu \partial^\mu\\
K_{\mu} & = & x^2 \partial_\mu - 2x_\mu x_\nu \partial^\nu
\end{array}
\]

and they satisfy
\begin{align*}
[ M_{\mu\nu} , M_{\rho\sigma} ] & =
g_{\rho\nu} M_{\mu\sigma} + g_{\mu\rho} g_{\sigma\nu}
- g_{\mu\sigma} M_{\rho\nu} -  g_{\sigma\nu}  M_{\mu\rho} &&
[ D, P_\mu ]  = P_\mu\\
[ M_{\mu\nu}, P_{\rho} ]  & = g_{\nu\rho} P_{\mu} -g_{\mu\rho} P_{\nu} && [ D, K_\mu ]  = -K_\mu\\
[ M_{\mu\nu}, K_{\rho} ]  & =  g_{\nu\rho} K_{\mu} -g_{\mu\rho} K_{\nu} &&
[ K_{\mu}, P_{\nu} ]  =  - 2 M_{\mu\nu} + 2 g_{\mu\nu} D %
\end{align*}
with all other commutators vanishing.

\end{notation}

\begin{rmk}
 As in Remark \ref{conformal_embedding_rmk} the choice of decomposition of the conformal Lie algebra into these four subspaces is non-canonical: it depends on a choice of linear subspace $\RR^{p,q} \subseteq \RR^{p+1,q+1}$.
\end{rmk}

\subsection{The Superconformal Case}

We will focus on superconformal Lie {\em algebras}, as they play the key role in this paper and because they are simpler to describe in technical terms
(In particular, we will not discuss super analogs of the conformal compactification.)
Our approach is borrowed from Shnider's elegant and efficient discussion~\cite{Shnider}.
We begin by recalling how one formulates the super versions of the spacetimes $\RR^{p,q}$ and of their isometries.
That is, we review what super-Minkowski space is and what the super Poincar\'e group is (in arbitrary signature)
to elucidate the logic that motivates a formulation of superconformal algebras.

\begin{dfn}
 The \emph{Poincar\'e group} in signature $(p,q)$ is the group
 \[\mathrm{ISO}(p,q) = \RR^{p+q} \rtimes \mathrm{SO}(p,q)\]
or oriented isometries of $\RR^{p,q}$. The {\em Lorentz group} is the subgroup $\mathrm{SO}(p,q) \sub \mathrm{ISO}(p,q)$ that fixes the origin in $\RR^{p,q}$.
\end{dfn}

We can (and should) view $\RR^{p,q}$ as the homogeneous space given by the quotient of the Poincar\'e group by the Lorentz group.
There is a natural spin cover of the Poincar\'e group where we replace $\mathrm{SO}(p,q)$ by its spin cover.
We will typically focus on the common Lie algebra $\mf{so}(p,q)$ of these groups, which does not care about the spin cover.
The Lie algebra $\mf{so}(p,q)$ has an infinitesimal action on~$\RR^{p,q}$.

Super versions of these objects arise as extensions that depend on spinors as follows.
Pick a real (Majorana) spinorial representation $S$ of $\mathrm{Spin}(p,q)$ and a spin-equivariant symmetric bilinear map $\Gamma \colon S \times S \to \RR^{p+q}$.
(Here $S$ need not be irreducible; it can be a direct sum of spinor representations.
The pairing is unique up to rescaling if $S$ is irreducible.)

\begin{dfn}
 The {\em super Poincar\'e group} is the super Lie group whose odd component is $S$ and whose reduced (bosonic) Lie group is the Poincar\'e spin group; the group structure depends on the pairing $\Gamma$ on spinors.  We denote the super Poincar\'e group associated to this data by $\mathrm{ISO}(p,q|S)$.
\end{dfn}

\begin{rmk} \label{N_notation_rmk}
 In the physics literature it is typical for super Poincar\'e groups to be referred to by an index usually denoted by $\mc N$ combinatorially describing spinorial representations of $\mr{Spin}(p,q)$.  
If there is a single irreducible spinorial representation~$S$, then the super Poincar\'e algebra associated to $S^{\oplus k}$ is called the $\mc N=k$ super Poincar\'e algebra and denoted $\mathrm{ISO}(p,q|k)$.  
 If there are two irreducible spinorial representations $S_+,S_-$, then the super Poincar\'e algebra associated to $S_+^{k_+} \oplus S_-^{k_-}$ is called the $\mc N=(k_+,k_-)$ super Poincar\'e algebra and denoted~$\mathrm{ISO}(p,q|k_+,k_-)$.
\end{rmk}

There is a super manifold that is the quotient of this super group by the Lorentz spin group;
we denote it by $\RR^{p,q|s}$, where $s$ will denote the dimension of $S$.
The super Poincar\'e Lie algebra $\mf{iso}(p,q|S)$ acts infinitesimally on this superspace.  We will often wish to refer to the symmetries of $S$ that commute with the spin action.  We make the following definition.

\begin{dfn}
 The \emph{R-symmetry group} $G_R$ of the super Poincar\'e group $\mathrm{ISO}(p,q|S)$ is the group of outer automorphisms of $\mathrm{ISO}(p,q|S)$ that act trivially on the even part.
\end{dfn}

In looking for a super version of the conformal algebra,
we want to construct a super Lie algebra $\sconf(p,q|S)$ that depends on this choice of a Majorana spinor representation $S$ and the bilinear form.
It should include the super Poincar\'e Lie algebra as a subalgebra,
just as the Poincar\'e Lie algebra is a subalgebra of $\mathfrak{conf}(p,q)$.
Moreover, the even part of $\sconf(p,q|S)$ should contain the conformal algebra $\mathfrak{conf}(p,q)$ as a Lie subalgebra.
In other words, $\sconf(p,q|S)$ should be a common generalization of usual supersymmetry algebra and usual conformal algebra.

\begin{rmk}
When studying super Lie algebras $\gg$, 
we will use the notation $[-,-]$ for the full graded skew-symmetric Lie bracket
\[ [-,-] \colon \wedge^2(\gg) \to \gg.\]
In particular we will not use a different notation for the (symmetric) bracket of two odd elements $\sym^2(\gg_1) \to \gg_0$ compared to the notation the (skew-symmetric) bracket of two even elements. 
By contrast, in the physics literature these cases are sometimes distinguished in notation, for instance by writing the bracket of two odd elements $\mc Q$ and $\mc Q'$ as~$\{\mc Q, \mc Q'\}$.
\end{rmk}

It is useful to ask for something stronger: a common generalization {\it in how they act as symmetries} of the supersymmetry algebra and the conformal algebra.
We thus request super Lie algebras satisfying two properties:
\begin{dfn}[{\it cf.} \cite{Shnider}]
 A super Lie algebra $\gg$ is a \emph{superconformal algebra} of signature $p,q$ if $\gg$ acts on super-Minkowski space $\RR^{p,q|s}$ by infinitesimal derivations, and the following conditions hold:
 \begin{enumerate}
  \item There is an embedding $i \colon \mathfrak{iso}(p,q|S) \hookrightarrow \gg$ of super Lie algebras, and the restriction of the action along $i$ coincides with the standard action of infinitesimal super-isometries.
  \item There is an embedding $j \colon \mathfrak{conf}(p,q) \hookrightarrow \gg$, and the restriction of the action along $j$ coincides with the standard action of infinitesimal conformal transformations on $\RR^{p,q}$, acting trivially on the odd coordinates.
 \end{enumerate}
\end{dfn}

The remarkable theorem is that such a super Lie algebra exists {\em only} when $p+q \leq 6$, as discovered by Nahm \cite{Nahm}, although we emphasize here the clean mathematical approach of Shnider \cite{Shnider},
who articulates these properties as explicit hypotheses.
Shnider begins by studying the complexification of the situation above, since any such real super Lie algebras has a complexifiction and there is no dependence on a choice of signature after complexifying.
He then shows that if such a complex super Lie algebra $\gg$ exists,
there is a simple super Lie algebra, satisfying the same properties, which is a quotient or subalgebra of~$\gg$.
Moreover, its bosonic part splits as the sum of $\mathfrak{so}(p+q,\CC)$ and a complementary ideal.
He then uses Kac's classification of simple super Lie algebras to show that such complex super Lie algebras do {\em not} exist above dimension~6.
Identifying the relevant simple super Lie algebras in dimensions less than or equal to 6 is possible and already accomplished.

We will follow some of this strategy below, typically analyzing a problem in the complexification before turning to real forms.
As in the bulk of the literature, we restrict our attention to the simple cases.  

\begin{lst}
\label{list}
We record the simple super Lie algebras that appear as complexified superconformal algebras in dimensions 3 to~6.
\begin{itemize}
 \item In dimension 3, the superconformal algebras are $\mathfrak{osp}(k|4,\CC)$ for $k \ge 1$.  The even part is $\so(k,\CC) \oplus \sp(4,\CC) \iso \so(k,\CC) \oplus \so(5,\CC)$, containing the conformal algebra $\so(5,\CC) \iso \mathfrak{conf}(3,\CC)$.
 \item In dimension 4, the superconformal algebras are $\sl(4|k,\CC)$ for $k \ge 1, k \ne 4$, and $\mathfrak{psl}(4|4,\CC)$ in the special case $k=4$.  The even part is $\sl(4,\CC) \oplus \sl(k,\CC) \oplus \CC$ if $k \ne 4$ and $\sl(4,\CC) \oplus \sl(4,\CC)$ in the $k=4$ case, containing the conformal algebra $\sl(4,\CC) \iso \so(6,\CC) \iso \mathfrak{conf}(4,\CC)$.
 \item In dimension 5, there is a unique superconformal algebra, the exceptional super Lie algebra $\mathfrak f(4,\CC)$. The even part is $\so(7,\CC) \oplus \sl(2,\CC)$, containing the conformal algebra $\so(7,\CC) \iso \mathfrak{conf}(5,\CC)$.
 \item In dimension 6, the superconformal algebras are $\mathfrak{osp}(8|2k,\CC)$ for $k \ge 1$.  The even part is $\so(8,\CC) \oplus \sp(2k,\CC)$, containing the conformal algebra $\so(8,\CC) \iso \mathfrak{conf}(6,\CC)$.
\end{itemize}
\end{lst}

\begin{rmk}
 In each of these examples, the even part of the superconformal algebra includes $\mathfrak{conf}(n, \CC) \iso \so(n+2,\CC)$ as a summand.  We will denote the quotient algebra by $\gg_R$, and refer to it as the algebra of R-symmetries.  In each case -- as we can see directly from the classification -- $\gg_R$ coincides with the Lie algebra of the group $G_R$ of outer automorphisms of the superconformal algebra that act trivially on the even part (the \emph{group} of R-symmetries).
\end{rmk}

\begin{rmk}
As in Remark~\ref{N_notation_rmk} the superconformal algebras are often indexed following the combinatorial indexing for the spinorial representation in their super Poincar\'e summand.  
Hence one refers to $\mc N=k$ superconformal algebras in dimension 3,4 and $\mc N=(k,0)$ superconformal algebras in dimension~6.  
We will use this terminology below.
\end{rmk}

\begin{rmk}
The classification of simple super Lie algebras has played an essential role in the physics literature on the structure and properties of superconformal theories \cite{Mack, Minwalla, BHP}. In particular, Dolan and Osborn \cite{DolanOsborn} developed the representation theory of 4d superconformal algebras, and C\'ordova, Dumitrescu and Intriligator \cite{CDI} provided a thorough analysis of the representation theory of superconformal algebras in dimensions 2 to 6 and its applications to superconformal field theory.
\end{rmk}

\subsection{The Question of Classifying Twists}

Topological field theories have had an enormous impact in both theoretical physics and mathematics,
and the most important examples arise from supersymmetric theories by a twisting procedure,
introduced by Witten \cite{Witten_1988}.
We recall here a less sophisticated version of that procedure before we turn to superconformal field theories.
In between, we provide a quick survey of mathematical work on these notions,
mostly growing out of the pioneering but unpublished paper of Duflo and Serganova \cite{DufloSerganova},
as it provides some important structural results.

\subsubsection{How Twists Appear in the Supersymmetric Case}

Loosely speaking, a supersymmetric field theory is a field theory on $\RR^{p,q}$ with an action of a super Poincar\'e Lie algebra.
(One typically asks for more, but we will focus on this kind of symmetry for now.)
It is reasonable to explore how elements of the Lie algebra of supertranslations act on the theory.
Let $\mathfrak{stran}(p,q|S)$ be this super Lie algebra, where $S$ denotes the direct sum of spinor representations that provides the odd component and the non-trivial Lie bracket is given by a vector-valued pairing $\Gamma$ on~$S$.

\begin{rmk}
 For the moment, by $\mathfrak{stran}(p,q|S)$ we mean a \emph{real} Lie algebra whose even part is $\RR^{p,q}$.  Later on, and in the parallel superconformal setting, we will consider the less refined setting of theories with an action of the complexification of such a real Lie algebra.  The discussion of twists that follows makes sense either over the real or the complex numbers.
\end{rmk}

\begin{dfn}
A {\em translation twist} is an odd element $\cQ$ such that $[\cQ,\cQ] = 0$.
\end{dfn}

A choice of twist leads to some interesting structures:
\begin{enumerate}
\item A subspace $\mf b_{\cQ} \subset \RR^{p,q}$ defined by
\[\mf b_\cQ = \mr{Im}(\cQ)\]
where we view the twist as a map $\cQ \colon \Pi S \to \RR^{p,q}$ sending an odd element $s$ to the even element~$[Q,s]$.
Such a subspace determines a foliation $\cF^{\cQ}$ of~$\RR^{p,q}$ into affine leaves.
\item A deformation of a supersymmetric theory by adding $\cQ$ to its BRST differential (i.e., by adding a term to its Lagrangian that encodes how $\cQ$ acts on the theory). This deformed theory is ``trivial'' along leaves of the foliation $\cF^{\cQ}$ in the sense that BRST cohomology classes of observables are isomorphic under translation along these leaves.
\item A differential $\ZZ/2$-graded Lie algebra $(\mathfrak{stran}(p,q|S), [\cQ,-])$, which can be seen as acting on the deformed field theory.
\end{enumerate}
We call such a $\cQ$-deformation of a supersymmetric field theory a {\em twisted theory},
and we see that it admits a symmetry by this dg Lie algebra.

When $\mf b_\cQ$ contains all of $\RR^{p,q}$, we call $\cQ$ a {\em topological} twist as the associated twisted theories must be ``topological'' in the sense that we can move insertion points around without changing expected values.
But there are cases where $\mf b_\cQ$ is a proper subspace, leading to more subtle changes in the theory.
For instance, in the Euclidean setting, twists lead to theories that can be holomorphic in some directions and topological in others.

Recently, there has been a systematic classification of the space of twists (at least for complexified supertranslation Lie algebras), motivated by the desire to organize and analyze the behavior of twisted supersymmetric field theories.
For details of this classification, see~\cite{ElliottSafronov,ElliottSafronovWilliams, EagerSaberiWalcher} .

The following definition when applied to the super Lie algebra $\fg = \mf{stran}(p,q|S)$ characterizes the space of possible supertranslation twists.

\begin{dfn} \label{nipotence_var_def}
Let $\fg = \fg_0 \oplus \Pi \fg_1$ be a super Lie algebra, where $\fg_i$ for $i=0,1$ refer to its even and odd summands and $\Pi$ is the parity shift operator.  The (affine) {\em nilpotence variety} of $\fg$ is the affine quadric subvariety of $\fg_1$ defined by
\[\Nilp_{\fg} = \{\cQ \in \fg_1 \colon [\cQ,\cQ]=0\}.\]
This affine variety is preserved by the rescaling action, and the \emph{projective nilpotence variety} $\bb P\Nilp_{\fg} \sub \bb P(\fg_1)$ of $\fg$ is its projectivization.
\end{dfn}

The affine and projective nilpotence varieties both carry an action of the Lie algebra $\fg_0$ inherited from the adjoint action. 
If $\fg$ is the super Lie algebra of a super Lie group $G$, 
then the even part $G_0$ of $G$ also acts on the nilpotence varieties by the group-level adjoint action.

In the case where $\fg = \mf{stran}(p,q|S)$ the nilpotence variety carries a natural action of the group $\mr{Spin}(p,q) \times G_R$ where $G_R$ is the group of R-symmetries.

\begin{dfn}
The \emph{moduli stack of twists} for the supertranslation algebra  is the quotient stack
\[\Twist_{\fg} = \left[\Nilp_{\fg} / (\mr{Spin}(p,q) \times G_R)\right].\]
\end{dfn}

\subsubsection{Associated Varieties of Super Lie Algebras}

The notions we just mention were already explored by mathematicians who study super Lie algebras,
so we quickly record the terminology used by this community and some of their pertinent results.

\begin{dfn}
For a finite-dimensional super Lie algebra $\gg$ over $\CC$, 
its {\em associated variety} consists of the square-zero odd elements:
\[
X_\gg = \{ x \in \gg_1 \,|\, [x,x] = 0\}.
\]
It is a Zariski-closed conical subset of $\gg_1$, viewed as an affine space.
\end{dfn}

Hence the associated variety of a supertranslation algebra is precisely the affine nilpotence variety introduced in the preceding section.
Note that a point $x \in X_\gg$ determines a dg Lie algebra $(\gg, [x,-])$,
and Duflo and Serganova studied both its cohomology $\gg_x = \mr H^*(\gg, [x,-])$ and the {\em Duflo-Serganova} functor
\[
\mr{DS}_x \colon {\rm Mod}(\gg) \to {\rm Mod}(\gg_x),
\]
which is nicely behaved. For instance, it is symmetric monoidal, preserves the superdimensions of finite-dimensional representations, and induces ring maps between the reduced Grothendieck groups of (well-behaved) abelian subcategories (such as finite-dimensional representations).

Let $G_0$ denote the simply-connected, connected algebraic group with Lie
algebra $\gg_0$.
It acts on $\gg_1$ by the adjoint action and $X_\gg$ is preserved by this action.
It is natural to study the stack quotient $[X_\gg/G_0]$,
and in the case that $\gg$ is a supertranslation algebra, 
it maps to the moduli stack of twists introduced in the preceding section.
(That moduli space is a further quotient, taking into account the natural isometries, like rotation, of super-Minkowski space.)

\def\tt{{\mathfrak t}}

When $\gg_0$ is reductive and $\gg_1$ is a semisimple $\gg_0$-module, 
the super Lie algebra $\gg$ is called {\em quasireductive}.
In this case there is a Cartan subalgebra $\tt \subset \gg_0$.
Let $\Delta_1$ denote the roots of $\tt$ acting on~$\gg_1$, and let $\Delta_0$ denote the (usual, even) roots of~$\gg_0$.

There is a remarkable theorem showing that the set $X_\gg/G_0$ of $G_0$-orbits in the associated variety is {\em finite} for an important class of Lie superalgebras (which does {\em not} include supertranslations).
We will use this result in the superconformal setting.

To state the theorem, we must introduce more terminology.

\begin{dfn}
An {\em iso-set} for $\gg$ is a subset $A$ of $\Delta_1$ such that
\begin{itemize}
\item its elements are linearly independent and 
\item if any pair of odd roots $\alpha, \beta \in \Delta_1$ satisfies $\pm \alpha, \pm \beta \in A$, then $\alpha + \beta \not\in \Delta_0$.
\end{itemize}
Let $\cS_\gg$ denote the collection of iso-sets. 
\end{dfn}

The Weyl group $W$ of $G_0$ preserves~$\cS_\gg$.

\begin{thm}[Theorem 5.1, \cite{GHSS}]
\label{thm DS}
There is a bijection between $X_\gg/G_0$ and $\cS_\gg/W$ if $\gg$ is a complex super Lie algebra from the list
\[
\{ \gl(m|n), \sl(m|n) \text{ when $m \neq n$}, \osp(m|2n), \mathfrak{d}(2|1;a), \mathfrak{f}(4), \mathfrak{g}(3)\}.
\]
In particular, there is a {\em finite} set of orbits.
\end{thm}

In \cite{GHSS} they push further and for such classical super Lie algebras, they classify the stabilizers and normalizers for each point $x \in X_\gg$, as well as the dimensions of the orbits.

We want to raise a cluster of natural questions. 
We mention physical motivations below.

\begin{qst}
\label{qst: real form}
For a real super Lie group (particularly the real simple ones), what are the real twists of its real super Lie algebra? That is, what is the associated variety? Moreover, what are the orbits?
\end{qst}

\subsubsection{How Twists Appear in the Conformal Case} \label{conf_twist_section}

We raise here the question of studying the analogue of twists by elements of a superconformal Lie algebra.

Let $\sconf(p,q|S)$ be the super Lie algebra, where $S$ denotes the spinor representation that provides the odd component and $\Gamma$ is the pairing on~$S$.

\begin{dfn}
A {\em conformal twist} is an odd element $\cQ \in \Pi S$ such that $[\cQ,\cQ] = \Gamma(\cQ,\cQ) = 0$.
\end{dfn}

A twist $\cQ$ determines a subspace of $\mathfrak{conf}(p,q)$ and hence a foliation of the conformal compactification $C(\RR^{p,q})$.
If we view a superconformal field theory as living on $C(\RR^{p,q})$,
a twist then determines a deformation to a theory that is ``trivial'' along the leaves of this foliation.
We call it the {\em $\cQ$-twisted} superconformal theory.
Moreover, the dg Lie algebra $(\sconf(p,q|S), [\cQ,-])$ acts on this twisted theory.
It is to be hoped that such twisted theories admit phenomena at least as special and interesting as twists of supersymmetric theories.  
Indeed, since many supertranslation algebras may be embedded as subalgebras of superconformal algebras, 
many supersymmetric twists arise as special examples of superconformal twists.  
In this sense twisting in the superconformal context may be viewed as an extension of twisting in the usual context. 
Hence we encounter a basic question.

\begin{qst}
{\it For each signature $(p,q)$ with $p + q > 2$, what are the twists of the corresponding superconformal Lie algebra?}
\end{qst}

In the bulk of this paper we begin to explore this question by studying a few explicit examples and discuss how twisting affects theories in qualitative terms.

Following \cite{ElliottSafronov, EagerSaberiWalcher}, we phrase our problem in the following way.  
Note that for the rest of this section we will discuss nilpotent elements of the \emph{complexified} superconformal algebra; 
we will discuss the role of real forms shortly.

\begin{dfn}
The {\em conformal nilpotence variety} for $\sconf(p,q|S)$ is the associated variety $X_{\sconf(p,q|S)}$ of the complex super Lie algebra $\sconf(p,q|S)_\CC$.
We denote it by~$\Nilp_{(p,q|S)}(\CC)$.
\end{dfn}

Note that for each embedding of super Lie algebras from a supertranslation algebra to the corresponding superconformal algebra, there is a map of nilpotence varieties.
Every translation twist determines a conformal twist, in other words.

There are natural symmetries to take into account.
\begin{itemize}
\item Let ${\rm SConf}(p,q|S)_0$ is the simply-connected, connected algebraic group over $\CC$ whose Lie algebra is~$\sconf(p,q|S)_0$.
It acts on the associated variety, 
and if two twists are in the same orbit,
then the associated dg Lie algebras are isomorphic.
Thus, it is natural to consider the quotient stack
\[
[X_{{\sconf(p,q|S)}}/{\rm SConf}(p,q|S)_0],
\]
which we call the {\em maximal quotient stack of conformal twists}.
\item On the other hand, this group ${\rm SConf}(p,q|S)_0$ does {\em not} play nicely with the original superconformal geometry that motivated us.
Instead it has various {\em real} forms that are symmetries of real super manifolds that ``extend'' the conformal compactification $C(\RR^{p,q})$. 
If $G$ denotes one of these real forms, 
it is natural to consider the quotient stack
\[
[X_{\sconf(p,q|S)}/G],
\]
which we call the {\em $G$-quotient stack of conformal twists}.
\end{itemize}
One goal of this paper is to describe the geometric properties of these spaces,
with the following possibility as motivation.

\begin{qst}
Are there orbits among the conformal twists that do {\em not} come from the translation twists?
These would lead to novel twisted field theories.
\end{qst}

In light of List~\ref{list} and Theorem~\ref{thm DS}, 
Duflo and Serganova have given essentially a complete treatment of the maximal quotient stacks: 
they showed that for (nearly) every conformal nilpotence variety,
its ${\rm SConf}(p,q|S)_0$-orbits consist of a finite set of points.
That result implies each maximal quotient stack is a disjoint union of classifying stacks $\ast/{\rm Stab}(x)$ where $x \in X_{\sconf(p,q|S)}$ is a point on the associated orbit and where ${\rm Stab}(x)$ denotes its stabilizer inside~${\rm SConf}(p,q|S)_0$.

The only open case is $\psl(4|4)$, which we examine below.
For this algebra, the orbits do {\em not} form a finite set;
thus the stack is more interesting.
This case is particularly interesting because it corresponds to $\mathcal{N}{=}4$ supersymmetric theories in four dimensions,
which includes many special field theories.

But for physical questions, it is also important to examine the quotients by {\em real} groups.
Because the group is smaller, the quotient space is a much richer object: 
there are many more orbits.
For example, the work of \cite{Beem_2015} examines a complex twist but working in a fixed signature, and hence with an associated real form.
We revisit that example in Section~\ref{Schur_section}.

Finally, we wish to point out a third important situation.
It is natural 
\begin{itemize}
\item to consider twists in the {\em real} super Lie algebras arising from the real superconformal groups,
(i.e., the real points $\Nilp_{(p,q|S)}(\RR)$ of the conformal nilpotence variety)
and 
\item to identify the real orbits of these real twists.
\end{itemize}
In other words, we raise Question~\ref{qst: real form} here specifically for the real forms of the superconformal algebras.
In this situation, we obtain Lie subalgebras of vector fields on the conformal manifold $C(\RR^{p,q})$, namely $\mf z_\cQ = \mr{Ker}(\cQ)$ and $\mf b_\cQ = \mr{Im}(\cQ)$. These foliations have rich geometric content,
and unpacking their geometry would undoubtedly be fruitful.

This third situation also has clear physical motivation:
an action functional of a genuine physical theory should be real-valued,
and a Hamiltonian should be self-adjoint.
A superconformal field theory has an action of a real superconformal group,
so it is natural to look for twists that determine deformations as a genuine theory.
That pushes us to look at the real twists.  We discuss this further in Section \ref{reality_section}.

\section{The Observables of a Twisted Superconformal Theory}
\label{sec:twistedobsphys}

Given a superconformal theory with $\sconf(p,q|S)$ as a symmetry algebra, 
a nilpotent supercharge $\cQ$ can be added to the action as a BRST operator
leading to a new theory that has the $\cQ$-cohomology of $\sconf(p,q|S)$ as a symmetry algebra.
For supersymmetric theories, this process (often dubbed ``twisting'') is well-known and produces new theories with remarkable properties,
such as the topological field theories encoding Donaldson invariants of 4-manifolds \cite{Witten_1988}.
A crucial step is to understand the $\cQ$-cohomology of $\mathfrak{siso}(p,q|S)$ and its geometric meaning.
For superconformal theories, the twists can have more exotic geometric consequences and the associated twisted theories have surprising behavior, including the localization discovered by Beem {\it et al} \cite{Beem_2015}.

\begin{rmk}
We observe that while we introduced the notion of twisting as a modification of the action functional by the introduction of terms proportional to the nilpotent supercharge $\cQ$, we can, alternatively work directly with the algebra of operators that appear in the twisted theory by analyzing the $\cQ$-cohomology of the untwisted algebra of operators.  As a consequence, our derivations will be valid also in the cases of inherently strongly coupled theories, defined in the absence of a classical action functional.
\end{rmk}

In this section we will give a physical description of how conformal twists affect a theory,
and in section~\ref{sec:twistedobsmath} we offer detailed mathematical formulations using the language of disc-algebras and factorization algebras.

\subsection{Subalgebras of \texorpdfstring{$\cQ$}{Q}-Closed and \texorpdfstring{$\cQ$}{Q}-Exact Elements}

Let $\cQ$ be an odd square-zero element of $\sconf(p,q|S)$.  
As we hinted above, the properties of the $\cQ$-twist of a superconformal field theory defined on $C(\RR^{p,q})$ -- e.g., whether the twisted theory becomes topological or holomorphic upon restriction to certain affine subspaces -- are determined by the cohomology of $\sconf(p,q|S)$ under the differential $[\cQ, -]$.
We will make this idea more precise as follows.

\begin{dfn}
Let $Z_{\cQ} \sub \mathrm{Conf}(p,q) \times G_R$ denote the stabilizer group of the element $\cQ$ under the adjoint action.  
Let $\mathfrak z_{\cQ}$ denote the (real) Lie algebra of $Z_{\cQ}$. In other words, it is the even part of the centralizer of $\cQ$ in the superconformal algebra.  We will also write $\mf z_{\cQ}^\CC = \mf z_{\cQ} \otimes_\RR \CC$ for its complexification.
\end{dfn}

\begin{dfn}
Let $\mathfrak b_{\cQ}^\CC$ denote the image of the operator
\[ 
[\cQ, -] \colon \sconf(n|S, \CC)_1 \to \sconf(n|S, \CC)_0,
\]
arising from the adjoint action of~$\cQ$.
\end{dfn}

These two Lie algebras are related, as the notation -- drawn from homological algebra -- suggests.

\begin{prp}
The subspace $\mathfrak b_{\cQ}^\CC \sub \mathfrak z^\CC_{\cQ}$ is a (complex) Lie subalgebra.
In fact, it is an ideal, and so the quotient $\mathfrak z^\CC_{\cQ}/\mathfrak b^\CC_{\cQ}$ is a Lie algebra.
\end{prp}

As a result, we can identify $\mathfrak b_{\cQ}^\CC$ as the Lie algebra of a Lie subgroup $B_{\cQ}^\CC \sub Z_{\cQ}^\CC$ inside the complexification of $Z_{\cQ}$.  
Write $B_{\cQ}$ for the real Lie subgroup of $Z_{\cQ}$ obtained with the Lie algebra $\mf b_{\cQ} = \mf b_{\cQ}^{\CC} \cap \mf z_{\cQ}$.

\begin{rmk}
As we discussed in Section~\ref{conf_twist_section} we will be particularly interested in the complexified Lie algebra (the superconformal nilpotence variety is embedded in its the odd component), 
as well as real forms of the superconformal group (acting on the nilpotence variety).  
We define $Z_\cQ$ intrinsically as a subgroup of the real superconformal group.  On the other hand, for the image $B_\cQ$ our must start with the (complex) Lie algebra, so we are forced to make our definition starting with the complex subgroup $B_\cQ^\CC$ and then intersecting with a chosen real form.
\end{rmk}

\begin{proof}
Let $X \in \mathfrak z_{\cQ}$ and $Y = [\cQ, \cQ'] \in \mathfrak b_{\cQ}$. Then
\[
[\cQ, [X, \cQ']] = [[\cQ,X],\cQ'] + [X, [\cQ, \cQ']]  = [X, Y]
\]
so $[X,Y]$ is also in $\mathfrak b_{\cQ}$. This computation shows $\mathfrak b_{\cQ}$ is an ideal.
\end{proof}

\subsection{Observables and BRST Operators}

There are two major types of manifolds on which a superconformal field theory might live.
It might be defined on $\RR^{p,q}$ (as is usually implicit when a Lagrangian is given in coordinates),
or it might be defined on the conformal compactification~$C(\RR^{p,q})$.

\subsubsection{The Maximal Case}

To start let us suppose the theory $\cT$ lives on the biggest possible space, namely $C(\RR^{p,q})$.
In that case, we may suppose that the superconformal group acts as symmetries of the theory;
in particular, its even part provides a real Lie group acting as symmetries of the theory.

Let $\cT^\cQ$ denote the theory obtained from $\cT$ by twisting by the supercharge $\cQ$.  The full group of conformal transformations no longer acts on $\cT^\cQ$, but the group $Z_{\cQ}$ and its subgroup $B_{\cQ}$ continue to act.
Similarly, there are the infinitesimal symmetries given by actions of the Lie algebras $\mathfrak z_{\cQ}$ and $\mathfrak b_{\cQ}$, which act on operators and hence also on operator products.

Note that group elements of $\mr{SConf}_0$ that do {\em not} fix $\cQ$ are {\em not} symmetries.
Nor do non-$\cQ$-closed elements of $\sconf_0$ act as infinitesimal symmetries.
Twisting the theory changes the symmetries.

In the twisted theory, we focus only on the $\cQ$-closed operators (i.e., we only care about operators $\cO$ such that $\cQ \cO = 0$).
Moreover, a $\cQ$-exact operator $\cO$ (i.e., $\cO = \cQ \cO'$ for some operator $\cO'$) has trivial expected value.
Note, in particular, that if $\cO_1$ is exact with $\cQ \cO'_1 = \cO_1$, 
its product with any $\cQ$-closed operators $\cO_1 \cO_2 \cdots \cO_n$ is also exact:
\[
\cQ(\cO'_1 \cO_2 \cdots \cO_n) = \cO_1 \cO_2 \cdots \cO_n,
\]
so that such products also have vanishing expected value.

These features of the twisted theory mean we should study cohomology classes of operators~$[\cO]_\cQ$: 
that is, if $\widetilde{\cO} = \cO + \cQ \cO'$ (i.e., the operators differ by a $\cQ$-exact operator), 
then they are identified in the twisted theory $[\cO]_\cQ = [\widetilde{\cO}]_\cQ$.
These properties have strong consequences for how the symmetries act on operators.
For instance, in a topological twist of a supersymmetric theory,
an $n$-point function
\[
\langle \cO_1(x_1) \cO_2(x_2) \cdots \cO_n(x_n) \rangle
\]
is independent of the insertion points $\{x_i\}$ because translations are $\cQ$-exact.

The same argument implies that for a conformal twist, the operator product is unchanged if we move an insertion point along a $B_\cQ$-orbit:
\[
\langle \cO_1(x_1) \cO_2(x_2) \cdots \cO_n(x_n) \rangle = \langle \cO_1(b(x_1)) \cO_2(x_2) \cdots \cO_n(x_n) \rangle
\]
for any $b \in B_\cQ$.
In formulas, we see this claim infinitesimally: let $\beta = [\cQ,\cQ']$ be a $\cQ$-exact element of the superconformal algebra and compute
\begin{align*}
\langle (\beta \cO_1(x_1)) \cO_2(x_2) \cdots \cO_n(x_n) \rangle 
&= \langle \cQ(\cQ'\cO_1(x_1)) \cdots \cO_n(x_n) \rangle + \langle \cQ'(\cQ\cO_1(x_1)) \cdots \cO_n(x_n) \rangle\\
&= 0 + 0 = 0.
\end{align*}
Thus, the twisted theory $\cT^\cQ$ will behave like it is ``topological'' along $B_\cQ$-orbits, in the sense that moving insertions along those orbits does not affect operator products.

In contrast, the action of $Z_\cQ$ looks more like the action of translation in a supersymmetric theory before twisting.
For instance, let $\cO_1$ and $\cO_2$ be $\cQ$-closed operators and $z \in Z_\cQ$. 
The operator product satisfies
\[
\langle \cO_1(x_1) \cO_2(x_2) \rangle = \langle \cO_1( z(x_1) ) \cO_2( z(x_2)) \rangle,
\]
so that it is preserved by ``translation'' in the $z$-direction of {\em both} insertion points $x_1$ and $x_2$.  This is the same behaviour as we see in overall coordinate transformations in a Lorentz-invariant theory, the only difference here is that only transformations along a $Z_{\cQ}$-orbit are possible.

On the other hand, if we translate just one insertion, the expected value changes:
if we vary $z$, the expected value $\langle \cO_1( z(x_1) ) \cO_2(x_2) \rangle$ is an interesting function in~$z$, 
much as a 2-point function typically depends on the relative position of the insertion.

To summarize, we should analyze $C(\RR^{p,q})$ (and configurations of multiple points in it) into $Z_\cQ$-orbits and, further, into $B_\cQ$-orbits.
The twisted theory $\cT^\cQ$ will be ``constant'' along $B_\cQ$-orbits and hence behave like it is ``topological'' in those directions.
It will be equivariant along $Z_\cQ$-orbits.

Here we have emphasized the geometry of how these groups act on spacetime, but these groups also act, in some sense, ``internally,'', since the subgroups $Z_{\cQ}$ and $B_{\cQ}$ usually involve elements embedded diagonally in the product of the group of R-symmetries and the group of conformal symmetries.

\begin{rmk}
While the above discussion has emphasized the role of the real groups $B_\cQ$ and $Z_\cQ$ (and therefore the real Lie algebras $\mathfrak b_\cQ$ and $\mathfrak z_\cQ$ and the vector fields they generate), 
the \emph{complex} Lie algebras  $\mathfrak b_\cQ^\CC$ and $\mathfrak z_\cQ^\CC$ are frequently substantially larger.  
Even though these complex Lie algebras do not act {\em geometrically} on the conformal compactification, 
they still have a strong impact on the observables of the $\cQ$-twisted theory.  
A useful analogy to bear in mind is that on $\RR^2$ viewed as the complex plane,
there is a complex-valued vector field $\partial/\partial \bar{z}$ that is not geometric in the sense that it does not produce integral curves,
but it acts on complex-valued functions $C^\infty(\RR^2)^\CC$ and the kernel of that action is the ring of holomorphic functions.
We discuss this kind of phenomenon further in Section~\ref{reality_section}.
\end{rmk}

\subsubsection{Working on an Affine Patch}

Now suppose we pick an ``affine patch'' $\RR^{p,q} \subset C(\RR^{p,q})$. 
We mean here that we pick the image $g \RR^{p,q}$ of the defining $\RR^{p,q}$ under the action of some element $g$ of the conformal group.
On such a patch, a theory has a nice coordinate description, and so it can be analyzed very explicitly.
Note, however, that orbits of the groups $Z_{\cQ}$ and $B_{\cQ}$ are not contained in the affine patch.

Nonetheless, our above arguments about operators and their products carry over to this situation, with small modifications. 
For instance
if $\cO_1$ and $\cO_2$ are $\cQ$-closed operators and $z \in Z_\cQ$, 
then the operator product satisfies
\[
\langle \cO_1(x_1) \cO_2(x_2) \rangle = \langle \cO_1(z x_1) \cO_2( z x_2) \rangle,
\]
so long as the points $x_1, x_2$ and their translates $z(x_1), z(x_2)$ all live in the affine patch.
Similarly,
the operator product is constant if we move an insertion point along a $B_\cQ$-orbit intersected with the affine patch.

In the next subsection we explain how to recover vertex algebras, associative algebras, and other higher algebraic structures by picking an affine patch based on the behavior of the supercharge.
In particular we will choose a patch so that the closed and exact elements in the superconformal algebra include (complexified) translations of the affine patch.

We are, however, free to choose the affine patch independently from our choice of supercharge~$\cQ$,
so that translations may not closed or exact.

In general the exact Lie algebra $\mf b_{\cQ}$ defines a foliation of the conformal compactification $C(\RR^{p,q})$, 
which we may restrict to any affine patch.  
A $\cQ$-twisted theory will be locally constant along the leaves of this foliation.  
In the examples below these leaves are affine subspaces: 
for other affine patches, 
they will instead be subspaces generated by a subgroup of the group of conformal transformations.

For example, the conformal compactification possesses an automorphism that interchanges translations and special conformal transformations in the superconformal algebra.  
Given a topological twist by $\cQ$ as in Section~\ref{topological_section}, 
we may consider the twist by the image of $\cQ$ under this automorphism
or, equivalently, consider the $\cQ$ twist in the ``opposite'' affine patch.  
This twist will no longer be globally topological;
instead it will be locally constant along the foliation generated by the action of special conformal transformations.
(Since such a foliation might be singular,
there may be ``defects'' that appear.)

\subsection{Structures on Observables on Affine Subspaces}

Let us summarize some of the structures that can be realized on the algebra of local operators of a twisted theory in an affine subspace (i.e., an affine subspace of an affine patch).
For more detailed proofs we refer the reader to Section~\ref{sec:twistedobsmath}.

\subsubsection{Vertex Algebras} \label{vertex_subsection}

We begin with the structure that originally motivated the authors of \cite{Beem_2015} to study the observables in a conformal twist, that of a \emph{vertex algebra}.  
This structure occurs when there is a two-dimensional affine subspace, 
which we identify with a copy of the complex plane $\CC \hookrightarrow C(\RR^{p,q})$ with holomorphic coordinate~$z$.  
We require the following specific conditions on the twisting supercharge~$\cQ$:
\begin{enumerate}
 \item The group of isometries of $\CC$ should be closed (i.e. contained in $Z_{\cQ}$).
 \item The complexified translation $\dd/\dd \ol z$ in the antiholomorphic direction should be exact (i.e. contained in $\mf b_{\cQ}^\CC$).
\end{enumerate}
Note something interesting about the second condition: 
we have an infinitesimal symmetry of the observables (which form a complex vector space) that does {\em not} generate a flow on the real manifold.
We discuss this further in Section~\ref{reality_section}.

Such a twisting leads to vertex algebra.

\begin{thm}[See Theorem \ref{VA_theorem}]
If $\cQ$ satisfies conditions (1) and (2) for an affine subspace $\CC$,
then the algebra of local operators in a $\cQ$-twisted theory $\mc T^{\cQ}$, restricted to this affine subspace, has the structure of a vertex algebra, under some additional functional analytic conditions.
\end{thm}

We refer to Section \ref{VA_section} for a more precise statement.

\begin{rmk} \label{su112_rmk}
In the introduction to \cite{Beem_2015c}, it is observed that one way to obtain the desired structure of a vertex algebra in the twist is by choosing an embedding of $\su(1,1|2)$ into the superconformal algebra, 
so that $\su(1,1)$ acts on an affine subspace $\CC$ by antiholomorphic M\"obius transformations.  
From this point of view, the supercharge $\cQ$ is identified with an odd element of the complexification $\sl(2|2,\CC)$, 
which is chiral of rank $(2,0)$, 
and the subalgebra of $\cQ$-exact elements is a copy of $\sl(2,\CC)$ embedded diagonally in the even part of $\sl(2|2,\CC)$. 
The authors of \emph{loc. cit.} observe that such subalgebras only occur in a small number of examples of superconformal algebras (at least in Lorentzian signature), occurring in dimension 2, 4 and~6.
\end{rmk}

\begin{rmk}
Given a vertex algebra of local operators along an affine subspace $\CC$ in such a twisted theory, 
\cite{ALW} enhance the vertex algebra to a vertex algebra that includes \emph{extended operators} in the directions transverse to the chosen complex plane (though not extended \emph{within} the complex plane).
In their examples, these extended operators arise by using {\em real} even elements that generate the directions.
We do not prove general results about such algebras of extended operators in this paper.
\end{rmk}

\begin{rmk}
The examples of conformal twists studied by Beem {\it et al}, as discussed in Remark~\ref{su112_rmk} above, are designed so that the vertex algebra carries an action of $\sl(2,\CC)$.  
In fact, in their examples this action can be enhanced to an action of the Virasoro algebra (an infinite-dimensional Lie algebra).  
That is, the algebra of local operators is not just of a vertex algebra, but a vertex \emph{operator} algebra.  
It would be interesting to analyze exactly what conditions on the superconformal theory and a chosen conformal twist guarantee that this enhancement is possible.
\end{rmk}

\subsubsection{Topological Field Theories}
\label{topological_section}

In contrast, we can instead consider an affine subspaces $\RR^k \hookrightarrow C(\RR^{p,q})$ on which the $\cQ$-twisted theory is fully topological,
i.e., the entire algebra of translations of $\RR^k$ is $\cQ$-exact.  
Examples of this nature generalize the topological twists of supersymmetric field theories first studied by Witten \cite{Witten_1988}, 
but conformal twists allow us to expand the range of possible supercharges to which we may apply the twisting construction.

The structure of the local observables in a topological twist can be described naturally using the structure of an \emph{$\bb E_k$-algebra}.  
Let us briefly overview the idea behind these structure, 
though we will not give a complete definition here (see for instance \cite[Chapter 3.6.4]{CG1} and \cite[Chapter 4.1]{MarklShniderStasheff} for more details and context).

Let $A$ be a complex vector space, or maybe more generally a cochain complex.  
To give $A$ the structure of an $\bb E_k$-algebra is to define a multiplication operator $m_e \colon A^{\otimes j} \to A$ for each choice of embedding of $j$ disjoint $k$-balls into a bigger $k$-ball:
\[e \colon B_k(x_1,r_1) \sqcup B_k(x_2,r_2) \sqcup \cdots \sqcup B_k(x_j,r_j)  \hookrightarrow B_k(0,1),\] 
with $B_k(0,1)$ the $k$-dimensional open unit ball in $\RR^k$, and $B_k(x_i,r_i)$ is a sub-ball of radius $r_i$ and center $x_i$.  These multiplication operators should vary smoothly as we vary the centers $x_i$ and radii $r_i$, 
and they should be ``associative,'' 
meaning compatible under (rescaled) inclusions of $k$-balls.

Let us consider the simplest non-trivial example, that of an $\bb E_1$-algebra.  
We will see that this is just a ``homotopical'' version of an associative algebra.

\begin{ex}
Let $k=1$, so $B_1(x,r)$ is the open interval $(x-r,x+r) \sub \RR$.  If we fix a positive integer $j$, the space of embeddings of $j$ open intervals in the unit interval has connected components given by orderings on the set $\{1, \ldots, j\}$, with each component being contractible.

For example, if our multiplication operators $m_e$ are constant on each connected component, then we may view $m_2$ as describing an ordinary \emph{associative algebra} structure on $A$: the compatibility of the operations $m_e$ under concentric inclusion is equivalent to the ordinary associativity condition.
\end{ex}

\begin{rmk}
More generally, one may think of an $\bb E_k$-algebra as an associative algebra that is ``part-way'' to commutative. 
In other words, $\bb E_k$ interpolates between ``associative'' (corresponding to $\bb E_1$) and ``commutative'' (corresponding to $\bb E_\infty$).
For example, when we consider an $\bb E_2$ algebra, it has a binary product associated to a pair of two disjoint disks in the unit disk.
There is a path in the space of embeddings that swaps these two disks, which can be seen as ``commuting'' the two inputs of the product.  
However, there is still a residue of non-commutativity: if one moves one disk all the way around the other (i.e. applies the swap twice), this path describes a loop in the space of embeddings that is non-contractible. (i.e. a non-trivial element of $\pi_1$ of the space of possible binary multiplications).  
More generally in an $\bb E_k$ algebra the space of binary multiplications will have non-trivial topology represented by a non-trivial $(k-1)^\text{st}$ homotopy group.
\end{rmk}

\begin{rmk}
There is a variant version of an $\bb E_k$-algebra called a \emph{framed} $\bb E_k$-algebra relevant to oriented manifolds: 
instead of embeddings of $k$-balls, one studies framed embeddings.  
This amounts to keeping track of the action of the rotation group $\SO(k)$ on $A$.  
We will not go into detail here, except to point out that a framed $\bb E_k$-algebra may be used to construct a topological field theory on a general {\em oriented} $k$-manifold by the process of factorization homology as developed by Ayala and Francis~\cite{AyalaFrancis}.
See \cite{ScheimbauerThesis} for a full development of this idea.
\end{rmk}

\begin{thm}[{See Theorem \ref{E_k_theorem} and Corollaries \ref{E_k_cor_1}, \ref{E_k_cor_2}}]
If the algebra of translations of an affine patch $\RR^k$ is $\cQ$-exact and the action of dilations of $\RR^k$ is $\cQ$-exact, 
then the local operators on $\RR^k$ have the structure of an $\bb E_k$-algebra.  

If the metric restricted to the affine patch is positive definite and the action of rotations is also $\cQ$-exact, 
then the local operators on $\RR^k$ have the structure of a framed $\bb E_k$-algebra.
\end{thm}

\begin{ex}
For instance, if $k=1$, so we are restricting to a line along which the translation is exact, then the local operators naturally have the structure of an ordinary associative algebra.
\end{ex}

\subsubsection{Higher Dimensional Non-Topological Examples}

Finally, let us briefly consider a more expansive class of examples, where we study an affine patch of dimension $>2$ that is nevertheless not fully topological.  
\emph{Holomorphic} examples occur when there is an even real-dimensional affine subspace that we identify as a copy of $\CC^d \sub C(\RR^{p,q})$, all of whose isometries are $\cQ$-closed and whose anti-holomorphic complexified isometries are $\cQ$-exact. 
One may also study affine subspace of the form $\CC^d \times \RR^k \sub C(\RR^{p,q})$ where the $\cQ$-twist is holomorphic in the first factor --- in the sense we just outlined --- and topological in the second factor --- in the sense of the previous subsection.

Let us say more about one specific example, namely the case of an affine patch of the form $\CC \times \RR \hookrightarrow C(\RR^{p,q})$ with one holomorphic dimension and one topological dimension.  
Let us use $z$ for a holomorphic coordinate on $\CC$ and $t$ for a coordinate on $\RR$. 
Suppose we find a twisting supercharge $\cQ$ satisfying the following conditions:
\begin{enumerate}
 \item The group of isometries of $\CC \times \RR$ should be closed (i.e. contained in $Z_{\cQ}$).
 \item The complexified translation $\dd/\dd \ol z$ in the antiholomorphic direction of $\CC$ and the translation $\d/{\d t}$ in the topological direction should both be exact (i.e. contained in $\mf b_{\cQ}^\CC$).
\end{enumerate}
We expect then --- under the function-analytic conditions of Theorem~\ref{VA_theorem} in the $\CC$ direction --- that the local observables in the $\cQ$-twisted theory will have the structure of a \emph{raviolo vertex algebra} in the sense recently described by Garner and Williams~\cite{GWraviolo}.
See \cite{OhYagi} for a discussion in the language of Poisson vertex algebras, with many physical applications.

\subsubsection{Reality Conditions} \label{reality_section}

The superconformal Lie groups and algebras have real forms as well as complexifications, 
and it is useful to track what phenomena depend upon (or manifest) the real form or the complex form.

For a given {\em real} twisting supercharge~$\cQ$, 
we can study its \emph{real} Lie algebras of closed and exact elements $\mathfrak z_\cQ$ and $\mathfrak b_\cQ$.  
These have geometric significance: these real vector fields generate flows, and $\cQ$-closed observables are locally constant along these flows.  
This feature leads to strong constraints on expected values, correlation functions, and operator product expansions.
As a rich but rather simple example, we explore  4d $\mathcal N =2$ chiral supercharges in Section~\ref{Schur_section}.  
It would be interesting to explore more examples, 
but we do not address any in the present work.

More generally, for any twisting supercharge, there is still interesting structure on the observables of a twisted theory 
coming from closed and exact {\em complex} Lie algebra elements. 
As the observables form a complex vector space,
it is natural to work with this complexification of the superconformal Lie algebra.
The $\cQ$-closed observables are now invariant under the action of $\cQ$-closed or exact elements of the superconformal algebra,
and such a condition can still lead to nontrivial and powerful constraints.
We saw above in Section~\ref{vertex_subsection}, for example, 
how $\partial/\partial \bar{z}$ can appear as a $\cQ$-exact element and how that imposes holomorphicity on the $\cQ$-twisted observables,
leading to a vertex algebra.
This kind of phenomenon --- an analogue of holomorphicity --- is available in many of the situations we explore in this paper
but to fully characterize it requires developing analogues of vertex algebras in novel situations,
a problem well worth studying.

\subsection{Connections to the Physics Literature}

Twists of supersymmetric theories have been studied for a long time.
One of the first examples of twists, giving rise to a ``topological'' theory, 
was introduced with the work of Witten \cite{Witten_1988}, 
which provided a quantum field theoretic interpretation of Donaldson invariants of 4-manifolds.
This twist on $\RR^4$ is a nilpotent supercharge $\cQ$ where all the translation generators $P^\mu$ are in the exact subalgebra~$\mf b_{\cQ}$ and hence gives rise to a purely topological theory.

In superconformal theories, there are more complicated examples of square-zero odd elements $\cQ$ in the superconformal algebra,
involving linear combinations of supersymmetric and conformal supercharges.
Thus, there is a more complicated structures of $\cQ$-closed and $\cQ$-exact generators in the even part $\sconf(n|S,\CC)_0$, giving rise to more interesting subvarieties where the protected sector of operators lives.
For a physicist, the essential point is this:
\begin{quotation}
{\it The type of generators that appear in the closed and exact subalgebras determine how the twisted theory will look and behave.}
\end{quotation}
As an example, there is a conformal twist of 4d $\mathcal{N}{=}2$ superconformal theories, described in \cite{Beem_2015},
that has been the launching pad for a slew of works studying algebraic structures corresponding to protected sectors within superconformal theories.
In this example, the nilpotent element is a combination of a super Poincar\'e charge and a superconformal charge, so it only exists in superconformal theories (but not in supersymmetric ones).
The structure of the closed and exact subalgebras implies that only some of the translations are closed and only some are exact, 
giving rise to an interesting ``mixed'' structure, 
where local operators from the 4d theory are restricted to a 2d complex plane in space-time.
The holomorphic coordinate on this plane corresponds to a closed (complex) translation, whereas the antiholomorphic coordinate, together with one of the real transverse translations, are in the exact subalgebra. The quotient $\mf z_{\cQ}/\mathfrak b_{\cQ}$ is an $\mathfrak{sl}_2$ algebra, that is then extended to a Virasoro algebra by identifying a candidate stress-energy tensor, whose modes $\{L_m\}$ give rise to an infinite chiral symmetry algebra.
The restriction of local operators to the complex plane and the action of this algebra provides a correspondence between (protected sectors of) local operators of $4d \,\mathcal{N}{=}2$ SCFTs and 2d vertex operator algebras.

Since this seminal work, the lore has been that the correspondence between $4d ~ \mathcal{N}{=}2$ SCFTs and vertex operator algebras is one-to-one: 
that is, given a vertex operator algebra, one can reconstruct the parent SCFT. 
This claim, however, cannot be generically true. 
The $4d ~ \mathcal{N}{=}4$ theories have a conformal manifold, 
and each point in it corresponds to a different SCFT, 
while the VOA is insensitive to exactly marginal couplings.
In fact, the VOA obtained after the twist is the same for all points in the conformal manifold.
Furthermore, the VOA data is only sensitive to the local operator data. Again in the case of $4d ~ \mathcal{N}{=}4$, theories are distinguished by global forms of gauge groups, which depend on the spectrum of line operators~\cite{Aharony_2013}.
(These perturbative descriptions live at corners of the aforementioned conformal manifolds.)
Furthermore, by including line operators in the conformal twist of $4d ~ \mathcal{N}{=}2$ SCFTs (new cohomology classes relative to the ones obtained from local operators), 
one obtains a bigger, richer vertex algebra structure that contains the local operator VOA \cite{ALW}. 
The map from 4d SCFT to VOA can be further modified by imposing additional restrictions on the local operators allowed after the twist, 
and this gives rise to a vertex algebra that is a ``square root'' of the original VOA~\cite{Buican}.

\begin{rmk}
If one considers extended operators in this setting, it is also important to consider that these operators can stretch to infinity. 
In this case the signature of the underlying space-time may have a considerable impact on the behavior of the twisted theory, 
compared to the case when one only considers local operators \cite{ALW}.
We will not focus on these subtle features in the present paper and work as much as possible with complex algebras and the largest embedding spaces possible. 
It is worth noting, however, that when writing explicitly a physical theory, these subtleties ought to be addressed.
\end{rmk}

Enlarging the amount of supersymmetry, relative to the Beem {\it et al.} case, gives rise to additional examples, 
notably the Kapustin-Witten twist\cite{KapustinWitten}. 
In this case, although the nilpotent element is a combination of Poincar\'e supercharges, 
an identification with the enlarged R-symmetry is used, giving rise to a 2d topological theory.
Other types of twists of $\mathcal{N}{\geq}2$ SCFTs have been introduced by putting the theory on a non-trivial background and twisting by it (i.e., an $\Omega$-background) effectively localizing the theory on a lower dimensional locus in space-time~\cite{Nekrasov}.
Nevertheless, in the case of 4d $\mathcal{N}{=}4$ SYM, 
a twist of this kind can be equivalently realized without specifying any background, by simply using a particular combination of supercharges. 
This localizes the theory on a 2d slice, as in the Beem {\it et al.} case, 
but with an additional 2d transverse quasi-topological sector~\cite{DedGai}.

Table~\ref{tab:example_twists} provides a sample of the conformal twists that have appeared in the physics literature.
See \cite{DedGai} for a more thorough discussion (and many pertinent references).
We exclude those twists of SCFTs that are generated by nilpotent elements constructed purely from Poincar\'e supercharges, since these are better known and have already been classified (for a similar table of examples in the super Poincar\'e case, see Tables 1--4 of~\cite{ElliottSafronovWilliams}).

\begin{rmk}
Notice that the conformal twists appearing in the table are all studied in the context of the algebra of operators after restriction to either a real line $\RR$ or a complex line $\CC$.  From the point of view taken in this paper there are also interesting structures to investigate on the operators of subspaces of larger dimension, for instance the structure of an $\bb E_k$ algebra on the operators on a topological real $k$-plane.
\end{rmk}

\begin{table}
\begin{tabular}{c|c|c|c}
{\it Bulk theory} & {\it Name} & {\it Geometry of the twist} & {\it References}\\
\hline
3d $\mathcal N=4$ on $\RR^3$  & 1d protected associative algebra & $\RR \times\RR^2$ & \cite{Beem_2017} \\
4d $\mathcal N=2$ on $\RR^4$  & 2d Vertex Operator Algebra & $\CC\sim \RR^2 \subseteq \RR^4$ & \cite{Beem_2015} \\
4d $\mathcal N=4$ on $\RR^4$  & A/B 2d TQFTs & $\CC \times \Sigma$ & \cite{KapustinWitten} \\
4d $\mathcal N=4$ on $\RR^4$  & A/B 2d quasi-topological sector & $\CC \times \RR^2$ & \cite{DedGai} \\
4d $\mathcal N=2$ on $\RR^{1,3}$  & 2d Extended VOA & $\CC\sim \RR^2 \subseteq \RR^{1,3}$ & \cite{ALW}\\
6d $\mathcal N=(2,0)$ on $\RR^6$  & 2d VOA & $\CC\sim \RR^2 \subseteq \RR^6$ & \cite{Beem_2015c}
\end{tabular}
\caption{Occurences of superconformal twists of superconformal theories in the prior literature, classified by the amount of superconformal symmetry and the class of superconformal supercharge by which the twisting is performed.}
\label{tab:example_twists}
\end{table}

\section{4d Superconformal Twists}

In this section we explore the 4-dimensional case.
We begin by describing each complex nilpotence variety and then analyze its orbits under the action of the associated complex superconformal group,
including the closed and exact transformations for many twists.
These results are available in \cite{DufloSerganova} but we give the arguments here,
both to demonstrate how concrete and accessible they are and also to describe them in a way convenient for later results.

Note, however, that the case of the $\mathcal{N}{=}\hspace{.1em}4$ superconformal algebra is not included in Theorem~\ref{thm DS}, it provides some novel twists of $\mathcal{N}{=}\hspace{.1em}4$ super Yang--Mills theory that might be fruitful to explore.

We then turn to analyzing the orbits under the action of real forms of the  superconformal groups,
which depend upon the signature of the spacetime.
For these results we leverage work of Fels, Huckleberry, and Wolf.

\subsection{Complex Nilpotence Variety for \texorpdfstring{$\mc N \neq 4$}{N=4}}
\label{4d_complex_section}

Consider the complexified 4d $\mc N=k$ superconformal algebra $\sconf(4|k,\CC)$ for $k \ne 4$.
(The $k = 4$ case is distinct, and we discuss in the next subsection.)  
The superconformal algebra can be identified with the simple super Lie algebra $\sl(4|k)$.  
The even part of $\sconf(4|k,\CC)$ is isomorphic to
\begin{align*}
(\sconf(4|k,\CC))_0
&\iso \sl(4,\CC) \oplus \sl(k,\CC) \oplus \CC \\
&\iso \so(6,\CC) \oplus \sl(k,\CC) \oplus \CC
\end{align*}
where we use the exceptional isomorphism between $\sl(4,\CC)$ and $\so(6,\CC)$.  
The odd part of $\sconf(4|k,\CC)$ is isomorphic to
\begin{align*}
(\sconf(4|k,\CC))_1
&\iso V_4 \otimes W_k \oplus V_4^* \otimes W_k^* \\
&\iso S^{(6)}_+ \otimes W_k \oplus S^{(6)}_- \otimes W_k^*
\end{align*}
where $W_k$ denotes the defining $k$-dimensional representation of the factor $\sl(k,\CC)$ 
and $V_4$ denotes the four-dimensional defining representation of $\sl(4,\CC)$, 
which is isomorphic to the positive semispin (or Weyl spinor) representation $S^{(6)}_+$ of~$\so(6,\CC)$.  
The one-dimensional factor in the center of $(\sconf(4|k,\CC))_0$ acts trivially on $(\sconf(4|k,\CC))_0$ but it acts nontrivially on~$(\sconf(4|k,\CC))_1$, 
as we will see below.

The Lie bracket of even elements is given by the direct sum of the natural brackets,
and the even part acts on the odd part as determined by the representation structure just stated.
It remains to describe the bracket between odd elements in explicit terms.

For any $k$, there is a $\sl(k,\CC)$-invariant decomposition 
\[
\hom(W_k,W_k) \iso W_k \otimes W_k^* \iso \sl(k,\CC) \oplus \CC.
\]  
Denote the linear projections onto the two factors by 
$$\tr \colon W_k \otimes W_k^* \to \CC,$$ 
which correspond to matrix operation of trace $M \mapsto \tr(M)$, 
and 
$$\mr{red} \colon W_k \otimes W_k^* \to \sl(k,\CC),$$
which corresponds to ``making a matrix trace-free'' $M \mapsto M - (\tr(M)/k) \mr{Id}$.
There is an analogous $\sl(k,\CC)$-invariant decomposition involving~$V_4$.

These decompositions lead to a convenient description of the bracket between odd elements.

\begin{lmm}
Under the isomorphism
\[
V_4 \otimes W_k \oplus V_4^* \otimes W_k^* \iso \hom(V_4^*, W_k) \oplus \hom(W_k, V_4^*)
\]
expressing an odd element $\cQ$ as sum of matrices $\cQ_+ + \cQ_-$,
the bracket is 
\begin{align*}
[\cQ,\cQ'] 
&= [\cQ_+ + \cQ_-, \cQ'_+ +\cQ_-] \\
&= \mr{red}(\cQ_- \circ \cQ'_+ + \cQ'_- \circ \cQ_+) + 
\mr{red}(\cQ_+ \circ \cQ'_- + \cQ'_+ \circ \cQ_-) \\
&\quad\quad+ (\tr((\cQ_- \circ \cQ'_+ + \cQ'_- \circ \cQ_+) + \tr(\cQ_+ \circ \cQ'_- + \cQ'_+ \circ \cQ_-))
\end{align*}
in $\sl(4,\CC) \oplus \sl(k,\CC) \oplus \CC$.
\end{lmm}

\begin{proof} 
Any odd element decomposes into a sum of terms from each summand.
A pure tensor in the first summand has the form $Q_+ \otimes w_+$ with $Q_+ \in S^{(6)}_+$ and $w_+ \in W_k$.
A pure tensor in the second summand has the form $Q_- \otimes w_-$ with $Q_+ \in S^{(6)}_+$ and $w_- \in W_k^*$.
Pairs of elements in the same summand of $(\sconf(4|k,\CC))_1$ have trivial bracket.
The Lie bracket between a pair of elements in opposite summands $[Q_+ \otimes w_+ , Q'_- \otimes w'_-]$ is given by
\begin{align*}
\mr{red}(Q_+ \otimes Q'_-) \tr(w_+ \otimes w'-) + \tr(Q_+ \otimes Q_-') \mr{red}(w_+ \otimes w'_-) + \tr(Q_+ \otimes Q'_-) \tr(w_+ \otimes w'_-).
\end{align*}
It is then direct to verify the two descriptions of the bracket match under the isomorphism.
\end{proof}

It is now straightforward to compute the superconformal nilpotence variety.

\begin{prp} 
\label{4d_complex_nilpotence_prop}
The complex nilpotence variety $\Nilp_{(4|k)}(\CC)$ for the 4d $\mc N=k$ superconformal algebra with $k \ne 4$ is
\[  \{ \cQ_+ + \cQ_- \in \hom(V_4^*, W_k) \oplus \hom(W_k, V_4^*) \colon \cQ_+ \circ \cQ_- = 0 \text{ and } \cQ_- \circ \cQ_+ = 0\}. \]
\end{prp}

\begin{proof}
In terms of the linear maps $\cQ_+, \cQ_-$ we have
\[ [\cQ, \cQ] = 2[\cQ_+, \cQ_-], \]
which equals zero if
\begin{align*}
\mr{red}(\cQ_- \circ \cQ_+) &= 0 \in \sl(V_4^*) \\
\mr{red}(\cQ_+ \circ \cQ_-) &= 0 \in \sl(W_k) \\
\tr(\cQ_- \circ \cQ_+) + \tr(\cQ_+ \circ \cQ_-) &= 0 \in \CC.
\end{align*}
The first two conditions require that $\cQ_- \circ \cQ_+$ and $\cQ_+ \circ \cQ_-$ are both multiples of an identity matrix,
and the last condition requires the sum of their traces to be zero.

Now we use the condition that $k \ne 4$.  
This tells us that either $\cQ_- \circ \cQ_+$ has rank less than 4 (if $k < 4$) or $\cQ_+ \circ \cQ_-$ has rank less than $k$ (if $k > 4$).  
In either case, the only scaled identity matrix with less than full rank is the zero matrix, so we must have $\tr(\cQ_- \circ \cQ_+) = -\tr(\cQ_+ \circ \cQ_-) =0$, 
and hence $\cQ_- \circ \cQ_+$ and $\cQ_+ \circ \cQ_-$ both equal zero.
\end{proof}

\subsection{The \texorpdfstring{$\mc N=4$}{N=4} Case}
\label{4d_N4_section}

The 4d $\mc N=4$ superconformal algebra is slightly different,
because  the super Lie algebra $\sl(4|4)$ is not simple: 
it has a one-dimensional center generated by the element $\mr{diag}(1,1,1,1|1,1,1,1)$.  
In order to obtain a {\em simple} super Lie algebra, 
we should take the quotient by this element to obtain the super Lie algebra $\psl(4|4)$.  
Thus, we identify the 4d $\mc N=4$ complex superconformal algebra $\sconf(4|4,\CC)$ as~$\psl(4|4)$.

\begin{rmk}
If one is willing to work with non-simple superconformal algebras, 
then one can consider $\sl(4|4)$,
which we treated in the preceding section.
\end{rmk}

The even part of $\sconf(4|4,\CC)$ can be identified with
\[(\sconf(4|4,\CC))_0 \iso \sl(4,\CC) \oplus \sl(4,\CC) \iso \so(6,\CC) \oplus \sl(4,\CC).\]
The odd part of $\sconf(4|4,\CC)$ can be identified with
\[(\mf{sconf}(4|k,\CC))_1 \iso V_4 \otimes W_4 \oplus V_4^* \otimes W_4^* \iso S^{(6)}_+ \otimes W_4 \oplus S^{(6)}_- \otimes W_4^*\]
identically to the description from the previous subsection.
The Lie bracket between odd elements is similar to the $\mc N\ne 4$ case,
except that we do not need the sum of trace term.  

\begin{prp}
\label{4d_4_complex_nilpotence_prop}
The complex nilpotence variety $\Nilp_{(4|4)}(\CC)$ for the 4d $\mc N=4$ superconformal algebra is 
\[
C_{\tr} \cup_{\{0\}} C_{\mr{red}},
\]
the union of the two cones at the origin where
\[
C_{\tr} = \{Q_+ \otimes w_+ + Q_- \otimes w_- \colon \tr(Q_+ \otimes Q_-) = \tr(w_+ \otimes w_-) = 0\}
\]
and
\[
C_{\mr{red}} = \{Q_+ \otimes w_+ + Q_- \otimes w_- \colon \mr{red}(Q_+ \otimes Q_-) = \mr{red}(w_+ \otimes w_-) = 0\}.
\]
\end{prp}

\begin{proof}
 The right-hand side is contained in the left-hand side, this is immediate from the description of the Lie bracket.  If $Q = Q_+ \otimes w_+ + Q_- \otimes w_-$ is a square zero element then $\tr(Q_+ \otimes Q_-)$ if and only if $\tr(w_+ \otimes w_-) = 0$ by the same argument as Proposition \ref{4d_complex_nilpotence_prop}.  If neither trace term equals zero and $[Q,Q]=0$ then we must have $\mr{red}(Q_+ \otimes Q_-) = \mr{red}(w_+ \otimes w_-) = 0$.
\end{proof}

\begin{rmk}
Let us discuss the subspace of this nilpotence variety consisting of nilpotent elements in the super Poincar\'e algebra associated to a choice of embedding $\so(4,\CC) \sub \so(6,\CC)$.  Under this embedding the representation $V_4$ decomposes as $S^{(4)}_+ \oplus S^{(4)}_-$, where $S^{(4)}_\pm$ are the two semispin representations of $\so(4,\CC)$ (each of complex dimension~2). 
The complex nilpotence variety for the $\mc N=4$ super Poincar\'e algebra is isomorphic to
\[
\Nilp_{(4|4)}^{\mf{siso}}(\CC) 
= \{Q'_+ \otimes w_+ + Q'_- \otimes w_- \colon \mr{Tr}(w_+ \otimes w_-) = 0 \} \sub C_{\mr{Tr}},
\]
where $\mc Q'_+ \in S^{(4)}_+ \sub S^{(6)}_+ \iso V_4$ and $\mc Q'_- \in S^{(4)}_- \sub S^{(6)}_- \iso V_4^*$.  
Note that the condition $\mr{Tr}(Q'_+ \otimes Q'_-) = 0$ is automatically satisfied.  The decomposition of this super Poincar\'e nilpotence variety into $\spin(4,\CC) \times \SL(4,\CC)$-orbits is interesting, including for instance the one-parameter families of topological supercharges studied by Kapustin and Witten \cite{KapustinWitten}.  This decomposition is discussed in \cite[Section 4.4]{ElliottSafronov} and~\cite[Section 9]{EagerSaberiWalcher}.
\end{rmk}

\subsection{Complex Group Orbits}

The complex superconformal group is isomorphic to $\SL(4|k,\CC)$ if $k \ne 4$, with even part isomorphic to
\[
\mr{SConf}(4|k,\CC)_0 \iso \{(A,B) \in \GL(4,\CC) \times \GL(k,\CC) \colon \det(A)\det(B) = 1\}.
\]
It might be helpful to see that there is a group map
\[
\SL(4,\CC) \times \SL(k,\CC) \times \CC^\times \to \mr{SConf}(4|k,\CC)_0
\]
sending $(M,N,\lambda)$ to $(\lambda M, \lambda^{-1} N)$.

\begin{rmk}
This map is locally a diffeomorphism.  Indeed, taking in a neighborhood of matrices $(A,B)$ where $\det(A)$ lies in a small disk in $\CC^\times$ and choosing a local fourth root in this disk,  
we can define a local inverse sending $(A,B)$ to $(A/\det(A), B/\det(B), \det(A)^{1/4})$.
\end{rmk}

In the $k=4$ case the superconformal group is identified with $\PSL(4|4,\CC)$, 
whose even part is isomorphic to~$\SL(4,\CC) \times \SL(4,\CC)$.

These super groups act on the super Lie algebra $\mf{sconf}(4|k,\CC)$ by the adjoint action.
In explicit terms,
an element $(A,B,\lambda) \in \SL(4,\CC) \times \SL(k,\CC) \times \CC^\times$ acts on
\[(\mf{sconf}(4|k,\CC))_1 \iso \hom(W_k^*, V_4) \oplus \hom(W_k, V_4^*)\]
by
\[
(\mc Q_+, \mc Q_-) \mapsto (\lambda^{-2}B\mc Q_+ A^{-1}, \lambda^2A \mc Q_- B^{-1}).
\]
This action manifestly preserves the {\em rank} $r_\pm$ of each operator $\mc Q_\pm$,
and for $k \neq 4$, these invariants classify the orbits in~$\Nilp_{(4|k)}(\CC)$.

\begin{prp}
For $k \neq 4$, there is a bijection
\[
\Nilp_{(4|k)}(\CC)/\mr{SConf}(4|k,\CC)_0 \iso \{ (r_+,r_-) \in \NN^2 \, : \, r_+ + r_- \le \min(4,k) \},
\]
where a twist $(\mc Q_+, \mc Q_-)$ maps to $(\mr{rank}(\mc Q_+), \mr{rank}(\mc Q_-))$.
\end{prp}

That is, the orbits of $\Nilp_{(4|k)}(\CC)$ are in bijection with pairs $(r_+,r_-)$ of non-negative integers where $r_+ + r_- \le \min(4,k)$.
Note that we use $\Nilp_{(4|k)}(\CC)/\mr{SConf}(4|k,\CC)_0$ to denote the quotient set, in contrast to the quotient stack.
(The quotient topology is not discrete (and so non-Hausdorff):
it is given by the poset structure, as having rank $\geq k$ is an open condition.)

This result suggests the following notation.

\begin{dfn}
Let $\mc O(4|k)_{r_+, r_-}$ denote the complex group orbit associated to a pair $(r_+,r_-)$ of ranks. 
\end{dfn}

\begin{proof}
Given a square-zero supercharge $(\cQ_+, \cQ_-)$, we assign to it a pair of non-negative integers by letting $r_\pm = \mr{rank}(\cQ_\pm)$.  
Since $\mr{Im}(\cQ_+) \sub \ker(\cQ_-)$,
the rank-nullity theorem tells us $r_+ \leq k - r_-$ so we have $r_+ + r_- \le k$.
Likewise, since $\mr{Im}(\cQ_-) \sub \ker(\cQ_+)$, we have $r_+ + r_- \le 4$.

Now, we may choose bases (or apply row/column operations) for $V_4^*$ and $W_k$ so that $\cQ_+$ and $\cQ_-$ are represented by matrices of the following form:
\begin{equation} \label{diag_matrices_2}
\cQ_+ = \pmat{\mr{id}_{r_+} &0 \\ 0 & 0}, \ \cQ_- = \pmat{0&0\\0&\mr{id}_{r_-}}.
\end{equation}
Note that these are not square matrices: we write 0 to refer to rectangular zero matrices of appropriate sizes.

In other words, the action of $\GL(4,\CC) \times \GL(k,\CC)$ on the locus of square-zero supercharges of ranks $(r_+, r_-)$ is transitive.  
To complete the proof, we just need to ensure that we can choose our two change of basis matrices to have the same determinant.  
Alternatively, it is enough to find an element $(X,Y) \in \GL(4,\CC) \times \GL(k,\CC)$ stabilizing the diagonal matrices \ref{diag_matrices_2} 
where $\det(X)/\det(Y) = \lambda \in \CC^\times$ for all values of~$\lambda$.

To do this, again use the fact that $k \ne 4$.  
If $k < 4$ then $r_+ + r_- < 4$, and we let $X = \mr{diag}(1, \ldots, 1, \lambda, 1, \ldots, 1)$ with $\lambda$ in the $r_+ + 1^{\text{st}}$ position and $Y = \mr{id}_k$.  
Likewise if $k > 4$ then $r_+ + r_- < k$, and we let $Y = \mr{diag}(1, \ldots, 1, \lambda^{-1}, 1, \ldots, 1)$ with $\lambda^{-1}$ in the $r_+ + 1^{\text{st}}$ position, and $X = \mr{id}_4$.
\end{proof}

Something interesting happens, however, when $k = 4$:
the orbits are not a finite set, but instead come in continuous families.
In other words, this case does {\em not} sit among the examples covered by Theorem~\ref{thm DS} of Duflo and Serganova.

\begin{prp}
There is an isomorphism
\[
\Nilp_{(4|4)}(\CC)/\mr{SConf}(4|4,\CC)_0 
\cong \{ (r_+,r_-) \in \ZZ_+^2 \, : \, r_+ + r_- \leq 4 \} \sqcup \CC^\times \sqcup \CC^\times.
\]
When both terms $\mc Q_\pm$ have positive rank,
the orbit of the twist $(\mc Q_+, \mc Q_-)$ is an isolated point.
When one term has rank 0 and the other has rank 4 (i.e., $(r_+,r_-) = (4,0)$ or $(0,4)$),
then the orbit sits in a one-parameter family labeled by the determinant of the full rank term (e.g., by $\det(\mc Q_+)$ in the $(4,0)$~case).
\end{prp}

We will use $\mc O(4|4)_{r_+, r_-}$ to denote these components,
but $\mc O(4|4)_{4,0} \iso \CC^\times$, for instance.
Note that the origin in a copy of $\CC^\times$ corresponds to the limit of a family where the determinant vanishes and hence the rank drops.
Thus one can view the $(4,0)$ case (or $(0,4)$ case) as $\CC$ along with some points added that ``thicken'' the origin.

\begin{proof}
The argument in the $k=4$ case carries over for matrices with positive ranks.
In the case of ranks $(4,0)$, the first matrix $\mc Q_+$ is full rank,
but the action of $\SL(4,\CC)$ from the left (respectively, right) does not change the determinant.
On the other hand, these actions by $\SL(4,\CC) \times \SL(4,\CC)$ by row and column operations can be used to put $\mc Q_+$ into a diagonal form with three entries of 1s and a final entry~$\det(\mc Q_+)$.
\end{proof}

\subsection{Closed and Exact Transformations}

Let us discuss some specific properties of the complex orbits in some of the low rank cases.

\begin{enumerate}
\item  \textbf{Rank $(0,0)$}: 
The zero supercharge always squares to zero.   
The kernel of $[\cQ,-]$ is $\so(6,\CC) \oplus \sl(k,\CC) \oplus \CC$ (the entire even component), 
and the image of $[\cQ,-]$ is just~$\{0\}$.

\item \textbf{Rank $(r_+, 0)$ and $(0, r_-)$}: 
We refer to these as \emph{chiral} supercharges, 
since either $\cQ_+$ or $\cQ_-$ vanishes.  
Such supercharges automatically square to zero, 
so these strata in $\Nilp_{(4|k)}(\CC)$ are isomorphic to $V_4 \otimes W_4 \backslash \{0\}$ and $V_4^* \otimes W_4^* \backslash \{0\}$, respectively.

The stabilizer $Z_{\cQ} \sub \SL(4,\CC) \times \SL(k,\CC) \times \CC^\times$ can be described in the following way.  
Let $P_{r,k} \sub \SL(k,\CC)$ be the parabolic subgroup with block diagonal Levi subgroup $\SL(r,\CC) \times \SL(k-r,\CC)$.  
Write
\[F_r \colon P_{r,4} \times P_{r,k} \to \SL(r,\CC)\]
for the homomorphism given by projection onto $\SL(r,\CC) \times \SL(r,\CC)$ 
(i.e., projection onto the upper-left block in each factor) 
composed with the product map $\SL(r,\CC) \times \SL(r,\CC) \to \SL(r,\CC)$.

\begin{prp}
If $\cQ$ is a chiral supercharge of rank $(r,0)$, 
then the stabilizer $Z_{\cQ}$ is isomorphic to $F_r^{-1}(I) \times \CC^\times$.
\end{prp}

\begin{proof}
We can check this by choosing a basis in which $\cQ$ is represented by a block diagonal matrix of rank~$r$.  
The stabilizer $Z_{\cQ}$ is then readily computed by solving
\[
\left(\begin{array}{c|c} 
A_1 & A_2 \\ \hline A_3 & A_4
\end{array}\right)
    \left(\begin{array}{c|c}
     I_r & 0 \\ \hline 0&0
    \end{array}\right)
    \left(\begin{array}{c|c}
     B_1 & B_2 \\ \hline B_3 & B_4
    \end{array}\right) = \left(\begin{array}{c|c}
     I_r & 0 \\ \hline 0&0
    \end{array}\right).
\]
Here 
$A = 
\left(\begin{array}{c|c}
A_1 & A_2 \\ \hline A_3 & A_4
\end{array}\right)$ 
is an element of $\SL(4,\CC)$ and 
$B = 
\left(\begin{array}{c|c}
B_1 & B_2 \\ \hline B_3 & B_4
\end{array}\right)$ 
is an element of $\SL(k,\CC)$. 
(For instance, $B_1$ is an $r \times r$ block, $B_2$ is an $r \times (k-r)$ block, and so on.)  
We find $A_1 = B_1^{-1}$ and $A_3 = B_2 = 0$.  
In other words $(A,B)$ lies in~$F_r^{-1}$.
\end{proof}

In particular, the dimension of $Z_{\cQ}$ is given as
\[\dim Z_{\cQ} = k^2 - kr + r^2 - 4r + 14\]
if $r \ne k$ and
\[\dim Z_{\cQ} = k^2 - kr + r^2 - 4r + 15\]
if $r=k$.

Similarly, we can describe the Lie algebra $\mf b_{\cQ} = \mr{Im}([\cQ, -])$ as a subalgebra of $\mf z_{\cQ} = \mr{Lie}(Z_{\cQ})$,
by a similar direct computation.  
We find the following.

\begin{prp}
Let $\cQ$ be a chiral supercharge of rank~$r$.  
The Lie algebra 
\[
\mf b_{\cQ} \sub \sl(4,\CC) \oplus \sl(k,\CC) \sub \sl(4,\CC) \oplus \sl(k,\CC) \oplus \CC 
\]
is isomorphic to the Lie algebra of block matrices of the following form:
\begin{align*}
\mf b_r 
&= \Bigg\{\left(\left(
\begin{array}{c|c}
\alpha & \beta \\ 
\hline 
0&0
\end{array}\right), 
\left(\begin{array}{c|c}
\alpha & 0 \\ 
\hline 
\gamma & 0
\end{array}\right), \lambda \right)
\in \sl(4,\CC) \oplus \sl(k,\CC) \oplus \CC \colon \\
&\qquad \qquad \alpha \in \sl(r,\CC), \beta \in \mr{Mat}_{r \times (4-r)}(\CC),\gamma \in \mr{Mat}_{(k-r) \times r}(\CC)\Bigg\}.
\end{align*}
\end{prp}

In particular, we have
\[\dim \mf b_{\cQ} = kr - r^2 + 4r.\]

For reference, we give a tabular description of the dependence of the dimension of $\mf z_{\cQ}$ and $\mf b_{\cQ}$ on $r$.  
Recall that $r \le \min(4,k)$.  
Thus the dimensions for each $k$, in terms of $r$, are given in Table~\ref{4d_chiral_table}.
\begin{table}[!h]
\begin{tabular}{c|c|c|c}
$r$ &$\dim \mf z_{\cQ}$ &$\dim \mf b_{\cQ}$ &$\dim \mf z_{\cQ}/\mf b_{\cQ}$  \\ \hline
   1 &$k^2-k+11 + \delta_{kr} - \delta_{k4}$&$k+3$&$k^2-2k+8 + \delta_{kr} - \delta_{k4}$ \\
   2 &$k^2-2k+10 + \delta_{kr} - \delta_{k4}$&$2k+4$&$k^2-4k+6 + \delta_{kr} - \delta_{k4}$ \\
   3 &$k^2-3k+11 + \delta_{kr} - \delta_{k4}$&$3k+3$&$k^2-6k+8 + \delta_{kr} - \delta_{k4}$ \\
   4 &$k^2-4k+14 + \delta_{kr} - \delta_{k4}$&$4k$&$k^2-8k+14 + \delta_{kr} - \delta_{k4}$
\end{tabular}
\caption{For chiral supercharges $\cQ$ of rank $r$, we give the dimension of the kernel and image of $[\cQ,-]$ as well as the dimension of the quotient.  }
\label{4d_chiral_table}
\end{table}

We note from this computation that the inclusion $\mf b_{\cQ} \sub \mf z_{\cQ}$ is always proper, at least for $k \ne 4$. 
In the $k=\mc N=4$ case the computation remains the same, but $\dim \mf z_{\cQ}$ is decreased by one.  The inclusion $\mf b_{\cQ} \sub \mf z_{\cQ}$ is still proper, 
with the exception of the maximal rank $r=4$ case where $\mf b_{\cQ} = \mf z_{\cQ}$ and
they span the diagonal copy of $\sl(4,\CC) \sub \sl(4,\CC) \oplus \sl(4,\CC)$.

\item \textbf{Rank $(1,1)$}: 
For $k > 1$, let $\mc Q$ be a square zero supercharge of the form $Q_+ \otimes w_+ + Q_- \otimes w_-$.  
The square-zero condition in this case is equivalent to the conditions $\tr(w_+, w_-) = 0 = \tr(Q_+, Q_-)$.  
This determines an analysis of the stabilizer and the image of a supercharge similar to what we have done already.

\begin{prp}
Let $\cQ = Q_+ \otimes w_+ + Q_- \otimes w_-$ be a rank $(1,1)$ square zero supercharge. 
The image $\mf b_{\cQ} \sub \sl(4,\CC) \times \sl(k,\CC) \times \CC$ of the linear map $[\cQ, -]$ is isomorphic to
\[\mf b_{\cQ} = \{(A,B) \in M_{21} \times M_{12} \colon \tr(A) + \tr(B) = 0\},\]
where for $i \ne j$, $M_{ij} \sub \gl(n,\CC)$ is the Lie subalgebra of matrices whose non-zero entries all lie in row $i$ or column~$j$.
\end{prp}

\begin{prp}
Let $\cQ = Q_+ \otimes w_+ + Q_- \otimes w_-$ be a rank $(1,1)$ square zero supercharge.   
The kernel $\mf z_{\cQ}$ is the Lie algebra of the stabilizer subgroup $Z_{\cQ} \sub \SL(4,\CC) \times \SL(k,\CC) \times \CC^\times$, 
whose elements are given in block form as
\[\left\{ \left(\pmat{a&0&0&0\\ \ast&b&\ast&\ast\\ \ast&0&\ast&\ast\\ \ast&0&\ast&\ast}, \pmat{a^{-1}&\ast&\ast&\cdots&\ast\\0&b^{-1}&0&\cdots&0\\ 0&\ast&\ast&\cdots&\ast\\\vdots&\vdots&&&\vdots\\0&\ast&\ast&\cdots&\ast}   ,\lambda \right) 
\colon a,b \in \CC^\times \right\}.\]
\end{prp}

One can check both propositions using the same setup.  
If $\{Q_1, Q_2, Q_3, Q_4\}$ and $\{w_1, \ldots, w_k\}$ are bases for $S_+$ and $W$, respectively, 
without loss of generality we may let $\cQ = Q_1 \otimes w_1 + Q_2^* \otimes w_2^*$.  
In this basis $\cQ$ is represented by a $k \times 4$ matrix and a $4 \times k$ matrix with ones in the $(1,1)$ and $(2,2)$ positions respectively, 
and zeroes in all other positions.  
It is then an elementary computation to work out the image and kernel of the operator~$[\mc Q,-]$.

\end{enumerate}

\begin{rmk}
 Of course the classification does not need to end here: there are conformal twists of higher rank, including twists of rank $(2,2)$ generalizing the family of twists in the $\mc N=4$ super Poincar\'e algebra studied by Kapustin and Witten \cite{KapustinWitten}.  We leave a detailed analysis of all cases for future work.
\end{rmk}

\section{Chiral Twists in 4d Superconformal Algebras: Some Case by Case Analysis}

In this section we discuss the orbit structure of superconformal twists under the  action of real superconformal groups in various signatures.  
We stick to the chiral twists as the situation is already complicated there,
and we eventually specialize to the situation that Beem {\it et al} \cite{Beem_2015} showed to be strikingly rich.

\subsection{Real Orbits of Chiral Twists: General Features}
\label{4d_real_orbits_section}

A key ingredient is to recognize that the complex orbits are fiber bundles of a nice kind, 
where the base is often a familiar homogeneous space and the fibers admit clean descriptions via linear algebra.
We begin by explaining this idea, and its power for studying real orbits, in the simplest example
before gesturing at some general machinery.

Recall that $\mc O(4|k)_{r_+, r_-}$ denotes the complex orbit of rank~$(r_+, r_-)$.
We will focus on {\em chiral} twists, so without loss of generality, set $r_- = 0$, use $r$ to denote~$r_+$,
and consider~$\mc O(4|k)_{r,0}$.

\begin{ex} \label{rank_1_real_form_ex}
Let's start with the case that $k = 2$ and $r = 1$. 
(A moment's thought shows $r =0$ is too simple.)
There is a smooth map
\[
\pi \colon \mc O(4|2)_{1,0} \to \mr{Gr}(1, 2) \cong \CC\PP^1
\]
sending $\mc Q \colon \CC^2 \to \CC^4$ to $\ker \mc Q$.
An element $(A,B)$ in 
\[
\mr{SConf}(4|2,\CC)_0 \iso \{(A,B) \in \GL(4,\CC) \times \GL(2,\CC) \colon \det(A)\det(B) = 1\}
\]
acts on $\mc Q$ by $A \mc{Q} B$.
There is a natural action of the factor $\SL(2,\CC)$ on $\CC\PP^1$ by M\"obius transformations, 
so this map $\pi$ is equivariant for the action of $\mr{SConf}(4|2,\CC)_0$ since 
\[
\pi(A \mc{Q} B) = \ker(A \mc{Q} B) = \ker(\mc{Q} B) = \ker(\mc{Q}) \cdot B, 
\]
where on the far right, we use the natural action of $B \in \GL(2,\CC)$ on the projective line.
The action of the factor $A \in \SL(4,\CC)$ preserves the fiber $\pi^{-1}(L)$ for each line $L \subset \CC^2$,
since $A \mc{Q} B$ has the same kernel as $\mc{Q}B$.
(The image of $\mc{Q}$ changes, however, under the action of~$\mr{SConf}(4|2,\CC)_0$.)

When we pick a real form $G_\RR$ of $\mr{SConf}(4|2,\CC)_0$,
we can analyze its action on $\mc O(4|2)_{1,0}$ by studying how $G_\RR \cap \SL(2,\CC)$ acts on the base space $\CC\PP^1$ and how $G_\RR\cap \SL(4,\CC)$ acts on each fiber $\pi^{-1}(L)$ where $L$ is in the image of~$\pi$.
For instance, recall that the real subgroup $\SL(2,\RR) \subset \SL(2,\CC)$ breaks $\CC\PP^1$ into three orbits: the unit circle (or equator), the open hemisphere containing 0, and the open hemisphere containing~$\infty$.
By contrast, the real subgroup $SU(2) \subset \SL(2,\CC)$ has a single orbit on the base.
\end{ex}

This mode of analysis generalizes nicely.
When $k \geq 4$, it is perhaps more convenient to view $\mc Q$ as an element of $\hom(\CC^4, \CC^k)$,
so we take that perspective now.
Thanks to our observations above, we have the following.

\begin{prp}
\label{4d_fibration_prop}
For each rank $r \leq 4$,
there is a smooth map
\[\pi \colon \mc O(4|k)_{r,0} \to \mr{Gr}(4-r, 4) \]
sending $\mc Q$ to $\ker(\mc Q)$.
In terms of the map $\pi$, the group $\mr{SConf}(4|k,\CC)$ acts on the base through the $\SL(4,\CC)$ factor,
and it acts on the fiber through the $\SL(k,\CC)$ factor.
\end{prp}

In words, the image of $\pi$ is through the quotient $\mc O(4|k)_{r,0} /\SL(k,\CC)$.
We may also observe that $\pi$ determines a fiber bundle in the base by the following computation.

\begin{lmm}
The fibers of $\pi$ are all isomorphic as $\SL(k,\CC)$-spaces.
\end{lmm}

\begin{proof}
Given $X, Y \in \mr{Gr}(4-r, 4)$, choose $B \in \GL(4,\CC)$ so that $Y = BX$.  There is a natural isomorphism $\phi \colon \pi^{-1}(Y) \to \pi^{-1}(X)$ by $\phi(\mc Q) = \mc QB$.  This map is equivariant for the left $\SL(k,\CC)$-action.
\end{proof}

\begin{ex}
In the case where $k=2, r=1$ discussed in Example \ref{rank_1_real_form_ex} above we realized $\OO(4|2)_{(1,0)}$ as a fiber bundle over $\bb{CP}^1$ with fiber isomorphic to the space of rank one $2 \times 4$ matrices with fixed kernel.  We can identify this fiber with $\CC^4 \backslash \{0\}$ with its natural action of $\SL(4,\CC)$.  We will discuss the real forms of $\mr{SConf}(4|2,\CC)_0$ in general in a moment, but we may summarize here what happens in each of the three possible signatures for $\RR^{p,4-p}$.
\begin{enumerate}

\item In signature $(4,0)$ we have $G_\RR \cap \SL(4,\CC) \iso \SL(2,\bb H)$, and $G_\RR \cap \SL(2,\CC) \iso \SU(2)$.  
The action of $\SU(2)$ on $\bb{CP}^1$ is transitive, as is the action of $\SL(2,\bb H)$ on $\CC^4 \backslash \{0\} \iso \bb H^2 \backslash \{0\}$.  
In this case the real orbit space is a single point. 
(That is, there is a single orbit; all nilpotent supercharges are equivalent.)

\item 
In signature $(3,1)$ we have $G_\RR \cap \SL(4,\CC) \iso \SU(2,2)$, and $G_\RR \cap \SL(2,\CC) \iso \SU(2)$.  
The action of $\SU(2)$ on $\bb{CP}^1$ is transitive, but the action of $\SU(2,2)$ on $\CC^4 \backslash \{0\}$ is more interesting: 
there is a continuous family of orbits parameterized by $\RR^\times$ (corresponding to non-zero values of the split signature Hermitian form on $\CC^{2,2}$), 
but there is also a family of orbits with a finite non-Hausdorff locus (corresponding to the null locus for the Hermitian form).
That is, the quotient space by the $\SU(2,2)$-action is non-Hausdorff.
\item In signature $(2,2)$ we have $G_\RR \cap \SL(4,\CC) \iso \SL(4,\RR)$, and $G_\RR \cap \SL(2,\CC) \iso \SL(2,\RR)$.  
The action of $\SL(2,\RR)$ on $\bb{CP}^1$ has three orbits: 
two open orbits (the two open hemispheres) and one closed orbit (the equator), 
so the quotient space has three points with a connected non-Hausdorff topology in which a single point is dense. 
The action of $\SL(4,\RR)$ on $\CC^4 \backslash \{0\}$ is again interesting, 
because the quotient space parametrizing orbits is a connected non-Hausdorff space containing an open subspace isomorphic to~$\RR^\times$.
\end{enumerate}
\end{ex}

In general, consider a real form of the supergroup $\mr{SConf}(4|k,\CC)$ containing the real form $\SO(p+1, 5-p)$ of the complex 4d conformal group $\SO(6,\CC)$.
The real form is labeled by its signature $\RR^{p,4-p}$ so
we will write $G_{p,4-p}$ for the even part of this real supergroup.
Note that $G_{p,4-p}$ is a real form of a cover of $\SL(4,\CC) \times \SL(k,\CC) \times \CC^\times$, if $k \neq 4$, or of $\SL(4,\CC) \times \SL(4,\CC)$, if $k=4$.
Just as above, we will organize our analysis in terms of $SL(-,\CC)$-factors, 
so let
\[
G^{\text{base}}_{p,4-p} = G_{p,4-p} \cap \SL(4,\CC)
\]
and let
\[
G^{\text{fib}}_{p,4-p} = G_{p,4-p} \cap (\SL(k,\CC) \times \CC^\times)
\]
denote the corresponding real factors.

We now discuss the $G_{p,4-p}$-orbits in the complex nilpotence variety for the possible choices of signature.
In short, we want to understand the quotient space
\[
\mc O(4|k)_{r,0}/G_{p,4-p}
\]
of orbits for the action of the full group.
It is convenient to approach this space in stages using the subgroups we just described. 
For instance we have maps
\[
\begin{tikzcd}
\mc O(4|k)_{r,0} \arrow[d] \arrow[dr,"\pi"] & \\
\mc O(4|k)_{r,0}/G^{\text{fib}}_{p,4-p} \arrow[r] \arrow[d] & \mr{Gr}(4-r, 4) \arrow[d]\\
\mc O(4|k)_{r,0}/G_{p,4-p} \arrow[r] & \mr{Gr}(4-r, 4)/G^{\text{base}}_{p,4-p}
\end{tikzcd}
\]
and we know that the fibers of $\pi$ are all isomorphic with actions of $\SL(k,\CC)\times \CC^\times$,
and hence with actions of its real form~$G^{\text{fib}}_{p,4-p}$.

In principle one can work out these orbit decompositions by hand, 
but there are quite general results of Fels--Huckleberry--Wolf (going back to Wolf's work from the 1960s) that help here, notably in identifying the lower right corner of the diagram.

\begin{thm}[{\cite[Theorem 2.6]{Wolf}, \cite[Theorem 3.2.1]{FHW}}] 
\label{Wolf_thm}
Let $G$ be a real form of $\SL(n,\CC)$ of real rank~$\rho$.  
Let $P$ be a parabolic subgroup of $\SL(n,\CC)$.  
Then the following hold:
\begin{itemize}
\item There are finitely many $G$-orbits in the partial flag variety~$\SL(n,\CC)/P$.  

\item If the partial flag variety is a hermitian symmetric space, 
then the number of orbits is precisely $\pmat{\rho+2 \\2}$.  

\item For general partial flag varieties, an upper bound holds: 
there are at most $c|W|$ orbits, 
where $c$ is the number of conjugacy classes of Cartan subgroups in~$G$.  

\end{itemize}
In all cases there is a {\em unique closed orbit}.  The real dimension of this closed orbit equals the complex dimension (i.e. half the real dimension) of the partial flag variety.
\end{thm}

Note that a Grassmannian is a partial flag variety:
$\mr{Gr}(m,n) = \SL(n,\CC)/P_{m,n-m}$ where $P_{m,n-m}$ is the parabolic subgroup of block upper triangular matrices where the first diagonal block is $m \times m$ and whose second diagonal block is $(n-m) \times (n-m)$.
(These matrices are the stabilizer of the $m$-dimensional subspace spanned by the first $m$ elements of the standard basis.)

In our case of interest, $n=4$ and $m = 4-r$.
The theorem implies that there are {\em finitely} many orbits in the bottom right corner:
\[
D^{(r)}_{p,4-p} = \mr{Gr}(4-r,4)/G^{\text{base}}_{p,4-p}
\]
is a finite set with some interesting topology that is connected but non-Hausdorff.
After identifying this finite set, we then need to understand the fibers of the map
\[
\mc O(4|k)_{r,0}/G_{p,4-p} \to \mr{Gr}(4-r, 4)/G^{\text{base}}_{p,4-p} = D^{(r)}_{p,4-p}
\]
and how $G^{\text{fib}}_{p,4-p}$ acts on them.
The fibers are all isomorphic, so  we see
\[
\mc O(4|k)_{r,0}/G_{p,4-p} \cong D^{(r)}_{p,4-p} \times (\GL(r,\CC)/G^{\text{fib}}_{p,4-p})
\]
as a {\em set}.

\begin{rmk}
The number $c$ appearing in the upper bound for the number of orbits $|D^{(r)}_{*,*}|$ is computed for general real forms of $\SL(n,\CC)$ by Sugiura \cite{Sugiura}.  
In particular we find $c=1$ in Euclidean signature and $c=3$ in split signature.  
The remaining Lorentzian signature case is a hermitian symmetric space so we obtain the exact number of orbits as being~6.
\end{rmk}

\subsection{The Case \texorpdfstring{$r = 2$}{r=2}}

We now turn to discussing $r = 2$ for each signature in turn.

\subsubsection{Signature $(4,0)$}

Take $p=4$.
We can identify
\[\mr{SConf}(4,0|2\ell,\RR) \iso \SL(2|\ell,\bb H)\]
where $\bb H$ denotes the quaternions.
(Note that for Euclidean signature, we must have $k$ even. Thus take $k = 2 \ell$.)
In general, when we write $\SL(n|m, \bb H)$, 
it refers to the set of quaternionic matrices such that
\[\mr{Nm}(\mr{sdet}(A)) = 1\]
where $\mr{Nm}(q) \in \RR_{\ge 0}$ denotes the norm of a quaternion $q$.  
In particular $\dim_\RR \SL(n,\bb H) = 4n^2-1$.

The even part of this superconformal group can be identified with
\begin{align*}
\mr{SConf}(4,0|2\ell,\RR)_0 &\iso \SL(2,\bb H) \times \GL(\ell,\bb H).
\end{align*}
We begin by examining the action of $\SL(2,\HH)$ on the base space~$\mr{Gr}(2;4;\CC)$.

\begin{lmm}
There are two orbits for the action of $\SL(2,\HH)$ on the base space~$\mr{Gr}(2;4;\CC)$.
That is,
\[
D^{(2)}_{4,0} = \mr{Gr}(2,4; \CC)/\SL(2,\HH)
\]
has two elements.
\end{lmm}

\begin{proof}
As the quaternions $\HH$ forget to a copy of $\CC^2 = \CC \oplus j \CC$,
we can identify 
\[
\CC^4 \cong \CC^2 \otimes_\CC \HH \cong \HH^2.
\]
In explicit terms, equip $\CC^2$ with the basis $\{e_1,e_2\}$ and let $\HH^2$ have the same basis (under extension of scalars).
Then $\CC^4$ has a $\CC$-linear basis by $\{e_1,je_1,e_2,je_2\}$.

Given a subspace $V \subset \CC^4$ with $\dim_\CC(V) = 2$, representing a point $[V] \in \mr{Gr}(2,4)$,
we obtain an $\HH$-linear subspace $V_\HH \subset \HH^2$ 
by taking the $\HH$-span of $V$, i.e.,
\[
V_\HH = V \otimes_\CC \HH. 
\]
If $V$ has complex basis $\{v_1,v_2\}$, then
\[
V_\HH = \HH v_1 \oplus \HH v_2 \subset \HH^2.
\]
This subspace has
\[
0 < \dim_\HH(V_\HH) \leq 2,
\]
namely either 1 or~2.

Both values 1 and 2 are realized in practice:
the complex subspace 
\[
\HH e_1 = \CC e_1 \oplus \CC (je_1) \subset \CC^4
\]
has $\dim_\HH(\HH e_1) = 1$,
while the subspace
\[
W = \CC e_1 \oplus \CC e_2 \subset \CC^4
\]
has $\dim_\HH(W_\HH) = 2$.
Thus the function $\mr{Gr}(2,4) \to \{1,2\}$ sending $V$ to $\dim_\HH(V_\HH)$ is surjective.

It remains to show that the action of $\SL(2,\HH)$ is transitive on the fibers of this surjection,
as this will show that there are two orbits.

The subspaces with $\dim_\HH(V_\HH) =1 $ parametrize $\HH\PP^1$,
so we have an inclusion
\[
\HH\PP^1 \subset \mr{Gr}(2,4)
\] 
and $\SL(2,\HH)$ acts transitively on $\HH\PP^1$ 
(i.e., every such subspace is of the form $A(\HH e_1)$ for some $A \in \SL(2,\HH)$).

On the other hand, if $\dim_\HH(V_\HH) = 2$, there is a complex basis $\{v_1,v_2\}$ and these span $\HH^2$ over the quaternions.
Thus, there is some $A \in \GL(2,\HH)$ such that $A e_j = v_j$.
This matrix $A$ might not have determinant of norm~1,
but by rescaling the basis vectors to get a new basis $\{v'_1,v'_2\}$, we can find $A' \in \SL(2,\HH)$ such that $A'e_j = v'_j$.
In other words, $\SL(2,\HH)$ is transitive on this fiber.
\end{proof}

\begin{ex}
Consider the $\mc N=2\ell$ superconformal algebra with $2\ell \ne 4$. 
In the space of chiral supercharges of rank $(r,0)$, 
the space of orbits in signature $(4,0)$ can be identified with
\[(\GL(2\ell,\CC)/\GL(\ell, \bb H)) \times \{1,2\}\]
where $\{1,2\}$ labels the $\HH$-dimension of the $\HH$-span in~$\mr{Gr}(2,4)$.
\end{ex}

\begin{ex} \label{Euclidean_Schur_ex}
If $2\ell = 2$, we can identify 
\[\GL(2,\CC)/\GL(1,\bb H) \iso \SL(2,\CC)/\SU(2) \times \CC^\times/\RR^\times \iso \mf h^3 \times S^1\]
where $\mf h^3$ refers to hyperbolic 3-space.  Thus topologically, the space of orbits of rank $(2,0)$ chiral supercharges in the $\mc N=2$ superconformal algebra, in Euclidean signature, can be identified with
\[(\RR^3 \times S^1)\times \{1,2\}\]
where $\{1,2\}$ labels the $\HH$-dimension of the $\HH$-span in~$\mr{Gr}(2,4)$.
\end{ex}

\subsubsection{Signature $(3,1)$}

We can identify
\[\mr{SConf}(3,1|k,\RR) \iso \SU(2,2|k).\]
In particular, the even part can be identified with
\[\mr{SConf}(3,1|k,\RR)_0 \iso \SU(2,2) \times \SU(k) \times \mr U(1).\]

The real rank of $\SU(2,2)$ is two, so in a partial flag variety of the form $\SL(4,\CC)/P$ there are always at most six orbits for the left-action of~$\SU(2,2)$.

\begin{ex}
In the chiral case with $\mc N=2$ the $\SU(2,2)$-action on the complex Grassmannian $\mr{Gr}(2,4; \CC)$ has orbits that we can describe quite concretely.  They are given by the possible restrictions of the split signature Hermitian metric on $\CC^{2,2}$ to a plane.  There are six orbits: the restricted (possibly degenerate) Hermitian spaces are 
\[\CC^{2,0}, \CC^{1,1}, \CC^{0,2}, \CC_{\mr{null}} \times \CC^{1,0}, \CC_{\mr{null}} \times \CC^{0,1}, \text{ and } (\CC_{\mr{null}})^2,\]
where $\CC_{\mr{null}}$ refers to $\CC$ with the zero quadratic form.  That is, the space $D^{(2)}_{3,1}$ is a finite topological space with six points.
\end{ex}

\begin{ex}
If $k = 2$, we can again analyze the fiber:
\[\GL(2,\CC)/\mr U(2) \iso \SL(2,\CC)/\SU(2) \times \CC^\times/\RR^\times \iso \mf h^3 \times S^1.\]
Thus again, topologically, the space of orbits of rank $(2,0)$ chiral supercharges in the $\mc N=2$ superconformal algebra, in Lorentzian signature, can be identified with
\[(\RR^3 \times S^1)\times D^{(2)}_{3,1}.\]
\end{ex}

\subsubsection{Signature $(2,2)$}

We can identify
\[\mr{SConf}(2,2|k,\RR) \iso \SL(4|k,\RR).\]
In particular, the even part can be identified with
\[\mr{SConf}(2,2|k,\RR)_0 \iso \SL(4,\RR) \times \SL(k,\RR) \times \RR^\times.\]
We begin by examining the action of $\SL(4,\RR)$ on the base space~$\mr{Gr}(2;4;\CC)$.

\begin{lmm}
There are three orbits for the action of $\SL(4,\RR)$ on the base space~$\mr{Gr}(2;4;\CC)$:
for $V \subset \CC^4$ with $\dim_\CC(V) = 2$,
its orbit $\SL(4,\RR) V$  is characterized by
\[
\dim_\RR(V \cap \RR^4) \in \{0,1,2\}.
\]
That is,
\[
D^{(2)}_{2,2} = \mr{Gr}(2,4)/\SL(4,\RR)
\]
has three elements.
\end{lmm}

\begin{proof}
Let $\RR^4$ denote the real subspace of $\CC^4$, and let $i \RR^4$ denote the real subspace of purely imaginary vectors.

Let $V \subset \CC^4$ with $\dim_\CC(V) = 2$, and hence represent a point $[V] \in \mr{Gr}(2,4)$.
We will study its intersection with $\RR^4$ (the ``real intersection''). 

Observe that the action of $\SL(4,\RR)$ on $\CC^4$ is complex-linear, 
preserves $\RR^4$ in $\CC^4$, and commutes with conjugation. 
Hence this action preserves~$\dim_\RR( V \cap \RR^4 )$.

Now we claim 
\[
\dim_\RR(V \cap R^4) \leq 2
\]
because the intersection consists of the conjugation-fixed points in~$V$. 
Let $V^{\rm conj} \subset V$ consist of $v\in V$ such that $\overline{v} = v$.
Note that if $v \in V^{\rm conj}$, 
then $iv$ is in $V$ but $iv$ is {\em not} fixed by conjugation.
Hence $V^{\rm conj}$ and $iV^{\rm conj}$ are transverse real subspaces of~$V$,
and hence
\[
\dim_\RR(V^{\rm conj} \oplus iV^{\rm conj}) = 2 \dim_\RR(V^{\rm conj}) \leq \dim_\RR(V) = 4,
\]
yielding our dimension bound.

Notice that there is a subspace $V$ whose intersection has dimension 2, 1, or 0:
\begin{itemize}
\item[(2)] Take $V = \CC e_1 \oplus \CC e_3$.
\item[(1)] Take $V = \CC e_1 \oplus \CC(e_3 + i e_4)$  because 
a linear combination $z e_1 + w(e_3 + i e_4)$ is real if and only if $z$ real, $w$ real, and $i w$ is real (i.e., $w = 0$).
\item[(0)] Take $V = \CC (e_1 + i e_2) \oplus \CC(e_3 + i e_4)$ because a linear combination $z (e_1 + i e_2) + w(e_3 + i e_4)$ is real if and only if $z$ real, $w$ real, $iz$ is real, and $i w$ is real (i.e., $z = 0 = w$).
\end{itemize}
Hence we know the orbits surject onto the set $\{0,1,2\}$ under the map $V \mapsto \dim_\RR(V \cap \RR^4)$.

We now explain why the action of $\SL(4,\RR)$ is transitve on each fiber of this map.

In case the dimension of intersection is 2,
then $V \cap \RR^4$ admits a real basis with two elements $\{v_1,v_2\}$.
There is some element $A$ of $\SL(4,\RR)$ such that $Ae_1 = v_1$ and $A e_3 = v_2$.
(This does not uniquely determine $A$, of course.)
Thus $V$ is in the orbit of $\CC e_1 \oplus \CC e_3$.

In case the dimension of intersection is 1,
then $V \cap \RR^4$ admits a real basis with one element $\{v_1\}$.
Extend to a {\em complex} basis $\{v_1, v_2\}$ where $v_2 = x + i y$ with $x, y \in \RR^4$ and $y \neq 0$.
We want $A \in \SL(4,\RR)$ such that $Ae_1 = v_1$ and 
\[
A(e_3+ie_4) = x + i y,
\]
i.e., $Ae_3 = x$ and $Ae_4 = y$. 
But we know $v_1, x, y$ are linearly independent over $\RR$ so such an $A$ exists.

In case the dimension of intersection is 0,
then $V$ has a complex basis $\{v_1,v_2\}$ where $v_j = x_j + i y_j$ with $y_1 \neq 0$ and $y_2 \neq 0$.
We want $A \in \SL(4,\RR)$ such that  
\[
A(e_1+ie_2) = x_1 + i y_1
\]
and
\[
A(e_3+ie_4) = x_2 + i y_2.
\]
These vectors $\{x_1,x_2,y_1,y_2\}$ are linearly independent over $\RR^4$,
so we can find a matrix $B$ in $\GL(4,\RR)$ satisfying our condition:
take $x_1$ to be the first column of $B$, $y_1$ to be the second column, and so on.
This matrix is $B$ might have $\det(B) \neq 1$, however.
This issue is easily fixed by adjusting the basis $\{v_1,v_2\}$ by rescaling. 
\end{proof}

\begin{ex}
 In the chiral rank 2 case we can again identify the space of orbits.  We find the product of a finite non-Hausdorff space and the homogeneous space
 \[\GL(2,\CC)/(\SL(2,\RR) \times \RR^\times) \iso \mc \RR^2 \times (S^1)^2,\]
where now $\SL(2,\CC)/\SL(2,\RR) \iso \RR^2 \times S^1$ is a 3d hyperboloid of one sheet $\{x_1^2 + x_2^2 - x_3^2-x_4^2 = 1\} \sub \RR^4$.

Thus, overall,the set of orbits in this split signature case is identified with 
\[
\RR^2 \times (S^1)^2 \times \{0,1,2\}
\] 
where the finite set identifies the orbit of the $\SL(4,\RR)$ action.
\end{ex}

\subsection{The Case of 4d \texorpdfstring{$\mathcal N=2$}{N=2} Chiral Supercharges}
\label{Beemetal}

In this section we will analyze the specific example considered by Beem {\it et al} \cite{Beem_2015} in their derivation of vertex algebras from 4d $\mathcal N=2$ superconformal field theory.  
Beem {\it et al} are concerned with a chiral supercharge of rank $(2,0)$ that we will call the \emph{Schur supercharge}.  
The analysis above tells us that for this example
\[
 \dim \mf z^\CC_{\mc Q} = 11\quad\text{and}\quad\dim \mf b^\CC_{\mc Q} = 8.
\]
Let us discuss such chiral supercharges in the usual physics notation.  

We choose an affine patch $\RR^{p,4-p} \hookrightarrow C(\RR^{p,4-p})$, 
associated to which there is an embedding $\mr{ISO}(p,4-p) \hookrightarrow \mr{Conf}(p,4-p)$.  
Note that this embedding is non-canonical and so the decomposition that we will discuss below is also non-canonical.  
The summands of the odd part of the superconformal algebra then decompose as representations of $\so(4,\CC) \sub \so(6,\CC)$ as follows.  
If $V_4 \otimes W_2$ is the positive chirality piece of $\sconf(4|2,\CC)_1$, 
we have the decomposition
\[V_4 \otimes W_2 \iso S_+ \otimes W_2 \oplus S_- \otimes W_2,\]
where $S_\pm$ are the two semispin representations of $\so(4,\CC) \iso \sl(2,\CC)_+ \oplus \sl(2,\CC)_-$.  
The convention in physics is to choose bases for these summands: let
\[
S_+ \otimes W_2 = \mr{Span}(\mc Q_{\alpha \mc I}) 
\quad\text{and}\quad 
S_- \otimes W_2 = \mr{Span}(\wt {\mc S}^{\dot \alpha \mc I})
\]
where $\alpha\in \{+, -\}$, $\dot \alpha \in \{\dot +, \dot -\}$, and~$\mc I \in \{1,2\}$.

For example, in this basis, the supercharge $\mc Q_{+1} + \mc Q_{-2}$ is a {\em Donaldson--Witten} supercharge \cite{Witten_1988}. 
Note that it lies in the first summand $S_+ \otimes W_2$, 
and hence in an ordinary 4d $\mc N=2$ super Poincar\'e algebra.
Thus it acts on supersymmetric theories, not just superconformal theories, and famously twists $\mc N=2$ super Yang--Mills theory into a 4d topological field theory.

As another example, the supercharge
\[
\bb Q = \mc Q_{+1} + \wt {\mc S}^{\dot +2}
\]
provides a {\em Schur} supercharge,
which we will examine in more detail below.
 

Exploiting some features of the ``Schur limit'' of a 4d $\mc N=2$ superconformal theory~\cite{GRRY},
Beem {\it et al} \cite{Beem_2015} discovered how to obtain vertex algebras from 4d $\mc N=2$ superconformal theories, 
providing a beautiful and powerful relationship between 4d and 2d field theory.
We return to this aspect in Section~\ref{sec:twistedobsmath}, 
where we explain how conformal twists modify observables of superconformal field theories.
In particular, we discuss in Section~\ref{sec: Saberi Williams} some powerful results of Saberi and Williams on this topic that provide many {\it examples} of how factorization techniques and twisting combine to yield vertex algebras,
which are complementary to the structural work we do in Section~\ref{sec:twistedobsmath}.

In this section we will analyze two separate problems:
\begin{itemize}
\item For the Schur twist, how does the signature interact with the cohomology? 
In particular, which real conformal transformations are exact and which are closed?
\item Is there a best affine patch for any given supercharge?
We examine whether changes of affine patch allow one to recognize a twist as a Donaldson--Witten supercharge.
\end{itemize}
Analogs of these questions are relevant for other superconformal algebras too.

\subsubsection{The real forms of the cohomology of a Schur supercharge} \label{Schur_section}

Let us now discuss how a Schur supercharge interacts with the various real forms, along with the real groups $Z_{\mc Q}$ and $B_{\mc Q}$. 
For concreteness, let us work in the basis described above and consider the supercharge
\[\bb Q  = \mc Q_{+1} + \wt {\mc S}^{\dot +2}.\]
Alternatively, viewing this supercharge as a $2 \times 4$ matrix via the given bases for $S_{\pm}, W_2$, we can identify
\begin{equation}
\label{Schur twist matrix}
\bb Q = \pmat{1&0&0&0\\0&0&1&0}
\end{equation}
where the first $2\times 2$ block relates to $S_+$ and the second block to $S_-$.
In this basis we can identify the complex Lie algebras of $\bb Q$-closed and $\bb Q$-exact elements explicitly.  

We now describe the projections of these algebras onto the Lie algebra of complexified infinitesimal conformal transformations~$\sl(4,\CC)$,
as we are particularly interested in how the Schur twist interacts with conformal geometry.

Let $\pi: \sconf(4|2,\CC)_0 \to \sl(4,\CC)$ denote projection onto the conformal Lie algebra, i.e., forget the R-symmetries.
We will view elements of $\sl(4,\CC)$ as block matrices whose blocks are $2 \times 2$ matrices.

\begin{lmm}
The projection of the closed and exact subalgebras are
\begin{align*}
 \pi(\mf z_{\bb Q}^\CC)  &= \left\{\left(\begin{array}{c|c}
     A & B \\ \hline C&D
    \end{array}\right) \colon A,B,C,D \in \mf p \text{ and }\tr(A)+\tr(D)=0\right\},  \\
 \pi(\mf b_{\bb Q}^\CC) &= \left\{\left(\begin{array}{c|c}
     A & B \\ \hline C&D
    \end{array}\right) \colon A,B,C,D \in \mf k \text{ and }\tr(A)+\tr(D)=0\right\} ,
\end{align*}
where
\[
\mf k \sub \mf p \sub \gl(2,\CC)
\]
with 
\begin{itemize}
\item $\mf p$ the subalgebra of lower triangular matrices, and
\item $\mf k$ the subalgebra of matrices with non-zero elements in the left column only.
\end{itemize} 
\end{lmm}

This lemma identifies the cohomology, with the $B$ and $C$ factors sitting  in $\mf p/\mf k$, and with the $A$ and $D$ factors living in a quotient more involved to describe, due to the trace condition. 

Let us now discuss the real forms for these complex Lie algebras associated to each signature.\\

\paragraph{Signature $(4,0)$}

In Euclidean signature we are considering the real form $\sl(2,\bb H) \sub \sl(4,\CC)$.  
We may obtain it as the fixed points of the involution
\[\theta_{\mr{Euc}}(X) = -JXJ\]
where
\[J= \pmat{0&1&0&0\\-1&0&0&0\\0&0&0&1\\0&0&-1&0}.\]
The fixed point locus of $\theta_{\mr{Euc}}$ acting on $\pi(\mf z_{\bb Q}^\CC)$ is three complex-dimensional, or six real-dimensional.
It consists of matrices of the form
\[
\pi(\mf z_{\bb Q}) \cap \so(1,5) = 
\left\{
\left(
\begin{array}{c|c}
a \bb I & b\bb I \\ 
\hline 
c \bb I & d \bb I
\end{array}
\right) 
\colon a,b,c,d \in \CC, a+d=0
\right\}
\]
where $\bb I = \mr{diag}(1,1) \in \gl(2,\CC)$.  
This subalgebra is isomorphic to $\sl(2,\CC) \iso \so(1,3) \iso \mr{conf}(2)$, 
and indeed it acts via conformal transformations on an embedded plane $\RR^{2,0} \sub \RR^{4,0}$. 
We may see this by observing that in $\sl(2,\bb H) \iso \mr{conf}(4)$, the subalgebras of infinitesimal translations and special conformal transformations, respectively, may be identified as the subalgebras of strict upper and lower triangular matrices.

The fixed point locus of $\theta_{\mr{Euc}}$ acting on $\pi(\mf b_{\bb Q}^\CC)$ is the zero matrix,
so there are no nonzero real $\bb Q$-exact conformal transformations in Euclidean signature.

In particular, the cohomology is precisely~$\pi(\mf z_{\bb Q}) \cap \so(1,5)$.
This result tells us how the twists interact with the conformal geometry.\\

\paragraph{Signature $(3,1)$}

In Lorentzian signature we are considering the real form $\su(2,2) \sub \sl(4,\CC)$.  
We may obtain it as the fixed points of the involution
\[\theta_{\mr{Lor}}(X) = -KX^\dagger K\]
where
\[K = \mr{diag}(1,1,-1,-1).\]
The fixed point locus of $\theta_{\mr{Lor}}$ acting on $\pi(\mf z_{\bb Q}^\CC) \cap \sl(4,\CC)$ is seven real-dimensional.  
It consists of matrices of the form
\begin{align*}
\pi(\mf z_{\bb Q})
= 
\Bigg\{
\left(
\begin{array}{c|c}
A & B\\ 
\hline 
\ol{B} & D
\end{array}
\right) 
&\colon 
A = \mr{diag}(\alpha, \beta), 
D = \mr{diag}(\gamma, \delta), 
B = \mr{diag}(a,b), \\ 
&\qquad  \text{where } a,b \in \CC \text{ and } \alpha, \beta, \gamma, \delta \in i\RR, \alpha+\beta + \gamma + \delta = 0 
\Bigg\}.
\end{align*}
This subalgebra of $\su(2,2)$ has a semisimple quotient isomorphic to $\su(2) \oplus \su(2)$, and the solvable radical generates a dilation in the conformal group.
The two factors of this semisimple quotient act on the two factors of $\CC \oplus \CC \iso \RR^4$ by antiholomorphic vector fields.

The fixed point locus of $\theta_{\mr{Lor}}$ acting on $\pi(\mf b_{\bb Q}^\CC)$ is three real-dimensional.
It consists of matrices of the form
\[\mf b_{\bb Q} \cap \so(2,4) = \left\{\left(\begin{array}{c|c}
     A & B\\ \hline \ol{B} & -A
    \end{array}\right) \colon A = \mr{diag}(\alpha, 0) B = \mr{diag}(a,0), a \in \CC, \alpha \in i\RR \right\}.\]
This Lie algebra is isomorphic to $\su(2)$, viewed as a subalgebra of $\su(2) \oplus \su(2)$ via the map $X \mapsto (X,0)$, 
and therefore acts on an embedded complex line $\CC \sub \CC^2 \iso \RR^4$ by antiholomorphic vector fields.


\paragraph{Signature $(2,2)$}

We will only discuss the split signature case very briefly. 
Recall that we must consider the real form $\sl(4,\RR) \sub \sl(4,\CC)$, 
so the projections $\pi(\mf z_{\bb Q})$ and $\pi(\mf b_{\bb Q}) \cap \so(3,3)$ are very easy to understand: 
they are the subalgebras of $\pi(\mf z_{\bb Q}^\CC) \cap \sl(4,\CC)$ and $\pi(\mf b_{\bb Q}^\CC) \cap \sl(4,\CC)$ consisting of matrices with real coefficients.
In particular, if we identify the upper-right block matrices with translations, 
then there are three closed real translations, of which two are exact.

\subsubsection{Relationship with Donaldson--Witten supercharges and real twists}

As we have seen, all rank $(2,0)$ supercharges $\mc Q$ lie in the same orbit for the complex group $\SO(6,\CC) \times \Sp(2,\CC)$, 
and the complex Lie algebras $\mf z^\CC_{\mc Q}$ and $\mf b^\CC_{\mc Q}$ are conjugate for different choices of $\mc Q$. 
Hence, from this complexified point of view, and on any specific affine path, 
there is {\em no} distinction between twisting by a Donaldson--Witten supercharge and a Schur supercharge: we can act by the complex superconformal group to relate them.  

There is a distinction at the level of a {\em real} form of the superconformal group (i.e. for the actual conformal group acting on the conformal compactification space in any specified signature).
As we saw in the preceding subsection, supercharges can live in different orbits for such a real group. 
Indeed, if we choose a specific affine patch $\iota \colon \RR^{p,4-p} \hookrightarrow C(\RR^{p,4-p})$, and therefore a subalgebra $\so(4,\CC) \sub \so(6,\CC)$, as described above, 
the Schur supercharge does not lie in the same $\mr{Spin}(p,4-p)$ orbit as the Donaldson--Witten supercharges. 
It is worth remarking, however, that this distinction relied on \emph{fixing} an affine patch. 
 
On the other hand, given a {\em real} rank $(2,0)$ supercharge, we may always {\em choose} an affine patch for which it has Donaldson--Witten type.

\begin{prp}
Let $\mc Q$ be a rank $(2,0)$ nilpotent supercharge in a real form $\mathfrak{sconf}(p,4-p|\mathcal{N}{=}\hspace{.1em}2)$.
Then there exists an affine patch $\iota \colon \RR^{p,4-p} \hookrightarrow C(\RR^{p,4-p})$ such that $\mc Q$, viewed in the the complexification $\mathfrak{sconf}(4|\mathcal{N}{=}\hspace{.1em}2, \CC)$, is a Donaldson--Witten supercharge, i.e., so that it lies in the span of the generators~$\mc Q_{\alpha \mc I}$.
\end{prp}

In other words, any real supercharge on one ``spacetime'' (i.e., on one affine patch) looks like a Donaldson--Witten supercharge after moving to another ``spacetime'' by a conformal transformation.

\begin{proof}
View $\mc Q$ as a rank 2 linear map $V_4^* \to W_2$, and let $U = \ker(\mc Q) \sub V_4^*$.  
Note that $V_4$ carries a canonical $\so(6,\CC)$-invariant inner product, 
allowing us to decompose $V_4 = U \oplus U^\perp$ into the sum of orthogonal two-dimensional subspaces.  
Consider now the Lie subalgebra of $\so(6,\CC)$ that preserves this decomposition,
which provides an embedding $$\sl(2,\CC) \oplus \sl(2,\CC) \to \sl(4,\CC) \cong \so(6,\CC),$$
or equivalently $\so(4,\CC) \subset \so(6,\CC)$.
Each summand $U, U^\perp$ is identified with an $\so(4,\CC)$-irreducible summand $S_-, S_+$, respectively.

In terms of this decomposition, the odd part of the complexified superconformal algebra also decomposes as
\[
\hom(V^*_4, W_2) = \hom(S_+, W_2) \oplus \hom(S_-, W_2)
\]
and $\mc Q$ is a full rank element of the first summand.
Thus it is a Donaldson-Witten twist.

Consider now the image $\mf b_{\mc Q}^\CC$ of $[\mc Q,-]$ inside the even part of the complexified superconformal algebra.
It is isomorphic, as a Lie algebra, to $\sl(2,\CC) \ltimes (\CC^2 \otimes \CC^2)$,
and we are free to view the four dimensional abelian factor as the complexified translations of an affine slice,
so let us denote this subalgebra by $T_\CC$.
(There is another copy of $\CC^4$, orthogonal to this image, which we can then view as the special conformal transformations for this affine slice.)

Even better, as $\mc Q$ comes from a real form, it is a fixed point of the corresponding involution (i.e., of the appropriate version of complex conjugation).
The bracket is equivariant for this involution, so the image of $[\mc Q,-]$ restricted to the odd part of the real form gives us a 4-dimensional real subspace $T_\RR$ of $T_\CC$.
We define an affine patch $\iota \colon \RR^{p,4-p} \hookrightarrow C(\RR^{p,4-p})$ as the inclusion of the open orbit of the $T_\RR$-action on the conformal compactification~$C(\RR^{p,4-p})$.
\end{proof}

This result may appear quite surprising, as it seems to suggest that one can convert a Schur supercharge to a Donaldson-Witten supercharge by picking a different affine patch (i.e., linear spacetime).
But, in fact, {\em no} Schur charge is {\em real} in Lorentzian or Euclidean signature.
Thus the proposition does not apply to Schur-twisted theories.

\begin{lmm}
The only twist in the real super Lie algebra $\sconf(3,1|2)$ is the zero element.
\end{lmm}

\begin{proof}
In Lorentzian signature, we have $\sconf(3,1|2) = \su(2,2|2)$ where
\[
\su(2,2|2) = \{A \in \gl(4|2,\CC) \, :\, A \KK + (-1)^{|A|} \KK A^\dagger = 0,\, \mr{str}(A) = 0\}
\]
where
\[
\KK = {\rm diag}(1,1,-1,-1,1,1),
\]
where the first four columns are even and the last two columns are odd.
Let $K= {\rm diag}(1,1,-1,-1)$ denote the signature $(2,2)$ form (or matrix) acting on the even vector space.

An odd element $\QQ \in \su(2,2|2)$ thus has the block form
\[
\left(\begin{array}{c|c}
0 & KQ^\dagger \\ 
\hline 
QK & 0
\end{array}\right)
\]
with $Q \in \mr{Mat}_{2,4}(\CC)$ and $Q^\dagger$ its conjugate-transpose.
Then
\[
[\QQ,\QQ] =
2\left(\begin{array}{c|c}
KQ^\dagger Q K & 0 \\ 
\hline 
0 & QQ^\dagger
\end{array}\right)
\]
by direct computation.
Note that the diagonal entries of $Q Q^\dagger$ are the norms of each row of $Q$,
as a vector in $\CC^2$ with the standard sesquilinear pairing.
In other words, these diagonal entries are the norms as vectors in Euclidean $\RR^4$ and hence must be nonnegative.
We can then have $[\QQ,\QQ] = 0$ if and only if every row is zero, i.e., $Q = 0$.
\end{proof}

On the other hand, in Euclidean signature, there is a space of real twists.
Recall that $\sconf(4,0|2) = \sl(2|1,\HH)$.

\begin{lmm}
A twist in $\sconf(4,0|2)$ has the form
\[
\left(\begin{array}{c|c}
0 & 0 \\ 
\hline 
C & 0
\end{array}\right) 
\quad\text{or}\quad 
\left(\begin{array}{c|c}
0 & B \\ 
\hline 
0 & 0
\end{array}\right) 
\]
in $\sl(2|1,\HH)$.
\end{lmm}

This result lets us show the following.

\begin{cor}
The Schur supercharge does not lie in  $\sl(2|1,\HH) \subset \sl(4|2,\CC)$.
\end{cor}

\begin{proof}[Proof of corollary]
To see this, we merely need to identify which complex matrices appear in the embedding.
Observe that the odd component of $\sl(4|2,\CC)$ is isomorphic to
\[
\hom_\CC(\CC^2, \CC^4) \oplus \hom_\CC(\CC^4, \CC^2)
\]
and we view the Schur supercharge as living in the second summand.
Recall the matrix~\eqref{Schur twist matrix} given above.

We need to describe how quaternionic matrices sit inside complex matrices.
In particular, we want to describe
\[
\hom_\HH(\HH^2, \HH) \hookrightarrow \hom_\CC(\CC^4, \CC^2)
\]
explicitly. 
As a first step, identify $\CC^2$ with $\HH$ by sending a column vector $(a,b)^T$ of complex numbers to $a + jb$ (note that $j$ precedes $b$).
Then an $\HH$-linear map $f: \HH^2 \to \HH$ is determined by a $2 \times 1$ matrix 
\[
(a+jb \; c +jd)
\] 
for some complex numbers $a,b,c,d$.
Thus we have an inclusion
\[
\hom_\HH(\HH^2, \HH) \hookrightarrow \hom_\CC(\CC^2, \CC^4)
\]
sending $f$ where $f(1) = (a+jb, c+jd)$ to the matrix
\[
\begin{pmatrix}
a & b & c & d \\
-b & a & -d & c
\end{pmatrix}.
\]
The Schur matrix~\eqref{Schur twist matrix} manifestly does not have this form.
(By contrast, the usual Donaldson--Witten twist has $a = 1$ and all other entries zero.)
\end{proof}

\begin{proof}[Proof of lemma]
An odd element has the block form
\[
\QQ=
\left(\begin{array}{c|c}
0 & B \\ 
\hline 
C & 0
\end{array}\right)
\]
where $B$ is any $2 \times 1$-matrix and $C$ is any $1\times 2$-matrix.
Then
\[
[\QQ,\QQ] =
2\left(\begin{array}{c|c}
BC & 0 \\ 
\hline 
0 & CB
\end{array}\right).
\]
Note that
\[
BC = 
\begin{pmatrix}
bc & b c'\\
b'c & b'c'
\end{pmatrix}
\]
where $B^T = (b\; b')$ and $C = (c\; c')$.
But if $BC = 0$, then we need either $B = 0$ or $C = 0$.
\end{proof}

\section{3d Superconformal Twists} 
\label{3d_section}

We examine the three-dimensional setting.

\subsection{Complex Nilpotence Variety}

Over $\CC$, the 3d $\mc N=k$ superconformal algebra $\sconf(3|k,\CC)$ can be identified with the super Lie algebra $\osp(k|4,\CC)$.  The even part of $\sconf(3|k,\CC)$ is isomorphic to
\[(\sconf(3|k,\CC)_0 \iso \so(k, \CC) \oplus \sp(4, \CC) \iso \so(k, \CC) \oplus \so(5,\CC).\]
The odd part of $\sconf(3|k,\CC)$ can be identified with
\[(\sconf(3|k,\CC))_1 \iso W_{k} \otimes V_4 \iso W_{k} \otimes \bb S_4,\]
where $W_{k}$ is the $k$-dimensional fundamental representation of $\so(k,\CC)$, and $V_4$ is the 4-dimensional fundamental representation of $\sp(4,\CC)$, or equivalently the (Dirac) spinor representation of $\so(5,\CC)$.  The Lie bracket between two odd homogeneous elements $w \otimes Q, w' \otimes Q'$ is given as follows.  Write $\omega(,)$ for the symplectic pairing on $V_4$, and write $g(,)$ for the symmetric pairing on $W_{2k}$.  We have
\[[w \otimes Q, w' \otimes Q'] = (w \wedge w')\omega(Q,Q') + g(w, w')(Q \cdot Q')\]
using the isomorphism of $\so(5,\CC)$ representations $\sym^2 (\bb S_4) \to \so(5,\CC)$ on the second factor.

Let us compute the associated superconformal nilpotence variety.

\begin{prp} \label{3d_complex_nilpotence_prop}
 The complex nilpotence variety for the 3d $\mc N=k$ superconformal algebra is
 \begin{align*}
 \mc N^{(3)}_{k} \iso (\{(w, w') \in W^2_{k} \colon &g(w,w) = g(w',w') = g(w,w') = 0\}\times (\bb{CP}^1)^2)/ \\
 &(((0,w'), (c,c')) \sim ((0,w'), (0,c')), ((w,0), (c,c')) \sim ((w,0), (c,0))).
 \end{align*}
\end{prp}

\begin{proof}
 Let $Q_1, Q_2, Q_3,Q_4$ be a Darboux basis for $\bb S_4$, such that $\omega(Q_1,Q_2) = \omega(Q_3,Q_4) = 1$.  Let
 \[\cQ = (w_1,w_2,w_3,w_4) = w_1 \otimes Q_1 + w_2 \otimes Q_2 + w_3 \otimes Q_3 + w_4 \otimes Q_4\]
 be an arbitrary element of $(\sconf(3|k,\CC))_1$.  Suppose this element squares to zero.  If we first consider the $\so(5,\CC)$ summand of $[\mathcal Q, \mathcal Q]$, this means that
 \begin{align*}
  \sum_{i,j=1}^4 g(w_i, w_j) (Q_i \cdot Q_j) &= 0 \\
  \implies g(w_i, w_j) &= 0 \text{ for all } i,j.
 \end{align*}
 Now, considering the $\so(2k,\CC)$ summand of $[\mathcal Q, \mathcal Q]$, we have
 \begin{align*}
  \sum_{i,j=1}^4 (w_i \wedge w_j) \omega(Q_i, Q_j) &= 0 \\
  \implies w_1 \wedge w_2 &= 0 \\
  \text{and } w_3 \wedge w_4 &= 0.
 \end{align*}
 Thus let $w_1 = w$, $w_2 = cw$, $w_3 = w'$ and $w_4 = c'w'$ for complex constants $c,c'$.
\end{proof}

\begin{rmk}
How does the translation nilpotence variety embed inside the conformal nilpotence variety?  
Two copies of the 3d $\mc N=k$ supersymmetry algebra embed inside $\sconf(3|k,\CC)$.  
We can understand the embedding by considering the restriction of the adjoint representation of $\sconf(3|k,\CC)$ to a representation of the subalgebra $\so(k,\CC) \oplus \so(3,\CC)$.  
At the level of super vector spaces,
this restricted representation takes the form  
\[(\so(k,\CC) \oplus \so(3,\CC) \oplus (\CC^3)^2 \oplus \CC) \oplus \Pi(W_k \otimes (\bb S_2 \oplus \bb S_2)),\]
where $\bb S_2$ is the (Dirac) spinor representation of $\so(3,\CC)$.  
From this point of view, the usual nilpotence variety embeds either by considering the locus $w'=0$ or the locus $w=0$ inside the conformal nilpotence variety.
\end{rmk}

In the next sections we analyze the complex and real orbits in $\mc N^{(3)}_k$ under the action of the even part of the superconformal group.  

\subsection{Complex Group Orbits \& Closed and Exact Transformations}

The even part of the superconformal group 
\[\mr{SConf}(3|k,\CC)_0 \iso \Spin(5,\CC) \oplus \SO(k,\CC)\]
acts on $\mc N^{(3)}_k$,
and each complex orbit has a well-defined \emph{rank}: 
that is, the rank of a square-zero supercharge $\cQ \in W_k \otimes V_4$ viewed as a linear map $W_k \to V_4^*$.  
The image of the linear map associated to $\cQ$ must be isotropic, 
so there will only be three complex orbits, associated to isotropic linear maps of rank 0, rank 1, and rank~2.
(This classification is another example of the orbit classification of Duflo--Serganova \cite[Theorem 4.2]{DufloSerganova}.)  

For the rank $r$ complex orbit we will describe the kernel and the image of the operator $[\cQ,-]$ for a supercharge $\cQ$ of rank $r$.
\begin{itemize}
\item \textbf{Rank 0}: The zero supercharge always squares to zero.  If $k < 2$ this is the only point in $\mc N^{(3)}_k$.  The kernel of $[\cQ,-]$ consists of the entirety of $\so(5,\CC) \oplus \so(k,\CC)$, and the image of $[\cQ,-]$ is just $\{0\}$.

\item \textbf{Rank 1}: Rank one nilpotent supercharges $\cQ = Q \otimes w$ exist for all $k \ge 2$.  The action of $\mr{SConf}(3|k,\CC)_0$ on the space of rank one nilpotent supercharges is transitive.

Let us write $\mf p_k \sub \so(k,\CC)$ for the maximal parabolic subalgebra with block diagonal Levi quotient $\so(2,\CC) \oplus \so(k-2,\CC)$.  Write $f_k \colon \mf p_k \to \CC \iso \so(2,\CC)$ for the projection onto the $\so(2,\CC)$ summand in this quotient.

\begin{prp}
The kernel $\mf z_{\cQ}$ of the operator $[\cQ,-]$ is isomorphic to the kernel of the map
\[f_5 - f_k \colon \mf p_5 \oplus \mf p_k \to \CC\]
as a subalgebra of $\so(5,\CC) \oplus \so(k,\CC) \iso \sp(4,\CC) \oplus \so(k,\CC)$.
\end{prp}

\begin{proof}
Without loss of generality let $\cQ = Q_1 \otimes w$ where $w = (1,i,0,\ldots,0) \in W_k$.  
The subalgebra $\mf z_{\cQ}$ is given as
\[\mf z_{\cQ} = \{(A,B) \in \sp(4,\CC) \oplus \so(k,\CC) \colon AQ_1 = \lambda Q_1, Bw = -\lambda w \text{ for some } \lambda \in \CC\}.\]

We first check that the reductive quotient $\mf l$ of $\ker(f_5-f_k)$ is contained in $\mf z_{\cQ}$.  
Here $\mf l$ is the subalgebra
\[
\mf l = \so(2,\CC) \oplus \so(3,\CC) \oplus \so(k-2,\CC) \sub \so(5,\CC) \oplus \so(k,\CC)
\]
embedded via the map $(A,X,Y) \mapsto (\mr{diag}(A,X), \mr{diag}(-A,Y))$.  
If $\lambda \mapsto A_\lambda$ under the natural isomorphism $\CC^\times \iso \SO(2,\CC)$, certainly
\[\mr{diag}(-A_\lambda,Y)w = -\lambda w.\]
Let
\[X = \pmat{a&b\\c&-a} \in \sl(2,\CC) \iso \so(3,\CC).\]
Under the exceptional isomorphism $\so(5,\CC) \to \sp(4,\CC)$, 
the matrix $\mr{diag}(A_\lambda, X)$ maps to
\[\pmat{\lambda&0&0&0\\0&a&0&b\\0&0&-\lambda&0\\0&c&0&-a} \in \sp(4,\CC).\]
Thus in particular $\mr{diag}(A_\lambda, X) Q_1 = \lambda Q_1$ as required.

The parabolic subalgebra $\mf p_k \sub \so(k,\CC)$ has nilradical of dimension $k-2$.  
To complete the argument we must show that the nilradical of $\mf p_5 \oplus \mf p_k$ is contained in $\mf z_{\cQ}$.  
In other words, it is sufficient to show that the quotient $\mf z_{\cQ}/\mf l$ is an abelian subgroup of the same dimension 
as the nilradical of $\mf p$, which is exactly $(5-2) + (k-2) = k+1$.  
We can compute this directly.  
The quotient $\mf z_{\cQ}/\mf l$ is spanned by matrices of the form
\[\pmat{0&\alpha&\beta&\gamma\\0&0&\gamma&0\\0&0&0&0\\0&0&-\alpha&0} \in \sp(4,\CC)\]
together with matrices of the form
\[\pmat{0&0&x_1&\cdots&x_{k-2}\\0&0&ix_1&\cdots&ix_{k-2}\\-x_1&-ix_1&0&\cdots&0\\ \vdots &\vdots & \vdots &\ddots &\vdots\\-x_{k-2}&-ix_{k-2}&0&\cdots&0} \in \so(k,\CC).\]
 The ideal generated by these elements is indeed an abelian subalgebra of dimension $k+1$ as required.
\end{proof}

The image $\mf b_{\cQ} \sub \mf z_{\cQ}$ of $[\cQ, -]$ is four-dimensional, 
spanned by a three-dimensional subspace of $\sp(4,\CC)$ 
and a fourth generator contained in the diagonally embedded copy of $\so(2,\CC)$ in the kernel described above.

\item \textbf{Rank 2}: 
Rank two nilpotent supercharges exist if $k \ge 4$, and this is the maximal possible rank.  
There are two orbits in the space of rank two nilpotent supercharges.  
We can use the $\so(5,\CC)$ action to scale the values of $c,c'$ in Proposition~\ref{3d_complex_nilpotence_prop} to zero, 
so we are studying the orbits in the space of pairs of orthogonal non-zero null vectors in~$W_k$.

\begin{prp}
If $k \ge 4$, there is a one parameter family of $\SO(k,\CC)$ orbits in the space of pairs of orthogonal non-zero null vectors in~$W_k$:
\[\mr{Orb}_x = \{A \cdot ((1,i,0,\ldots,0), (x,ix,1,i,0, \ldots, 0)) \colon A \in \SO(k,\CC)\}\]
where $x \in \CC$.
\end{prp}

\begin{proof}
Given a pair $(w_1,w_2)$ of orthogonal null-vectors, 
we may use the transitive $\SO(k,\CC)$ action to pass to a point in the orbit of the form $((1,i,0,\ldots,0), w)$, 
where $w \ne 0$ is orthogonal to $(1,i,0,\ldots,0)$.  
Thus, this means that $w = x(1,i,0,\ldots,0) + w'$, 
where $x \in \CC$ and $w' = (0,0,z_3,z_4,\ldots, z_k)$, 
but not all the $z_j$ are equal to zero.

Let us now consider the $\SO(k-2,\CC)$ action on the last $k-2$ coordinates.  
This action is transitive on the space of null vectors of the form $(0,0,z_3,z_4,\ldots, z_k)$, 
so we may pass to a point in the same orbit where $z_3=1, z_4=i$ and $z_j=0$ for $j > 4$.

The subgroup $\SO(k-2,\CC)$ is precisely the stabilizer of the element $(1,i,0,\ldots,0) \in W_k$, 
so these elements lie in different orbits for different values of $x \in \CC$.
\end{proof}

\end{itemize}

\begin{ex}
In the $\mc N=2$ case there is a single non-zero orbit of nilpotent supercharges 
given by elements $\cQ = Q \otimes w$, 
where $Q$ is a non-zero element of $V_4$ and $w$ is a non-zero null vector in $W_2$.  
A representative point is given by $Q = (1,0,0,0)$ and $w = (1,i)$.  
The kernel $\mf z_{\cQ}$ is given by the subalgebra
\[\mf p = \left\{\pmat{a&b&c&d\\0&e&d&f\\0&0&-a&0\\0&g&-b&-e} \right\} \sub \sp(4,\CC) \oplus \so(2,\CC),\]
embedded via the map $X \mapsto (X, -a) \in \sp(4,\CC) \oplus \so(2,\CC).$
The image $\mf b_{\cQ} \sub \mf z_{\cQ}$ is spanned by the four elements
\[\left\{\left(\pmat{0&0&1&0\\0&0&0&0\\0&0&0&0\\0&0&0&0}, 0\right), \left(\pmat{0&0&0&1\\0&0&1&0\\0&0&0&0\\0&0&0&0}, 0\right), \left(\pmat{1&0&0&0\\0&0&0&0\\0&0&-1&0\\0&0&0&0}, -1\right), \left(\pmat{0&1&0&0\\0&0&0&0\\0&0&0&0\\0&0&-1&0}, 0\right)\right\}\] in $\sp(4,\CC) \oplus \so(2,\CC)$.
\end{ex}

\subsection{Real Group Orbits}

We may analyze the orbits in the complex nilpotence variety
under the adjoint action of a real form of the even part of the superconformal group
using the same technique as in four dimensions.
In other words, for a complex orbit $\mc O_r$,
we first use the following fibration.

Let $\mr{SpGr}(r,4)$ denote the Grassmannian of isotropic $r$-dimensional subspaces in $\CC^4 \iso T^*\CC^2$,
equipped with its standard holomorphic symplectic structure.

\begin{prp}
There is a fibration
\[\pi \colon \mc O_r \to \mr{SpGr}(r,4) \]
that sends $\mc Q$, viewed as a rank $r$ linear map $W_k^* \to V_4$, to its image~$\mc Q(W_k^*) \subset V_4$.
\end{prp}

Now, we pursue the same approach as in Section \ref{4d_real_orbits_section}.
That is, we observe that the two factors $\SO(k,\CC)$ and $\Sp(4,\CC)$ of the even part of the superconformal group act only on the fiber and only on the base of the fibration $\pi$, respectively.
That is, $\SO(k,\CC)$ acts on $\GL(r,\CC)$ by left multiplication,
viewing $\GL(r,\CC)$ as the subgroup of $\GL(k,\CC)$ consisting of matrices with zeroes outside the upper-left $r \times r$ block.
Meanwhile $\Sp(4,\CC)$ acts on the homogeneous space $\mr{SpGr}(r,4) \iso \Sp(4,\CC) /P$, again by left multiplication.
Thus we may separate the question of orbits into the base and fiber directions.
In the base we may use the analog of Theorem~\ref{Wolf_thm} for general semisimple groups.
\begin{thm}[{\cite[Theorem 2.6]{Wolf}}]
Let $G_\CC$ be a complex semisimple group, let $G$ be a real form of $G_\CC$ and let $P$ be a parabolic subgroup of $G_\CC$.  Then there are finitely many $G$-orbits in the partial flag variety~$G_\CC/P$. There is a unique closed orbit of real dimension equal to $\dim_\CC G_\CC/P$.
\end{thm}
We may therefore observe that there are finitely many orbits for any real form of $\Sp(4,\CC)$.

In both signature $(3,0)$ and $(2,1)$, the real form of $\SO(k,\CC)$ is just $\SO(k,\RR)$.
Thus we may conclude that the space of real sub-orbits of the complex orbit $\mc O_r$ for the nilpotence variety in dimension 3 are isomorphic to finitely many copies of the homogeneous space
\[\GL(r,\CC)/\SO(r,\RR).\]
For the case $r=0$, this space is just a point.
If $r=1$, this homogeneous space is just $\CC^\times$.
If $r=2$ it is the seven real-dimensional manifold $\GL(2,\CC)/S^1 \iso \CC^3 \times \RR_{> 0}$.

\section{5d and 6d Superconformal Twists}

We give a brief examination of the five- and six-dimensional settings.

\subsection{5d Twists}

In dimension 5 there is a unique complex superconformal algebra, 
namely the exceptional super Lie algebra $\sconf(5|2,\CC) = \mf f(4)$.  
This algebra contains the 5d $\mc N=1$ supersymmetry algebra as a subalgebra.  
Indeed, the even part of $\sconf(5|2,\CC)$ is given by
\[(\sconf(5|2,\CC))_0 = \so(7,\CC) \oplus \sl(2,\CC),\]
and the odd part of $\sconf(5|2,\CC)$ is given by
\[(\sconf(5|2,\CC))_1 = S \otimes W\]
where $S$ is the 8-dimensional (Dirac) spinor representation of $\so(7,\CC)$ 
and $W$ is the 2-dimensional defining representation of~$\sl(2,\CC)$.

The bracket between two odd elements is given as follows.  
The 64-dimensional representation $S \otimes S$ of $\so(7,\CC)$ decomposes as
\begin{align*}
 S \otimes S &\iso \wedge^2(S) \oplus \sym^2(S) \\
 &\iso (V_7 \oplus \so(7,\CC)) \oplus (\CC \oplus \wedge^3 V_7),
\end{align*}
where $\wedge^3 V_7$ is the 35-dimensional irreducible representation 
arising as the third exterior power of the defining representation.  
In particular, there is an anti-symmetric equivariant linear map
\[F \colon S \otimes S \to \so(7,\CC),\]
and a symmetric equivariant linear map
\[\langle, \rangle \colon S \otimes S \to \CC.\]
The bracket is given by
\[[Q \otimes w, Q' \otimes w'] = \big(F(Q \otimes Q')\cdot \tr(w \otimes w') , \langle Q, Q'\rangle \cdot \mr{red}(w \otimes w')\big)\]
where we use again the notation $\tr, \mr{red}$ from Section~\ref{4d_complex_section}.

When we restrict $(\sconf(5|2,\CC))_1$ to a representation of $\so(5,\CC) \sub \so(7,\CC)$, 
we do indeed obtain two copies of $S_{(5)} \otimes W$, 
where $S_{(5)}$ is the four-dimensional (Dirac) spinor representation of $\so(5,\CC)$  
(that is, two copies of the spinors in the 5d $\mc N=1$ supersymmetry algebra).

Let us compute the associated superconformal nilpotence variety.

\begin{prp}
\label{5d_complex_nilpotence_prop}
The complex nilpotence variety for the 5d superconformal algebra is isomorphic to
\begin{align*}
\Nilp_{\CC} &\iso \{(Q_1,Q_2) \in S^2 \colon F(Q_1 \otimes Q_2) = 0 \text{ and } \langle Q_i,Q_j\rangle = 0, \text{ for any } i,j=1,2\}.
\end{align*}
\end{prp}

The complex adjoint orbits in the nilpotence variety are easy to analyze here: 
the action on $\Nilp_{\CC} - \{0\}$ is transitive, 
so there is only a single non-zero orbit in the nilpotence variety.

\begin{proof}
Let $\{w_1,w_2\}$ be a Darboux basis for $W$ with respect to the trace pairing.  
A general element of $(\sconf(5|2,\CC))_1$ can be written as $Q = Q_1 \otimes w_1 + Q_2 \otimes w_2$ for some $(Q_1,Q_2) \in S^2$.  
Let us compute the bracket $[Q,Q]$.  
According to our description of the bracket, the $\so(7,\CC)$ summand of $[Q,Q]$ is
\begin{align*}
[Q,Q]_{\so(7,\CC)} &= \sum_{i,j=1}^2 F(Q_i \otimes Q_j) \tr(w_i \otimes w_j) \\
&= 2F(Q_1 \otimes Q_2).
\end{align*}
The $\sl(2,\CC)$ summand of $[Q,Q]$ is
\begin{align*}
[Q,Q]_{\sl(2,\CC)} &= \sum_{i,j=1}^2 \langle Q_i, Q_j \rangle \mr{red}(w_i \otimes w_j) \\
&= \langle Q_1,Q_1 \rangle e + \langle Q_2, Q_2 \rangle f + \langle Q_1, Q_2 \rangle h
\end{align*}
where $\{e,f,h\}$ is the standard basis for $\sl(2,\CC)$
Therefore $[Q,Q]=0$ if and only if
\[F(Q_1 \otimes Q_2) = \langle Q_1,Q_1 \rangle = \langle Q_2,Q_2 \rangle = \langle Q_1,Q_2 \rangle = 0,\]
finishing the characterization of the nilpotent elements.
\end{proof}

\subsection{6d Twists}

In dimension 6 there exist $\mc N=(k,0)$ superconformal algebras for all positive integers $k$.  
We can identify superconformal algebras in dimension $6$ with the simple super Lie algebras $\osp(8|2k)$ 
using the triality isomorphism to view the defining representation of $\so(8,\CC)$ as a (Weyl) spinor representation.  
Thus, writing $\sconf(6|k,\CC) = \osp(8|2k)$ for the $\mc N=(k,0)$ superconformal algebra, 
we can identify the even part with
\[(\sconf(6|k,\CC))_0 = \so(8,\CC) \oplus \sp(2k,\CC),\]
and the odd part with
\[(\sconf(6|k,\CC))_1 = V_8 \otimes W_{2k} \iso S_+ \otimes W_{2k},\]
where $V_8$ is the defining representation of $\so(8,\CC)$, 
which is transformed into the Weyl spinor representation $S_+$ by the triality automorphism of $\so(8,\CC)$.  
As before, $W_{2k}$ denotes the $2k$-dimensional defining representation of~$\sp(2k,\CC)$.

The Lie bracket between two odd elements of $\sconf(6|k,\CC)$ is defined as follows,
in a similar manner to the previous examples:
\[[v \otimes w, v' \otimes w'] = (\omega(w,w') v \wedge v',\, g(v,v') F(w \otimes w')),\]
where $F \colon W_{2k} \otimes W_{2k} \to \sp(2k,\CC)$ is the canonical projection, $\omega$ is the symplectic pairing on $W_{2k}$, and $g$ is the symmetric pairing on~$S_+$.

We note that, because the relevant super Lie algebra is of the type $\osp$, the techniques from Section~\ref{3d_section} can be applied to this situation.

\begin{ex}
Let $k=1$, so $\sconf(6|k,\CC) = \sconf(6|1,\CC) \iso \osp(8|2)$. 
In this case, let $\{w_1,w_2\}$ denote a basis for $W_2$ such that $\omega(w_1,w_2)=1$. 
An arbitrary odd element then has the form
\[
Q = v_1 \otimes w_1 + v_2 \otimes w_2
\]
with $v_1,v_2 \in V_8$.
Then 
\[
[Q,Q] =  (v_1 \wedge v_2,\, \sum_{i,j} g(v_i,v_j) F(w_i \otimes w_j))
\]
so the first component vanishes only if $v_1 \wedge v_2 = 0$.
In other words, we need $v_2 = t v_1$ for some scalar $t$.
That means the second component vanishes only if $g(v_1,v_1) = 0$.
Hence the complex nilpotence variety for the 6d $\mc N=(1,0)$ superconformal algebra is
\[
\{0\} \sqcup \mc (N^{(6)}_{(1,0)} \setminus \{0\}) \times \CC^\times 
\]
with 
\[\mc N^{(6)}_{(1,0)} = \{v \in V_8 \colon g(v,v) = 0\}\]
the null cone in $V_8$.
In words, there is the zero point and for any nonzero $v$ in the nullcone, we can set $v_1 = v$ and we pick a scalar $t$ so that $v_2 = t v_1$.
There is a single non-zero orbit under the action of the complex group~$\Spin(6,\CC)$.
\end{ex}

\begin{rmk}
It is interesting to note that the $\mc N=(2,0)$ complex superconformal algebra in six dimensions and the $\mc N=8=k$ complex superconformal algebra in three dimensions are actually isomorphic; 
the roles of the conformal and R-symmetry transformations are interchanged.  
Both are isomorphic to the simple super Lie algebra $\osp(8|4,\CC)$.  
In particular, in the $\mathcal{N}{=}\hspace{.1em} (2,0)$ case
the classification of complex orbits for the even part of the complexified superconformal group 
is identical to the classification from Section~\ref{3d_section}.  
\end{rmk}

\section{Twisted Observables and Factorization Algebras}
\label{sec:twistedobsmath}

There are many ways to approach QFT mathematically,
and here we will use a framework that captures the spacetime-dependence of operators and operator products via operads.
In fact, we will use two approaches drawn from \cite{CG1, CG2}, freely referencing those books below.
The first approach uses prefactorization algebras and is highly flexible, 
allowing one to work on arbitrary manifolds and with operators whose support is an arbitrary open subspace.
A key result of \cite{CG2} is that a Lagrangian field theory on the manifold $M$ naturally produces a prefactorization algebra on $M$,
so we have a wealth of examples.
(That book also examines the perturbative quantization of such theories in Euclidean signature and shows these also yield prefactorization algebras.)
The second approach restricts to $M = \RR^n$ and only considers operators supported on open disks and how they multiply.
It matches closely with standard manipulations in the physics literature and has the additional virtue that there are convenient comparison results letting one extract more familiar algebraic structures -- like vertex algebras or associative algebras -- efficiently.
It may be more intuitive for most readers.

Both approaches let us make mathematical statements that are sharp versions of the discussion in Section~\ref{sec:twistedobsphys}.
Our discussion here amounts, primarily, to mimicking ideas and results from \cite{CG1, ElliottSafronov},
which examine chiral conformal and supersymmetric theories, respectively.
The arguments carry over straightforwardly to superconformal theories,
so we often state the result and merely indicate what needs to be changed.

In the final subsection we discuss results of Saberi and Williams \cite{SaberiWilliams},
who take a different approach to the Beem {\it et al} twists and exhibit a huge collection of examples, including Lagrangian 4d $\mc N=2$ superconformal field theories, that yield vertex algebras.

\subsection{Superconformal Prefactorization Algebras}

We will follow the conventions of \cite{CG1, CG2}, so our prefactorization algebras take values in cochain complexes of differentiable vector spaces.
This setting allows us to talk about smooth action of a Lie group on a prefactorization algebra,
with a Lie algebra acting by differential operators.
The experienced reader, however, will see how the idea of such symmetry could be ported to other settings.

\begin{dfn}
A {\em conformal (pre)factorization algebra} is a (pre)factorization algebra on a compactification $C(\RR^{p,q})$ that is smoothly equivariant for the action of the conformal group~$\Conf(p,q)$.
\end{dfn}

Similarly, we can work in $\ZZ/2$-graded cochain complexes of differentiable vector spaces for the superconformal case.

\begin{dfn}
A {\em superconformal (pre)factorization algebra} is a conformal (pre)factorization algebra on a compactification $C(\RR^{p,q})$ for which the Lie algebra action of $\mf{conf}(p,q)$ is extended to an action of a superconformal algebra~$\mf{sconf}(p,q,|S)$.
\end{dfn}

A conformal twist $\cQ$ for $\sconf(p,q|S)$ then determines a deformation of such a superconformal factorization algebra.

\begin{dfn}
Given a $\sconf(p,q|S)$-superconformal prefactorization algebra $\cF$ on $C(\RR^{p,q})$ and a conformal twist $\cQ$, let $\cF^\cQ$ denote the {\em $\cQ$-twisted} prefactorization algebra with values in $\ZZ/2$-graded cochain complexes of differentiable vector spaces whose underlying super vector space are those of $\cF$ but whose differential is $\d_\cF + \cQ$.
\end{dfn}

This construction is a version of ``adding $\cQ$ as a BRST operator.'' 

We note the following, which follows by inspection (at least for the reader of sections 3.7 and 4.8 of \cite{CG1} or of section 2 of~\cite{ElliottSafronov}).

\begin{lmm}
The groups $Z_{\cQ}$ and $B_\cQ$ act smoothly-equivariantly on $\cF^\cQ$ by restricting the action of $\mr{SConf}(p,q)_0$ on~$\cF$.
Moreover, the dg Lie algebra $(\sconf(p,q|S), \cQ)$ acts by derivations on~$\cF^\cQ$.
\end{lmm}

This lemma already provides a precise encoding of much of the discussion in section~\ref{sec:twistedobsphys}.
It says, for instance, that for any element $z \in Z_\cQ$ and for any disjoint open sets $U_1,U_2$ inside a larger open set $V \subset M = C(\RR^{p,q})$, there is a commuting diagram
\[
\begin{tikzcd}
\cF^\cQ(U_1) \otimes \cF^\cQ(U_2) \arrow[r] \arrow[d, "\cong"] & \cF^\cQ(V) \arrow[d, "\cong"] \\
\cF^\cQ(zU_1) \otimes \cF^\cQ(zU_2) \arrow[r] & \cF^\cQ(zV)  \\
\end{tikzcd}
\]
where $zU_1$, for instance, denotes the translation of $U_1$ by~$z$.
(See definition 3.7.1 of \cite{CG1}.)
If we take $V$ to be the whole manifold and pick a conformal state (or expected value map) $\langle-\rangle: \cF^\cQ(M) \to \CC$, 
this equivariance implies
\[
\langle \cO_1(x_1) \cO_2(x_2) \rangle= \langle \cO_1(zx_1) \cO_2(zx_2) \rangle
\]
where $x_i \in U_i$ and $\cO_i$ is a point-supported observable.

We will discuss the possibility of restricting to observables on a submanifold of $C(\RR^{p,q})$ in section~\ref{vertex_En_section} below.

\subsection{Algebras of Operator Products}

We overview one way to encode the most important operators and their products for a field theory whose underlying manifold is~$\RR^n$.
The essential idea is that one specifies the operators supported in any disk 
\[
D_r(x) = \{ y \in \RR^n \, : \, |y-x| < r\}
\]
and how operators from disjoint disks multiply to give operators on a bigger disk.
We assume that the spaces of operators varies smoothly as one adjusts the location and radius of the disks,
and we assume that operator product varies smoothly in those parameters as well.
More explicitly, let $\mr{Disk}_{r_1, \ldots, r_k; R}$ denote the open subset of $\RR^{(k+1)n}$ given by points $(x_1, \ldots, x_k; y)$ such that the disks $D_{r_i}(x_i)$ are pairwise disjoint but each sits inside $D_R(y)$.
Let $\cA$, which we will call an {\em $n$-disks algebra}, denote the following data:
\begin{itemize}
\item For each positive real number $r$, there is a cochain complex $A_r$, which we view as the operators supported in the disk $D_r(0)$.
\item For each point $p \in \mr{Disk}_{r_1, \ldots, r_k; R}$, there is a cochain map
\[
m_p \colon A_{r_1} \otimes \cdots \otimes A_{r_k} \to A_R,
\]
which we view as encoding the operator product. 
We require that $m_p$ vary smoothly with the configuration $p$, i.e.,
there is a smooth function
\[
m \colon \mr{Disk}_{r_1, \ldots, r_k; R} \to \mr{Hom}(A_{r_1} \otimes \cdots \otimes A_{r_k}, A_R)
\]
from the space of configurations into the space of ``multiplications.''
\item These maps compose in a natural way based on embedding disks into bigger disks. 
To be precise, suppose we pick a point $p \in \mr{Disk}_{r_1, \ldots, r_k; R}$ and points $q_i \in \mr{Disk}_{s^i_1, \ldots, s^i_{j_i}; r_i}$ for every $i$  such that ``outgoing'' disk of radius $r_i$ from $q_i$ equals the incoming disk of radius $_i$ from $p$.
Then we have the equation
\[
m_p \circ (m_{q_1} \otimes \cdots \otimes m_{q_k}) = m_P
\]
where $P$ denotes the point in $\mr{Disk}_{s^1_1, \ldots, s^1_{j_1}, s^2_1, \ldots s^k_{j_k}; R}$ arising by ``composing" the disks.
\end{itemize}
A detailed treatment of this notion appears in section 4.8 of~\cite{CG1} and in section 2 of~\cite{ElliottSafronov}.

\begin{rmk}
A technical issue is to formulate in what sense the multiplication $m$ is a smooth function, which requires finding a setting where the codomain $\mr{Hom}$ has the necessary structure.
In \cite{CG1} it is shown that differentiable vector spaces provide a sufficient setting and it is shown that most topological vector spaces are differentiable vector spaces.
\end{rmk}

A dg Lie algebra can act by derivations on a $n$-disks algebra~$\cA$.
Here is the key example for this paper.

\begin{dfn}
An $n$-disks algebra $\cA$ with values in $\ZZ/2$-graded cochain complexes of differentiable vector spaces is {\em superconformal} if $\sconf(p,q|S)$ acts by derivations on~$\cA$, where $n = p+q$.
\end{dfn}

A conformal twist $\cQ$ for $\sconf(p,q|S)$ then determines a deformation of such a superconformal disks-algebra.

\begin{dfn}
Given a $\sconf(p,q|S)$-superconformal $p+q$-disks algebra $\cA$ and a conformal twist $\cQ$,
let $\cA^\cQ$ denote the {\em $\cQ$-twisted} $n$-disks algebra whose underlying super vector space are those of $\cA$ but whose differential is $\d_\cA + \cQ$, where $n = p+q$.
\end{dfn}

A natural source of examples is to take a (well-behaved) superconformal prefactorization algebra (or twist thereof) and ask what it assigns to disks inside $\RR^{p,q}$.
Lagrangian theories with superconformal symmetry, whether classical or quantum, thus give many examples.

\subsection{Obtaining Vertex Algebras or \texorpdfstring{$\bb E_k$}{Ek}-Algebras} \label{vertex_En_section}

Under some reasonable hypotheses, one can obtain a vertex algebra or an $\bb E_k$ algebra (i.e., algebra over the little $k$-disks operad) from the observables of a field theory,
whether encoded as a prefactorization algebra or as a disk-algebra.
Precise results appear in Chapter 5 of \cite{CG1} and Section 2 of \cite{ElliottSafronov}.
Here we will state their consequences.

We will explain the essential format by an example.
Let $\cQ$ be a square-zero element of the super Lie algebra $\mathfrak{sconf}(p,q|S)$.
Suppose that the stabilizer $Z_Q$ contains a subgroup of $\mr{ISO}(p,q)$ that is isomorphic to $\mr{ISO}(m,n)$.
For simplicity, suppose this subgroup is the isometries of an isometric embedding $j \colon \RR^{m,n} \hookrightarrow \RR^{p,q}$.

We can consider observables supported along this submanifold:
define $j^* \cF^\cQ$ on $\RR^{m,n}$ by
\[
j^* \cF^\cQ(U) = \lim_{U \subset V \subset \RR^{p,q}} \cF^\cQ(V),
\]
where $V$ runs over open subsets of $\RR^{p,q}$ and $\lim$ denotes the limit in the appropriate category.
(When working with cochain complexes, we mean an $\infty$-categorical limit and hence, in practice, a homotopy limit.)
This formula yields a precosheaf because if $U \subset U'$, then any open set $V \subset \RR^{p,q}$ that contains $U'$ also contains $U$, so we have a canonical map
\[
j^* \cF^\cQ(U) \to \cF^\cQ(V)
\]
and hence
\[
j^* \cF^\cQ(U) \to \lim_{U' \subset V \subset \RR^{p,q}} \cF^\cQ(V)
\]
by taking limits.
This formula also determines a prefactorization algebra on~$\RR^{m,n}$ where the structure map
\[
j^* \cF^\cQ(U) \otimes j^* \cF^\cQ(U') \to j^* \cF^\cQ(U \sqcup U') 
\] 
for any pair of disjoint open sets $U, U'$ in $\RR^{m,n}$ is defined as follows.
Fix an open set $W \subset \RR^{p,q}$ containing $U$ but disjoint from $U'$,
and fix an open set $W' \subset \RR^{p,q}$ containing $U'$ but disjoint from $W$ (and hence $U$).
Given an open set $V \subset \RR^{p,q}$ that contains $U \sqcup U'$, there is an inclusion of open sets 
\[
(W \cap V) \sqcup (W' \cap V) \subset V
\]
and hence a map
\[
j^* \cF^\cQ(U) \otimes j^* \cF^\cQ(U') \to \cF^\cQ(W \cap V) \otimes \cF^\cQ(W' \cap V) \to \cF^\cQ(V),
\]
where the second map is the structure map of $\cF^\cQ$ as a prefactorization algebra on~$\RR^{p,q}$.
Taking limits, we obtain a map 
\[
j^* \cF^\cQ(U) \otimes j^* \cF^\cQ(U') \to j^* \cF^\cQ(U \sqcup U') = \lim_{U \sqcup U' \subset V \subset \RR^{p,q}} \cF^\cQ(V)
\]
as claimed.

\begin{rmk}
Even if $\cF^\cQ$ is a factorization algebra, this pullback $j^* \cF^\cQ$ is likely {\em not} a factorization algebra,
as the limit need not play well with the colimits that appear in the local-to-global condition (i.e., codescent).
\end{rmk}

\subsubsection{Obtaining \texorpdfstring{$\bb E_k$}{Ek}-Algebras}
Let us discuss the circumstances under which the restricted twisted theory $j^* \cF^\cQ$ on an affine subspace $\RR^{m,n}$ given by the embedding $j$ is topological.  Corollary 2.30 of \cite{ElliottSafronov} gives two conditions that guarantee that $j^* \cF^\cQ$ determines an algebra over the little $m+n$-disks operad -- or more colloquially, that the twist looks like a topological field theory along $\RR^{m,n}$.
These conditions are
\begin{itemize}
\item the sub-Lie algebra of translations inside the Lie algebra of $\mr{ISO}(m,n)$ are $\cQ$-exact and
\item any inclusion of disks in $\RR^{m,n}$ determines a quasi-isomorphism of observables on the disks.
\end{itemize}
This second condition follows if dilation is also $\cQ$-exact.

Corollary 2.39 applies to the case that $n = 0$. 
It says that one gets a {\em framed} $\bb E_{m}$-algebra if, furthermore, the subalgebra $\so(m)$ is $\cQ$-exact.
Such a situation appears not infrequently in Euclidean signature.
(An analog of this result should hold for mixed signature as well, using the techniques of~\cite{ElliottSafronov}.)

Let us make this application of the results of \cite{ElliottSafronov} more precise.

\begin{dfn} \label{Q_compatible_def}
A conformal embedding $j \colon \RR^{a,b} \hookrightarrow C(\RR^{p,q})$ defines a \emph{$\mc Q$-compatible subspace} if
\begin{itemize}
  \item $Z_\cQ$ preserves the image of $j$ and
  \item There is a subgroup $T_{\cQ}$ of $Z_\cQ$ isomorphic to $\RR^{a+b}$ whose action on $j(\RR^{a,b})$ corresponds to the translation on~$\RR^{a,b}$.
 \end{itemize}
\end{dfn}

Let $\obs$ be a prefactorization algebra on $C(\RR^{p,q})$, and let $\obs^{\cQ}$ denote its twist by the superconformal supercharge $\cQ$.  Using a choice of $\mc Q$-compatible subspace given by the embedding $j$, we obtain a prefactorization algebra $j^*\obs^{\cQ}$ on $\RR^{a,b}$ as discussed above.  The conditions we have placed on the embedding $j$ have been set up in order to ensure the existence of the following structure.  We first give a general definition.

\begin{dfn} \label{potential_def}
Let $\mf h \sub \sconf(p+q|S,\CC)_0$ be a Lie subalgebra.  A \emph{$\mc Q$-potential} for $\mf h$ is a linear subspace $\mf h' \sub \sconf(p+q|S,\CC)_1$ such that
\begin{itemize}
\item $[\mc Q, -]$ defines a linear isomorphism $\mf h' \to \mf h$.
\item $[\mf h', \mf h'] = 0$.
\item $[\mf h, \mf h'] \sub \mf h'$, and as an $\mf h$-representation $\mf h'$ is isomorphic to the adjoint representation.
\end{itemize}
\end{dfn}

\begin{rmk}
If $\mf h$ is a  Lie algebra, let $\mf h_{\mr{dR}}$ denote the contractible dg Lie algebra with underlying graded  Lie algebra $\mf h[1] \oplus \mf h$ and differential given by the identity.  Definition \ref{potential_def} says that given $\mf h \sub \sconf(p+q|S,\CC)_0$ we can extend to a commutative triangle of dg Lie algebras
\[\xymatrix{
 \mf h_{\mr{dR}} \ar[d] \ar[dr] & \\
 \mf h \ar[r] &(\sconf(p+q|S,\CC), [\mc Q, -]).
}\]
\end{rmk}

Now, Corollary 2.30 and Theorem 2.38 of \cite{ElliottSafronov} can be applied to the prefactorization algebra $j^*\obs^{\cQ}$ in the presence of appropriate potentials.  We obtain the following as a direct application of these results.

\begin{thm} \label{E_k_theorem}
Let $j \colon \RR^{a,b} \hookrightarrow C(\RR^{p,q})$ be a $\mc Q$-compatible subspace, and suppose that the group $T_{\cQ}$ from Definition \ref{Q_compatible_def} of translations of $j(\RR^{a,b})$ is contained in $B_\cQ \sub Z_\cQ$.  Let $G \sub B_{\cQ}$ be a subgroup acting on the image of $j$.  Let $\gg \sub \mf b^\CC_{\cQ}$ denote the complexified Lie algebra of $G$, and let $\mf g'$ be a $\mc Q$-potential for $\mf g$.  Finally, suppose that the factorization map $j^*\obs^{\cQ}(B_r(0)) \to j^*\obs^{\cQ}(B_R(0))$ associated to the inclusion of concentric balls is a quasi-isomorphism for all $0 < r < R$.  Then $j^*\obs^{\cQ}(B_1(0))$ canonically carries the structure of a $G \ltimes \bb E_{a+b}$-algebra.
\end{thm}

\begin{cor}[c.f. {\cite[Corollary 3.40]{ElliottSafronov}}] \label{E_k_cor_1}
If there is an embedding $\RR^\times \hookrightarrow G$ whose image acts on the image of $j$ by dilations, then the condition that $j^*\obs^{\cQ}(B_r(0)) \to j^*\obs^{\cQ}(B_R(0))$ is a quasi-isomorphism holds automatically.
\end{cor}

In particular we obtain framed $\bb E_a$-algebras in the pure signature case.

\begin{cor}[c.f. {\cite[Corollary 2.39]{ElliottSafronov}}] \label{E_k_cor_2}
If $b=0$, $G = \SO(a)$ acts on the image of $j$ by rotations, and the hypotheses of Theorem~\ref{E_k_theorem} hold,
then  $j^*\obs^{\cQ}(B_1(0))$ canonically carries the structure of a $\bb E^{\mr{fr}}_{a}$-algebra.
\end{cor}

\subsubsection{Obtaining Vertex Algebras} \label{VA_section}
Now let us address the circumstances under which we obtain a vertex algebra under twisting.  Let $\mc Q$ be a nilpotent supercharge, and let $j \colon \RR^2 \to C(\RR^{p,q})$ be a $\mc Q$-compatible subspace of definite signature.  We will impose the following requirements.
\begin{enumerate}
 \item The group $Z_{\cQ}$ contains a subgroup isomorphic to $\mr{ISO}(j(\RR^2))$ acting by isometries on the image of $j$.
 \item For a choice of complex coordinate $z$ on $\RR^2$, the complexified translation $\dd_{\ol z}$ is in $\mf b_{\cQ}^\CC$.
\end{enumerate}

Under these conditions we obtain the following result.

\begin{thm} 
\label{VA_theorem}
Let $\mc V = j^*\obs^\cQ(D_1(0))$ for $j$ and $\cQ$ as above, where $D_1(0) \sub \RR^2$ is the unit disk.  Let
\[V = \bigoplus_{k \in \ZZ} \mr H^\bullet(\mc V^k),\]
where $\mc V^k$ is the $k$-eigenspace of $\mc V$ for the action of the group of rotations.  Then $V$ can be equipped with the structure of a vertex algebra under the following conditions:
\begin{enumerate}
\item[(i)] For each $0 < r < R$, the prefactorization map $j^*\obs^\cQ(D_r(0)) \to j^*\obs^\cQ(D_R(0))$ restricts to a quasi-isomorphism on every eigenspace for the $S^1$-action.
\item[(ii)] $\mr H^\bullet(\mc V^k) = 0$ for all $k > K$ for some integer $K$.
\item[(iii)] The eigenspaces $\mr H^\bullet(\mc V^k)$ satisfy a category-theoretic tameness condition.
\end{enumerate}

\end{thm}

This result is be a direct consequence of \cite[Theorem 2.2.1]{CG1}, and we refer the reader to that reference for a discussion of the meaning of the technical conditions that appear in the statement of this theorem.  In particular, the ``tameness'' condition that we have not spelled out here is condition (iii) in~\emph{loc. cit.}

\subsection{The Approach of Saberi--Williams to Schur twists}
\label{sec: Saberi Williams}

Saberi and Williams \cite{SaberiWilliams} take a different approach to understanding the occurrence of vertex algebras from Schur twists of 4d $\mc N=2$ theories (in Euclidean signature),
an approach that is fundamentally constructive and hence produces a number of explicit examples.
They focus on factorization algebras built with the methods of \cite{CG1,CG2},
and so their examples arise as the observables of perturbative supersymmetric theories or as current algebras (i.e., local symmetries) of such theories.
Here we will review their construction and mention some key results,
and then we will explain how their rich physical examples offer examples of the framework of this paper.

Saberi and Williams follow a two step process:
\begin{itemize}
\item first, twist by a chiral element $\cQ$ to obtain a holomorphic field theory on~$\CC^2$, and
\item second, deform the differential by the holomorphic vector field~$z_2 \frac{\dd}{\dd \eps}$,
which corresponds $\mc S^{\dot +2}$ in this complex structure determined by~$\mc Q$.
\end{itemize}
For factorization algebras that arise as observables of Lagrangian 4d $\mc N=2$ theories or as local symmetries,
they show explicitly that these deformed and twisted factorization algebras ``localize'' onto the plane 
\[
\{z_2 = 0\} = \CC \times \{0\}
\]
inside~$\CC^2$.
We now turn to describing some key examples in their work.

For step one, start with any holomorphic twist by a supercharge $\mc Q$, 
initially not in the superconformal sense but purely in the super translation sense.  
Then a first result is the following symmetry enhancement for the observables of supersymmetric field theory.

\begin{prp}[{\cite[Proposition 4.7]{SaberiWilliams}}]
For an $\mc N=2$ supersymmetric classical $\gg$-gauge theory on $\RR^4$,
its twist by a translation supercharge $\mc Q$ admits an action of the local Lie algebra
\[
\mf X_2 = \Omega^{0,\bullet}(\CC^2, T_{\CC^{2|1}})\]
of holomorphic vector fields on $\CC^{2|1}$,
enhancing the action of the (twisted) super Poincar\'e algebra.
\end{prp}

This local Lie algebra $\mf X_2$ generates a current algebra on $\CC^2$ that is a kind of two-dimensional analog of the Virasoro algebra;
there are local cocycles analogous to the central extension (or Schwinger term) that determines the Virasoro algebra.
(Perhaps some readers may prefer to view it as a two-dimensional analog of the $N=2$ superconformal algebra, rather than the Virasoro algebra itself, since it involves a derived direction.)
This current algebra is an example of an {\em enveloping factorization algebra},
and it is a central example for their work.
(See Section~5.4 of \cite{SaberiWilliams} for details.)
This proposition thus ensures that a twisted 4d $\mc N=2$ gauge theory (even after quantization) has an enormous symmetry algebra via a two-dimensional version of the Virasoro algebra.

There are also analogs on $\CC^2$ of the affine vertex algebras arising from the well-known central extensions of the loop Lie algebras~$\fg[z,z^{-1}]$.
Such higher Kac-Moody factorization algebras often appear, in twisted theories, as enhanced symmetries of global symmetries.

Saberi and Williams then consider the \emph{deformation} of this $\mc Q$-twisted theory, or its symmetries,
by the action of the holomorphic vector field~$z_2 \frac{\dd}{\dd \eps}$.  
That is, they modify the differential on these factorization algebras by including this vector field as part of the differential.
This further deformation corresponds to the Schur twist
\[\bb Q = \mc Q_{+1} + \wt{\mc S}^{\dot +2}\]
discussed by us in Section~\ref{Beemetal}.
We state first what this deformation does to the symmetries (i.e., current algebras).

\begin{thm}[{\cite[Theorem 1.1]{SaberiWilliams}}]
The deformation of the $\mc N=2$ Virasoro factorization algebra on $\CC^2$ is {\em stratified} so that 
\begin{itemize}
\item it is trivial in the complement of the $\{z_2 = 0\}$ plane and 
\item its restriction to the $\{z_2 = 0\}$ plane recovers the Virasoro vertex algebra.
\end{itemize}
An analogous result holds for the $\mc N=2$ Kac-Moody factorization algebras on~$\CC^2$:
the deformation is stratified and recovers the universal affine vertex algebras.
\end{thm}

Notice that this further deformation {\em localizes} a factorization algebra on $\CC^2$ so that it ``lives'' on the plane~$\{z_2=0\}$.
In the complement of the plane, there is no interesting structure.
Their proof boils down to explicit resolutions familiar to anyone who has taken an introductory course in algebraic geometry: 
these local Lie algebras can be described using Dolbeault complexes for (graded) holomorphic vector bundles on $\CC^2$,
and adding $z_2 \frac{\dd}{\dd \eps}$ to the differential turns these Dolbeault complexes into a Koszul resolution of the $z_1$-plane inside~$\CC^2$.
(One appealing aspect of their approach is that it's clear how to use similar resolutions to obtain localizations onto subvarieties of complex manifolds in many other situations with holomorphic field theories.
That is, their techniques generalize substantially.)

This gloss of their result suppresses an important accomplishment of Saberi-Williams: 
they match central charges and levels, recovering exactly the correspondence of Beem {\it et al}.

In Section~7, Saberi and Williams also analyze a field theory: 
$\mc N=2$ supersymmetric QCD.
Their main result is the following.

\begin{thm}
Consider the holomorphic twist of $\mc N=2$ supersymmetric QCD with group $SU(N_c)$ and $N_f$ flavors of quarks (i.e., the matter content is built using $N_f$ copies of the fundamental representation).
The deformation by $z_2 \frac{\dd}{\dd \eps}$ can be quantized only if $N_f = 2N_c$.
In this case, the quantum observables form a stratified factorization algebra on~$\CC^2$:
\begin{itemize}
\item it is trivial in the complement of the $\{z_2 = 0\}$ plane and 
\item its restriction to the $\{z_2 = 0\}$ plane recovers BRST reduction by $\sl(N_c)$ of the $\beta\gamma$ vertex algebra for the quark representation.
\end{itemize}
\end{thm}

Their techniques apply to essentially all the Lagrangian $\mc N=2$ supersymmetric field theories.
Throughout \cite{SaberiWilliams} there are detailed computations that match the results of Beem {\it et al}.

We now point out that their examples match the hypotheses of Theorem~\ref{VA_theorem},
where $j$ embeds $\CC$ into $\CC^2$ by $j(z) = (z,0)$:
the vertex algebras they produce by ``localization'' are examples of our abstract result.
In particular, we see that many examples of Beem {\it et al} fit into our framework,
thanks to~\cite{SaberiWilliams}.

\printbibliography

\end{document}